\documentclass[aps,superscriptaddress,nofootinbib,floatfix,notitlepage,preprintnumbers]{revtex4-2}
\usepackage[utf8x]{inputenc}
\pdfoutput=1
\usepackage{graphicx}
\usepackage{hyperref}
\usepackage{bm}
\usepackage{multirow}
\usepackage{array,booktabs,colortbl,multirow}
\usepackage{colortbl,colordvi,color}
\usepackage{rotating}
\usepackage{placeins}
\usepackage{amsmath}
\usepackage{epigraph}
\usepackage[normalem]{ulem}
\usepackage[table,dvipsnames]{xcolor}

\newcommand{\be}{\begin{equation}}
\newcommand{\ee}{\end{equation}}
\newcommand{\bea}{\begin{eqnarray}}
\newcommand{\eea}{\end{eqnarray}}

\usepackage[sort&compress]{natbib}

\begin{document}
\raggedbottom

\preprint{MITP-26-030}

%\title{Another Critical Look at Rare \boldmath$B$-meson Semileptonic Decays}
\title{A Dispersive Look at Rare \boldmath$B$-meson Semileptonic Decays}

\author{M.~Ciuchini}
\affiliation{INFN, Sezione di Roma Tre, Via della Vasca Navale 84, I-00146 Rome, Italy}

\author{M.~Fedele}
\affiliation{PRISMA$^{++}$ Cluster of Excellence \& Mainz Institute for Theoretical Physics,
Johannes Gutenberg University, D-55099 Mainz, Germany}

\author{A.~Paul}
\affiliation{The Institute of Experiential AI, Northeastern University, 360 Huntington Ave., Boston, MA 02115, USA}
\affiliation{Channing Division of Network Medicine, Harvard Medical School, 188 Longwood Ave., Boston, MA 02115, USA}

\author{J.~Scholze}
\affiliation{PRISMA$^{++}$ Cluster of Excellence \& Mainz Institute for Theoretical Physics,
Johannes Gutenberg University, D-55099 Mainz, Germany}

\author{L.~Silvestrini}
\affiliation{INFN, Sezione di Roma, Piazzale A. Moro 2, I-00185 Roma, Italy}

\author{S.~Simula}
\affiliation{INFN, Sezione di Roma Tre, Via della Vasca Navale 84, I-00146 Rome, Italy}

\author{M.~Valli}
\affiliation{INFN, Sezione di Roma, Piazzale A. Moro 2, I-00185 Roma, Italy}

\author{L.~Vittorio}
\affiliation{INFN, Sezione di Roma, Piazzale A. Moro 2, I-00185 Roma, Italy}
\affiliation{Physics Department, Universit\`a di Roma La Sapienza, Piazzale A. Moro 2, I-00185 Roma, Italy}

\begin{abstract}
Rare semileptonic $b \to s$ flavour-changing neutral current transitions provide stringent tests of the Standard Model. Their interpretation is limited by hadronic uncertainties, notably the $B \to K^{(*)}$ and $B_s \to \phi$ form factors (FFs) and the matrix elements of four-quark operators. We perform a global analysis of $b \to s \ell^+\ell^-$ transitions taking these uncertainties fully into account, determining the FFs through the Dispersive Matrix method and comparing a setup based solely on lattice QCD (LQCD) with one that also includes light-cone sum-rule (LCSR) inputs at low $q^2$. Compared to the case where both input are taken into account, using only LQCD substantially enlarges the FF uncertainties at large recoil. Combined with the latest LHCb and CMS angular measurements sensitive to strong phases, our global fit yields strengthened evidence in favour of long-distance hadronic effects rather than a short-distance shift in $C_9$. We further present new SM predictions for the theoretically clean $b \to s \nu\bar\nu$ modes, which depend only on local FFs, and a New Physics analysis of these transitions in the Weak Effective Theory, discussing their impact on the interpretation of the recent Belle~II measurement and on the available experimental upper bounds.
\end{abstract}

\maketitle

\section{Introduction}

Rare flavour-changing neutral current (FCNC) transitions of the type $b \to s \ell^+ \ell^-$ provide some of the most stringent tests of the Standard Model (SM). Being loop-suppressed, these processes are highly sensitive to short-distance virtual effects of heavy new particles, offering a privileged window onto physics beyond the electroweak scale. Over the past decade, increasingly precise measurements of branching fractions and angular observables in $B \to K^{(*)} \mu^+ \mu^-$ and $B_s \to \phi \mu^+ \mu^-$ decays have subjected the SM to stringent scrutiny. These data have indeed generated sustained theoretical and experimental attention and have often been interpreted as possible hints of New Physics (NP).

However, the theoretical interpretation of these processes relies on our still imperfect understanding of hadronic matrix elements. Non-perturbative QCD dynamics -- encoded in both local FFs and non-local hadronic effects such as charm-loop contributions -- can affect branching fractions and angular observables, thereby closely mimicking short-distance NP contributions. The role of power corrections to QCD factorization, and in particular of the so-called charming penguins, in $B\to K^{(*)}\ell^+\ell^-$ and $B_s \to \phi \ell^+ \ell^-$ has been analysed in detail in a series of papers \cite{Khodjamirian:2010vf,Jager:2012uw,Lyon:2014hpa,Jager:2014rwa,Descotes-Genon:2014uoa,Ciuchini:2015qxb,Ciuchini:2016weo,Ciuchini:2017mik,Chobanova:2017ghn,Bobeth:2017vxj,Ciuchini:2018anp,Arbey:2018ics,Ciuchini:2019usw,Ciuchini:2020gvn,Gubernari:2020eft,Hurth:2020rzx,Ciuchini:2021smi,Ciuchini:2022wbq,Bordone:2024hui,Hurth:2025neo,Hurth:2025vfx,Altmannshofer:2026cwk}. On the other hand, local FFs have often been considered a minor source of uncertainty, especially in the context of optimized observables \cite{Kruger:2005ep,Matias:2012xw,Descotes-Genon:2013vna}. However, in the presence of power corrections to the infinite $b$-mass limit, and in particular of charming penguins, FF uncertainties do not actually cancel \cite{Jager:2012uw} and may become sizeable. Therefore, to provide a comprehensive and conservative assessment of potential NP effects in $b \to s \ell^+ \ell^-$, this work investigates in detail the role of FF uncertainties in (optimized) angular observables, critically assessing the impact of Light-Cone Sum Rules (LCSR) calculations. Indeed, while for $B \to K$ FFs Lattice QCD (LQCD) provides a first-principle determination now available across the full kinematic range, for $B \to K^*$ and $B_s \to \phi$ LQCD results are restricted to low recoil (large dilepton invariant mass $q^2$) and neglect the decay of the final-state vector meson. Consequently, LCSR calculations remain pivotal in computing FFs at large recoil, where the tensions between model-dependent theoretical predictions and experimental results are most pronounced. 

In virtually all previous global analyses -- see, e.g., Refs.~\cite{Ciuchini:2022wbq,Greljo:2022jac,Alguero:2023jeh,Guadagnoli:2023ddc,Bordone:2024hui,Hurth:2025vfx} for the most recent studies -- both LCSR and LQCD inputs have been adopted for decays to vector meson final states. To conservatively assess the FF uncertainty, in this work we systematically compare this ``standard'' approach to an alternative strategy where only LQCD results are used and extrapolated to the full kinematic range by implementing analyticity and unitarity constraints~\cite{Meiman63, Okubo:1971jf, Okubo:1971wup, Okubo:1971my} via the Dispersive Matrix (DM) method~\cite{Bourrely:1980gp,Lellouch:1995yv,DiCarlo:2021dzg}.

As detailed below, relying solely on LQCD inputs considerably enlarges the FF uncertainties in the large-recoil region. Since these uncertainties do not cancel in optimized observables once charming penguins are properly accounted for, such an inflation further complicates the extraction of hadronic matrix elements from experimental data. A first look at the impact of this choice is provided in the left panel of Figure~\ref{fig:hm1_hm2}, where the role of the LCSR input in constraining the charming-penguin contributions -- including what can be interpreted as a universal NP effect -- is clearly visible, resulting in a significantly larger allowed space for hadronic effects.

On the experimental side, an extensive suite of measurements by the LHCb, Belle, ATLAS, and CMS collaborations~\cite{LHCb:2013zuf,LHCb:2013tgx,LHCb:2013ghj,CMS:2013mkz,LHCb:2014cxe,CMS:2014xfa,LHCb:2015wdu,LHCb:2015svh,Belle:2016fev,LHCb:2017rmj,ATLAS:2018cur,Belle:2019oag,CMS:2019bbr,LHCb:2020lmf,CMS:2020oqb,LHCb:2020gog,LHCb:2021zwz,LHCb:2021xxq,LHCb:2021vsc,LHCb:2022qnv,LHCb:2022vje,CMS:2022mgd} has already provided a detailed picture of branching fractions, angular observables, and lepton-flavour universality ratios in $b \to s \ell^+\ell^-$ transitions. However, two recent comprehensive updates call for an updated phenomenological analysis. The CMS collaboration~\cite{CMS:2024atz}, analysing $140\,\mathrm{fb}^{-1}$ of data, has provided updated measurements of angular observables in $B \to K^* \mu^+ \mu^-$. Furthermore, the LHCb collaboration~\cite{LHCb:2025mqb} has released a milestone analysis of $B^0 \to K^{*0} \mu^+ \mu^-$ based on the full $8.4\,\mathrm{fb}^{-1}$ Run~1 and Run~2 dataset. This study delivers the most precise determinations to date of $CP$-averaged angular observables, $CP$ asymmetries, and differential branching fractions, while systematically accounting for $K\pi$ $S$-wave contributions and low-$q^2$ lepton-mass effects. As we demonstrate below, these comprehensive datasets provide several hints of sizeable penguin matrix elements, for example in the low -$q^2$ bins of the $B \to K \mu^+ \mu^-$ differential branching fraction or in those $B \to K^* \mu^+ \mu^-$ angular observables that are sensitive to strong phases, as recently noticed in Ref.~\cite{Altmannshofer:2026cwk}.

The Belle~II collaboration recently found evidence for the invisible decay $\mathcal{B}(B^+ \to K^+ \nu \bar{\nu}) = (2.3 \pm 0.7) \times 10^{-5}$ based on $362\,\mathrm{fb}^{-1}$ of data~\cite{Belle-II:2023esi}, which was obtained by treating the long-distance contribution $B^+\to\tau^+(\to K^*\bar\nu)\nu$~\cite{Kamenik:2009kc} as a background. This value corresponds to a $3.5\sigma$ signal with an excess of approximately $2.7\sigma$ over earlier SM expectations~\cite{Parrott:2022zte,Becirevic:2023aov}, prompting us to extend our analysis to include $b \to s \nu \bar{\nu}$ transitions as well. Unlike the charged-lepton modes, these purely electroweak decays are entirely free from charming-penguin contamination and depend exclusively on local FFs. We find that the dispersive FFs yield a shifted and more conservative SM prediction for the $B \to K^* \nu \bar{\nu}$ decays.

By adopting the DM method, incorporating a conservative parameterization of non-local charm dynamics, and leveraging the unprecedented precision of the latest measurements in both $b \to s \ell^+ \ell^-$ and $b \to s \nu \bar{\nu}$ transitions, this work provides a systematic and conservative analysis of rare semileptonic $B$ decays. The remainder of this paper is organized as follows. In Sec.~\ref{sec:DM_FF} we detail our computation of $B\to K$, $B\to K^*$ and $B_{s}\to \phi$ FFs obtained employing the DM approach. 
The phenomenological impact of these results on $b\to s\nu\bar\nu$ and $b\to s\ell^+\ell^-$ transitions is discussed in Sec.~\ref{sec:SM_updates}, where we present updated SM predictions, 
and in Sec.~\ref{sec:NP_updates}, where we investigate their implications for NP scenarios. Our conclusions are given in Sec.~\ref{sec:conclusions}, while technical details 
regarding the FF determination are gathered in Appendices~\ref{app:AppA} and~\ref{app:AppB}, additional fit results presented in Appendices~\ref{app:AppC} and~\ref{app:AppD}.

\section{Dispersive Matrix Form Factors for \boldmath $B\to K,K^*$ and $B_{s}\to \phi$}
\label{sec:DM_FF}

\subsection{Concise review of the Dispersive Matrix method}

Let us start by briefly reviewing the main properties of the non-perturbative DM approach of Ref.~\cite{DiCarlo:2021dzg} to the description of the hadronic FFs, based on the pioneering works in Refs.~\cite{Bourrely:1980gp, Lellouch:1995yv}.

By looking at a generic FF $f$, we can write down a dispersion bound of the form~\cite{Bourrely:1980gp, Lellouch:1995yv} 
\be
\label{eq:JQ2z}
\frac{1}{2\pi i } \oint_{\vert z\vert =1} \frac{dz}{z}   \vert\phi(z) f(z)\vert^2 \leq \chi\, , 
\ee
where $\phi(z)$ is an outer function depending on the spin-parity quantum channel and $\chi$ is related to the derivative of the Fourier transform of a suitable Green function of bilinear quark operators~\cite{Boyd:1997kz}. Hereafter, we will refer to the quantity $\chi$ as \emph{susceptibility}. The conformal variable $z$ in Eq.\,(\ref{eq:JQ2z}) is defined as
\be
     \label{eq:z}
     z \equiv z(t, t_0) \equiv \frac{\sqrt{t_+ - t} - \sqrt{t_+ - t_0}}{\sqrt{t_+ - t} + \sqrt{t_+ - t_0}} ~ , ~
\ee
where $t \equiv q^2 $ is the squared momentum transfer, $t_\pm \equiv (m_{B_{(s)}} \pm m_{P,\,V})^2$ for $B_{(s)} \to \{P,\,V\}\ell^+\ell^-$ decays\footnote{ Let us stress that, contrary to what was done in Ref.\,\cite{Gubernari:2023puw}, we do not choose $t_+ = (m_{B_s}+m_{\pi^0})^2$ since we do not include isospin violating effects in our study. Furthermore, in the cases of semileptonic $B \to K^*$ and $B_s \to \phi$ decays, we note that the ratio $\delta \equiv (t_+ - t_{\rm th})/t_{\rm th}$ is equal respectively to $\sim$0.1 and $\sim$0.2, where we have defined $t_{\rm th} \equiv (m_B + m_K)^2$. Thus, by following the argument of Ref.\,\cite{Gopal:2024mgb}, we can argue that sub-threshold effects do not give here an important contribution as the ratio $\delta$ is never an $\mathcal{O}(1)$ number. Recently, a numerical study of sub-threshold effects in hadronic form factors has been performed in Ref.\,\cite{Gubernari:2026sqc} and an alternative approach to take into account sub-threshold effects has been proposed in Ref.\,\cite{Simula:2025fft}.} and, finally, $t_0 < t_+$ is an auxiliary variable, which fixes the value of $t$ at which $z(t_0, t_0) = 0$. For the purposes of the present discussion, we simply set $t_0 = t_-$. Let us finally note that hypothetical sub-threshold poles have to be properly taken into account by modifying the outer function as
\begin{equation}
\label{eq:poles}
\phi(z) \to \phi_{p}(z) \equiv \phi(z) \times \frac{z-z(m_{P1}^2)}{1-\bar{z}(m_{P1}^2)z}  \times \cdots  \times  \frac{z-z(m_{PN}^2)}{1-\bar{z}(m_{PN}^2)z}.
\end{equation}
where we are \textit{e.g.} assuming the existence of $N$ poles for $f$ (clearly $m_{Pj} < t_+$ $\forall j=1,\cdots,N$). %The values of the masses of these poles are the same ones contained in Table \ref{tab:poles}.

By defining the inner product~\cite{Bourrely:1980gp,Lellouch:1995yv}
\be
 \label{eq:inpro}
\langle g\vert h\rangle =\frac{1}{2\pi i } \oint_{\vert z\vert=1 } \frac{dz}{z}   \bar {g}(z) h(z)\, , 
\ee
where $\bar{g}(z)$ is the complex conjugate of the function $g(z)$, we thus write Eq.~(\ref{eq:JQ2z}) as
\be
\label{eq:JQinpro}
0 \leq \langle \phi f \vert \phi  f\rangle \leq \chi\, .
\ee
Following Refs.~\cite{Bourrely:1980gp,Lellouch:1995yv}, we then introduce the set of functions
\be
g_t(z) \equiv \frac{1}{1-\bar{z}(t) z}\, , \nonumber
\ee
where $\bar{z}(t)$ is the complex conjugate of $z(t)$, so that we can associate to the generic FF $f$ the matrix~\cite{Bourrely:1980gp, Lellouch:1995yv, DiCarlo:2021dzg}
\be
\label{eq:Delta}
{\tiny
\mathbf{M} \equiv \left(
\begin{array}{ccccc}
\langle\phi f | \phi f \rangle  & \langle\phi f | g_t \rangle  & \langle\phi f | g_{t_1} \rangle  &\cdots & \langle\phi f | g_{t_N}\rangle  \\[2mm]
\langle g_t | \phi f \rangle  & \langle g_t |  g_t \rangle  & \langle  g_t | g_{t_1} \rangle  &\cdots & \langle g_t | g_{t_N}\rangle  \\[2mm]
\langle g_{t_1} | \phi f \rangle  & \langle g_{t_1} | g_t \rangle  & \langle g_{t_1} | g_{t_1} \rangle  &\cdots & \langle g_{t_1} | g_{t_N}\rangle  \\[2mm]
\vdots & \vdots & \vdots & \vdots & \vdots \\[2mm] 
\langle g_{t_N} | \phi f \rangle  & \langle g_{t_N} | g_t \rangle  & \langle g_{t_N} | g_{t_1} \rangle  &\cdots & \langle g_{t_N} | g_{t_N} \rangle 
\end{array} \right)  ~ , ~
}
\ee
which is the starting point of the numerical implementation of the DM method. Here $t_1, \ldots, t_N$ are the values of the squared 4-momentum transfer at which the FF $f$ has been computed on the lattice or using light-cone sum rules. 

By recalling the property of positivity of the inner products defined in Eq.\,(\ref{eq:inpro}), one can demonstrate that the determinant of the matrix $\mathbf{M}$ is, by construction, positive semidefinite, namely\,$\det \mathbf{M} \geq 0$. Thus, noting also that $z$, $\phi_p(z)$ and $f(z)$ can assume only real values in the allowed kinematical region for semileptonic decays $4 m_{\ell}^2 \leq t \leq t_-$, the original matrix~\eqref{eq:Delta} can be replaced by
\be
\label{eq:Delta2}
{\small
\mathbf{M}_{\chi} = \left( 
\begin{tabular}{ccccc}
   $\chi$ & $\phi_p f$                            & $\phi_1 f_1$                             & $...$ & $\phi_N f_N$ \\[2mm]
   $\phi_p f$     & $\frac{1}{1 - z^2}$     & $\frac{1}{1 - z z_1}$      & $...$ & $\frac{1}{1 - z z_N}$ \\[2mm]
   $\phi_1 f_1$ & $\frac{1}{1 - z_1 z}$  & $\frac{1}{1 - z_1^2}$     & $...$ & $\frac{1}{1 - z_1 z_N}$ \\[2mm]
   $... $  & $...$                           & $...$                              & $...$ & $...$ \\[2mm]
   $\phi_N f_N$ & $\frac{1}{1 - z_N z}$ & $\frac{1}{1 - z_N z_1}$ & $...$ & $\frac{1}{1 - z_N^2}$
\end{tabular}
\right) ~ , ~
}
\ee
where $\phi_i f_i \equiv \phi_p(z_i) f(z_i)$ (with $i = 1, 2, ..., N$) are the known non-perturbative values of $\phi_p(z) f(z)$. 

The condition of positivity of the determinant of the matrix~\eqref{eq:Delta2} acts as a constraint for the form factor $f(z)$ at a generic (real) value of $z$.
The (uniform) band of the allowed values of $f(z)$ is explicitly given by~\cite{DiCarlo:2021dzg}
\be
  \label{eq:bounds}
  \beta(z) - \sqrt{\gamma(z)} \leq f(z) \leq \beta(z) + \sqrt{\gamma(z)} ~ , ~
\ee 
where we have introduced
\bea
      \label{eq:beta_final}
      \beta(z) & \equiv & \frac{1}{\phi_p(z) d(z)} \sum_{j = 1}^N \phi_j f_j d_j \frac{1 - z_j^2}{z - z_j} ~ , ~ \\
      \label{eq:gamma_final}
      \gamma(z) & \equiv &  \frac{1}{1 - z^2} \frac{1}{\phi_p^2(z) d^2(z)} \left( \chi - \chi_\text{DM} \right) ~ , ~ \\
      \label{eq:chi0_final}
      \chi_\text{DM} & \equiv & \sum_{i, j = 1}^N \phi_i f_i \phi_j  f_j d_i d_j \frac{(1 - z_i^2) (1 - z_j^2)}{1 - z_i z_j} ~ , ~ \\
      \label{eq:d0}
     d(z) & \equiv & \prod_{m = 1}^N \frac{1 - z z_m}{z - z_m}  ~ , ~ \\
     \label{eq:di}
     d_j & \equiv & \prod_{m \neq j = 1}^N \frac{1 - z_j z_m}{z_j - z_m}  ~ . ~ 
\eea
Unitarity is satisfied only when $\gamma(z) \geq 0$, which implies $\chi \geq \chi_\text{DM}$. In this way, the input data are filtered by unitarity and, by construction, only the subset of input data satisfying exactly the unitary filter $\chi \geq \chi_\text{DM}$ will be considered.

Operationally, starting from a given input dataset for the FF $f(z)$, one may generate a sample of events (of the order of $10^5$) through an appropriate multivariate distribution and, then, evaluate the quantities in Eqs.(\ref{eq:beta_final}-\ref{eq:di}) event-by-event. For each event, the FF can be considered uniformly distributed between the lower and the upper bounds, defined in Eq.(\ref{eq:bounds}) at any value of $z$. Then, it is necessary to average over the whole set of generated events to obtain the final mean values and uncertainties of $f(z)$. Since the lower and the upper bounds of $f(z)$ are, in general, strongly correlated, one may adopt a multivariate distribution to describe them, following the procedure described in Section V.C of Ref.~\cite{DiCarlo:2021dzg}. Let us emphasize (see also Ref.~\cite{Simula:2025lpc}) that the DM procedure is equivalent to the envelope of the results of all possible Boyd-Grinstein-Lebed (BGL)~\cite{Boyd:1997kz} $z$-expansions, which satisfy unitarity and at the same time reproduce  exactly the input unitary data\footnote{A rigorous procedure for the implementation of kinematical constraints among different FFs within the DM approach can be found in Ref.~\cite{DiCarlo:2021dzg}.}. This implies that the DM band is rigorously truncation-independent. 

In what follows, we will focus on three FFs, namely $f_+(q^2),\,f_0(q^2),\,f_T(q^2)$, for the $B \to K \ell^+ \ell^-$ channel and on seven FFs, $i.e.$ $A_0(q^2)$,\,$A_1(q^2)$,\,$A_{12}(q^2)$,\,$T_2(q^2)$,\,$T_1(q^2)$,\,$T_{23}(q^2)$,\,$V(q^2)$ for the $B \to K^* \ell^+ \ell^-$ and $B_s \to \phi \ell^+ \ell^-$ ones. The definitions of the hadronic FFs and of the kinematical functions associated to each FF, as well as the numerical values adopted for the susceptibilities $\chi$ and the sub-threshold pole masses, are collected in Appendix \ref{app:AppA}. 

\begin{figure}[!t!]
    \centering
    \includegraphics[width=0.32\linewidth]{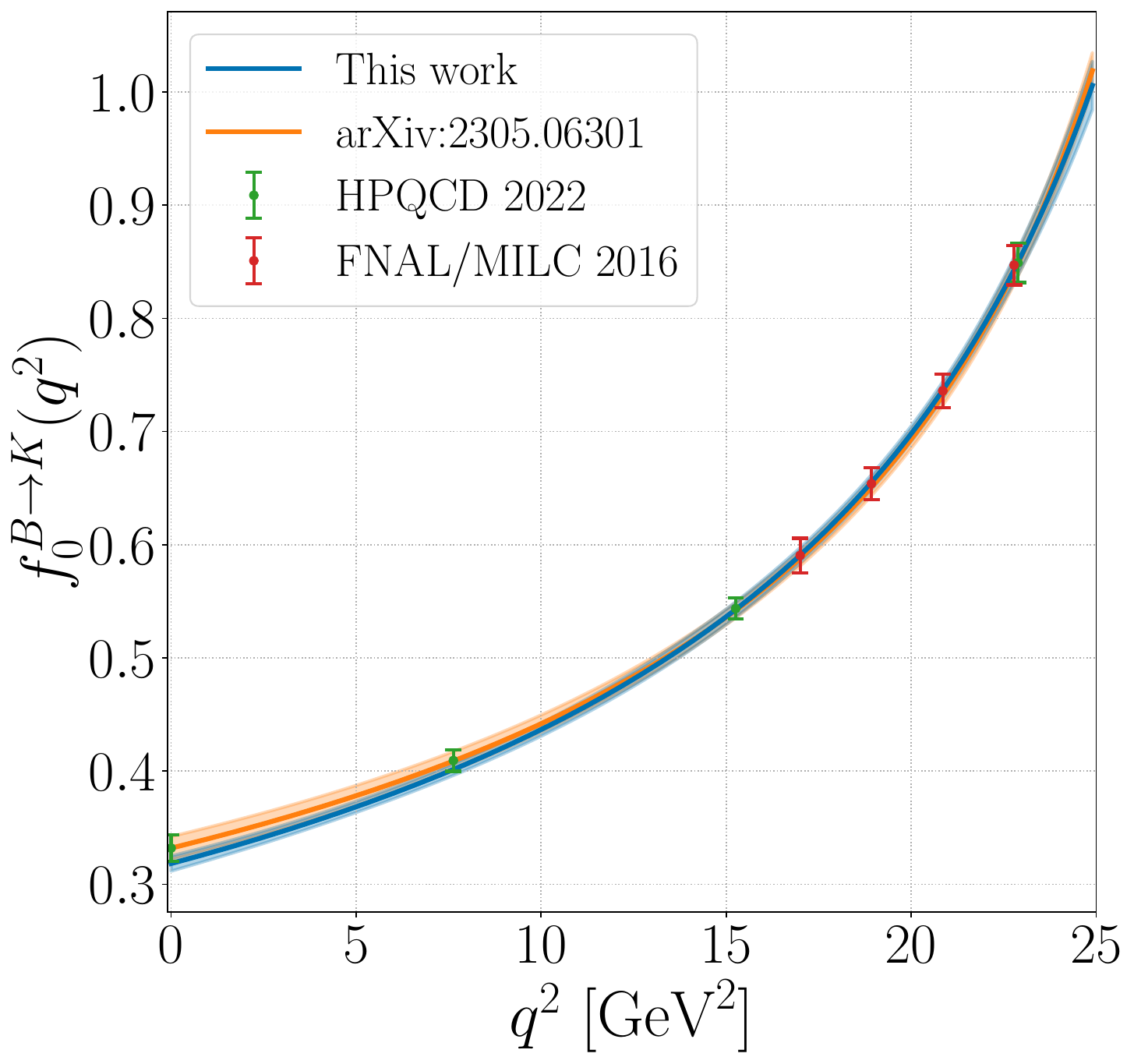}
    \hspace{0.008\linewidth}
    \includegraphics[width=0.32\linewidth]{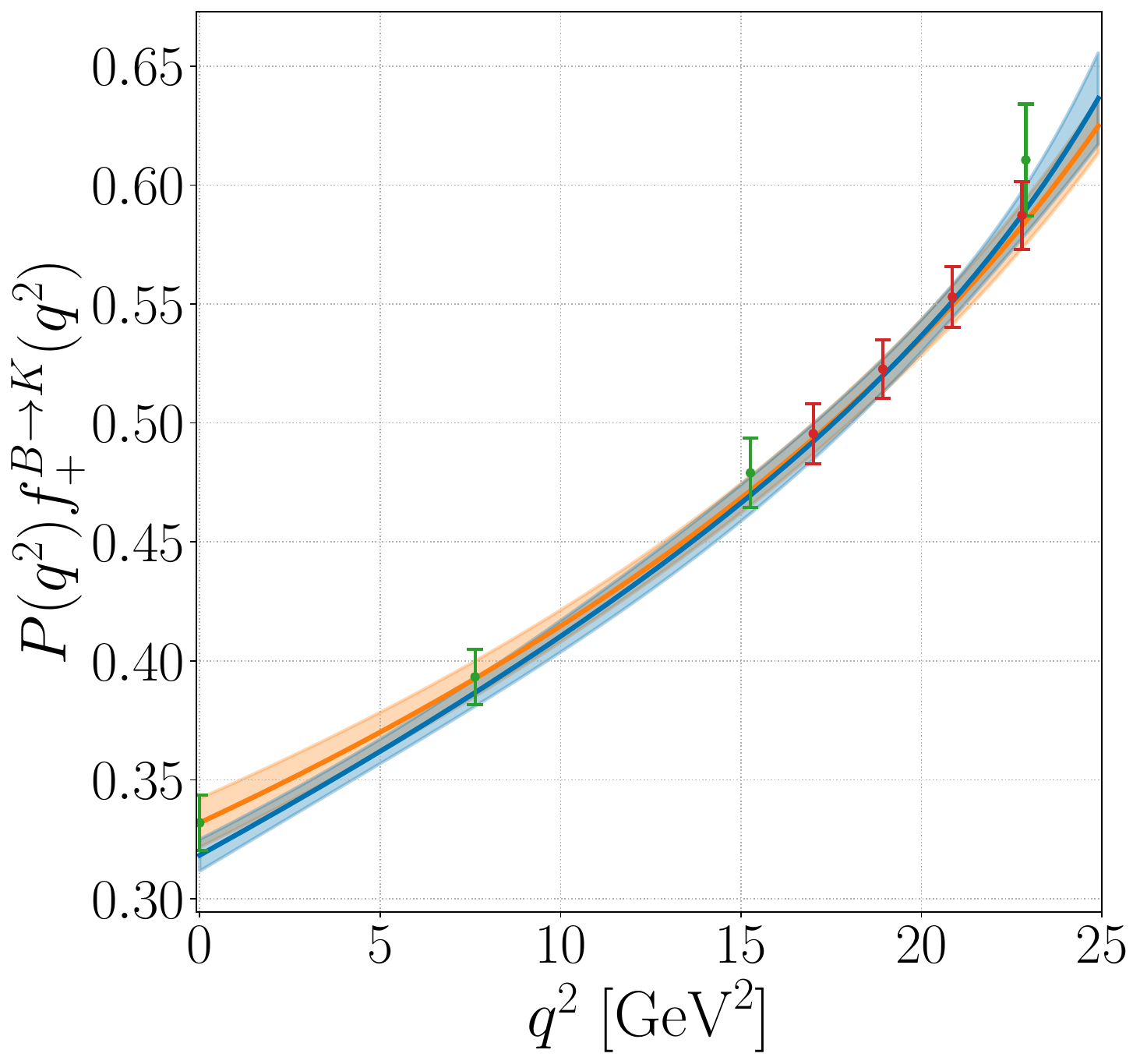}
    \hspace{0.008\linewidth}
    \includegraphics[width=0.32\linewidth]{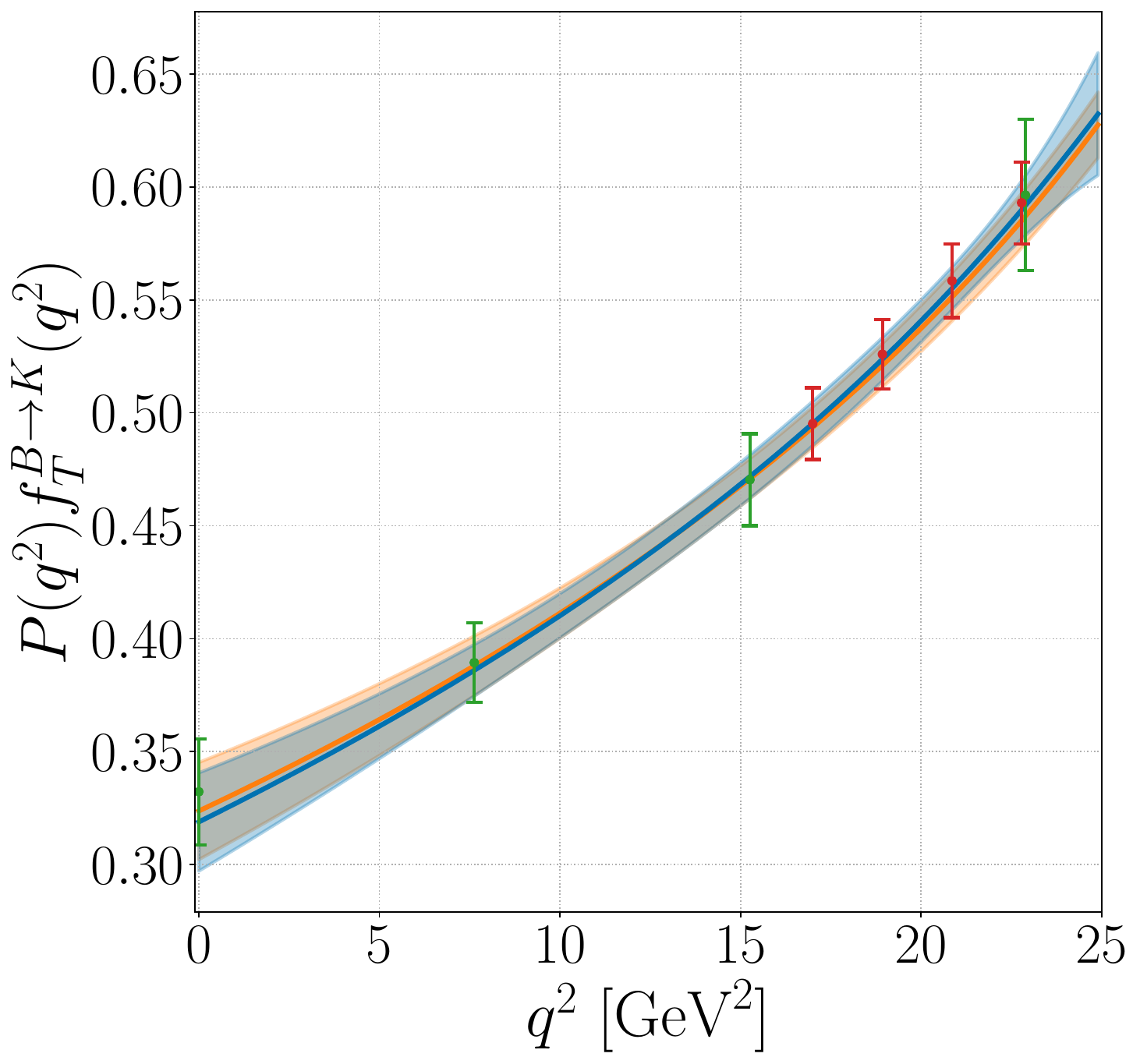}
    \caption{Comparison of the FFs entering the $B\to K$ transition. The blue band describes the results obtained in this work (see text for details), while the orange one reproduces the results from Ref.~\cite{Gubernari:2023puw}. For completeness, we report also the lattice data point from the HPQCD collaboration~\cite{Parrott:2022rgu} in green and from the FNAL/MILC collaboration~\cite{Bailey:2015dka} in red. The widths of the bands and the sizes of the error bars correspond to one standard deviation uncertainties. The FFs $f_+$ and $f_T$ are multiplied by the factor $P(q^2)=1-q^2/M_{B_s^*}^2$ to increase readability.}
    \label{fig:FF_BtoK}
\end{figure}

\subsection{Numerical results for the FFs}

\begin{figure}[!t!]
    \centering
    \includegraphics[width=0.32\linewidth]{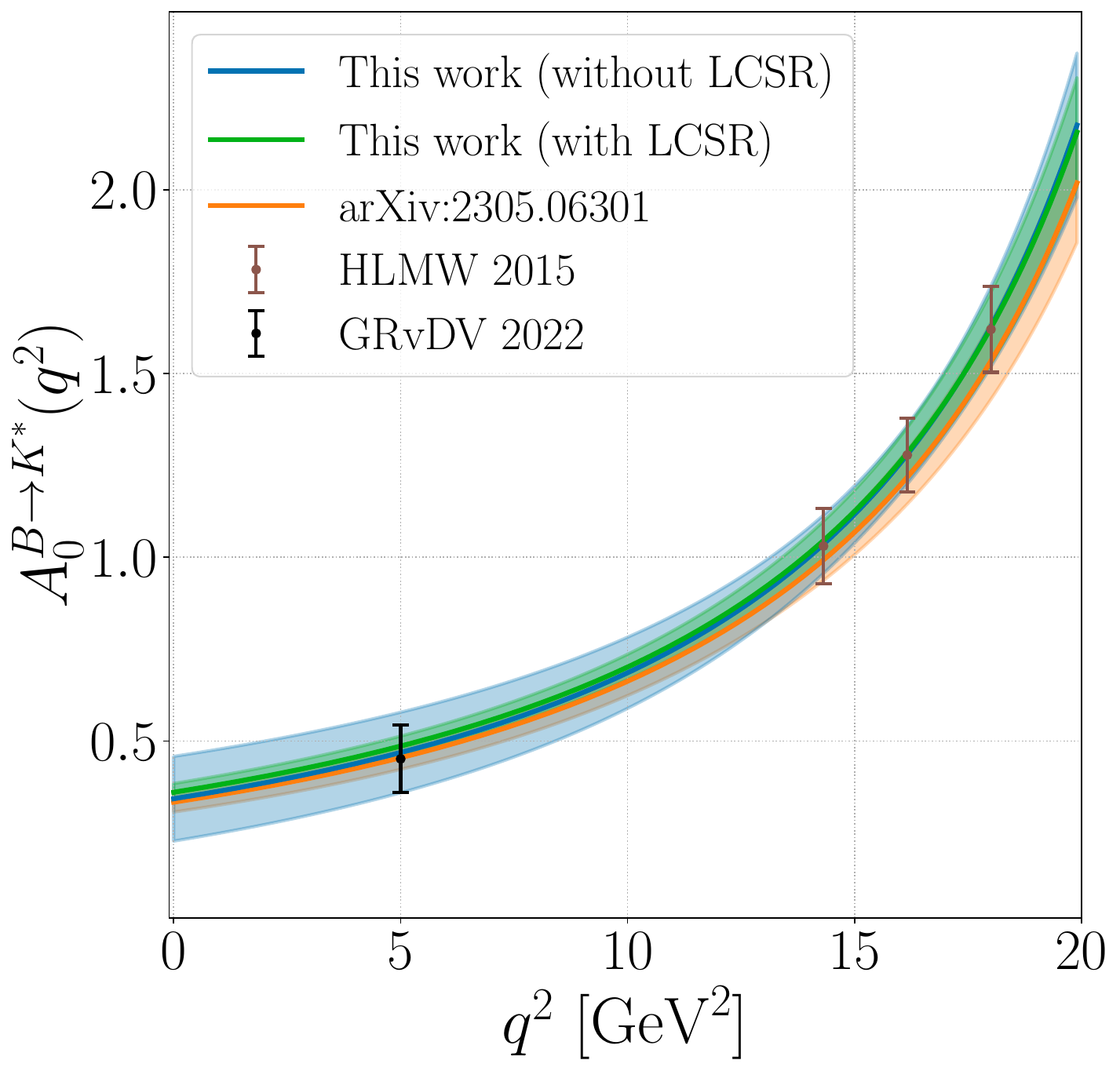}
    \hspace{0.008\linewidth}
    \includegraphics[width=0.32\linewidth]{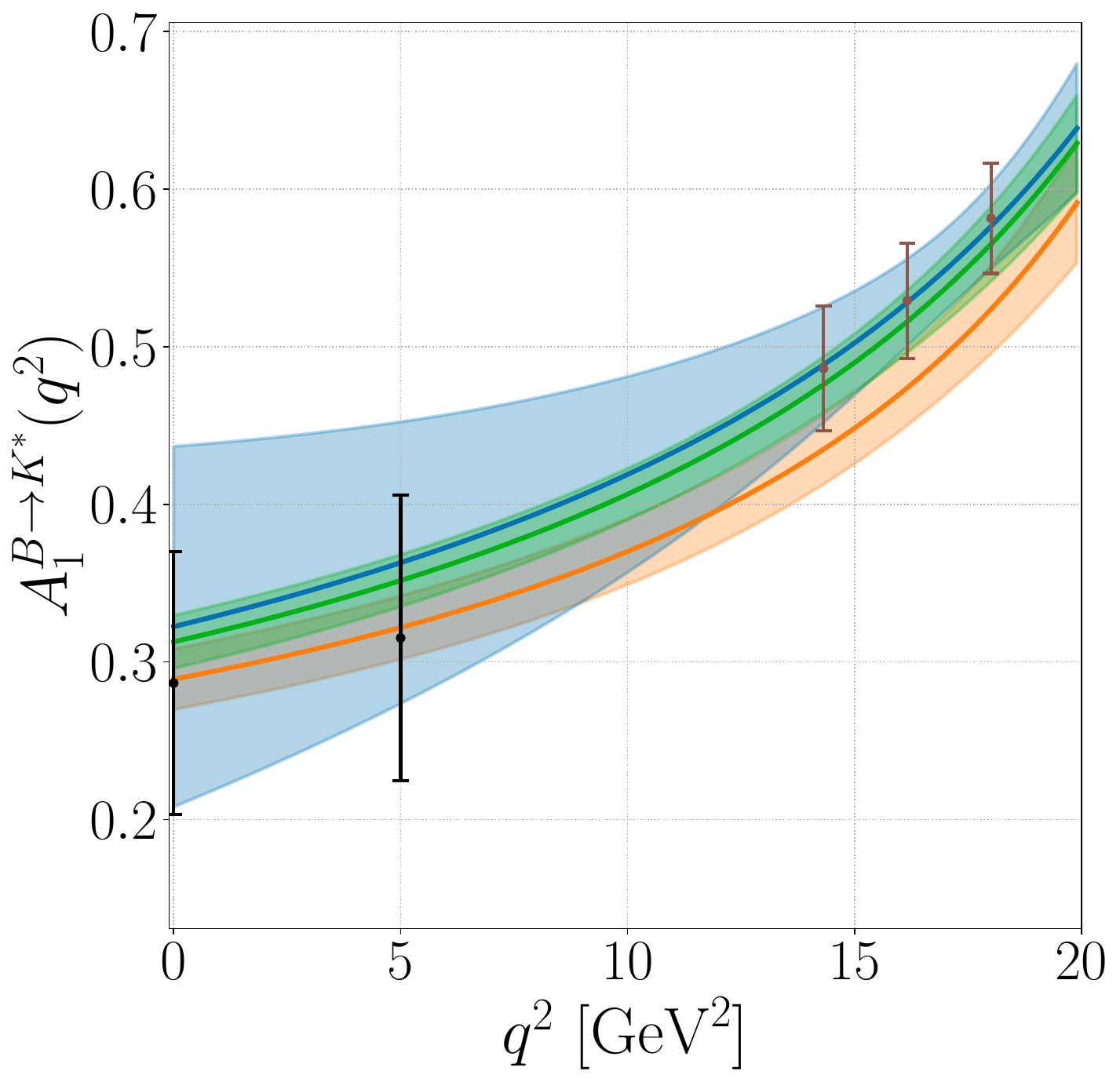}
    \hspace{0.008\linewidth}
    \includegraphics[width=0.32\linewidth]{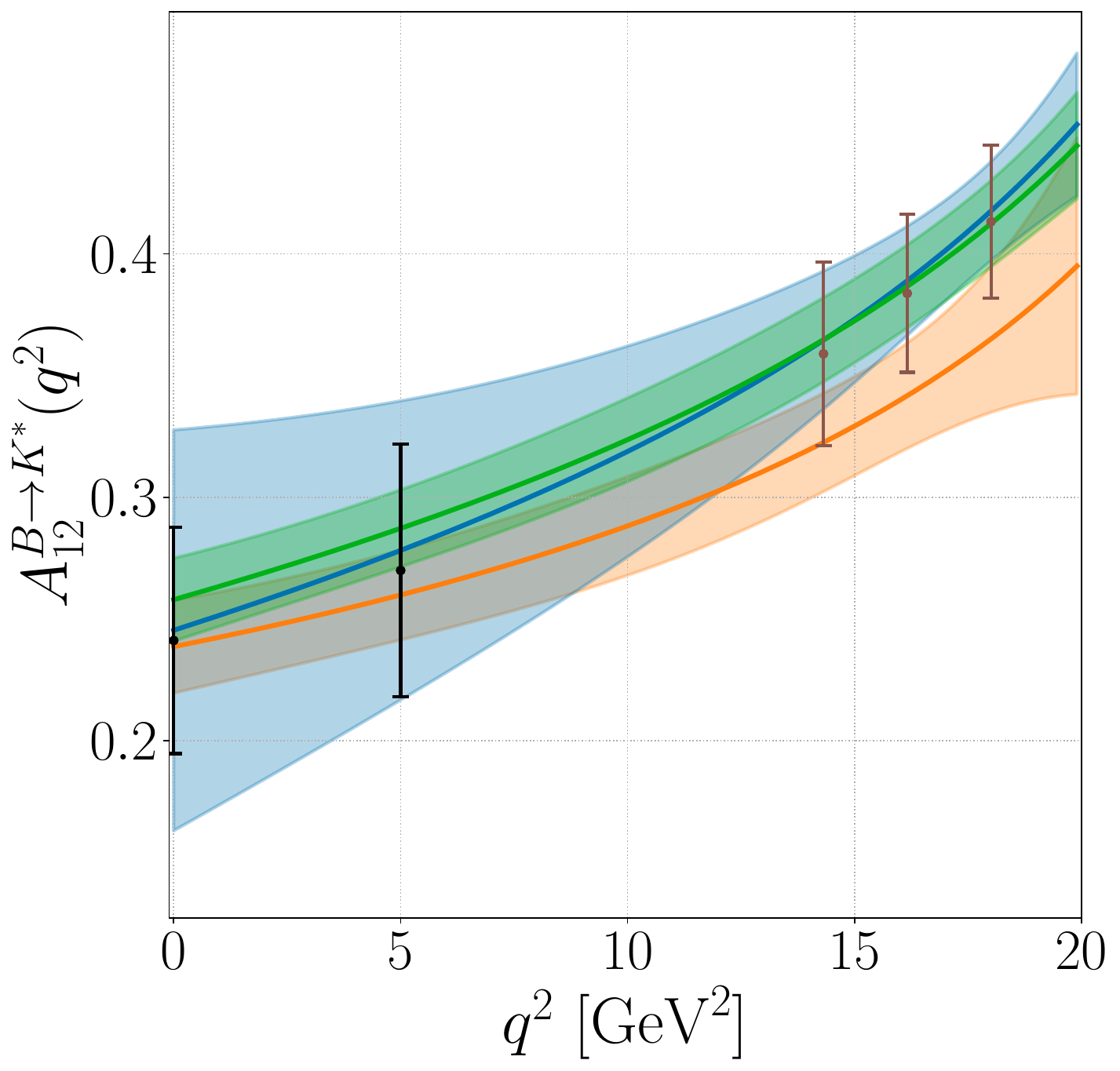}
    \includegraphics[width=0.32\linewidth]{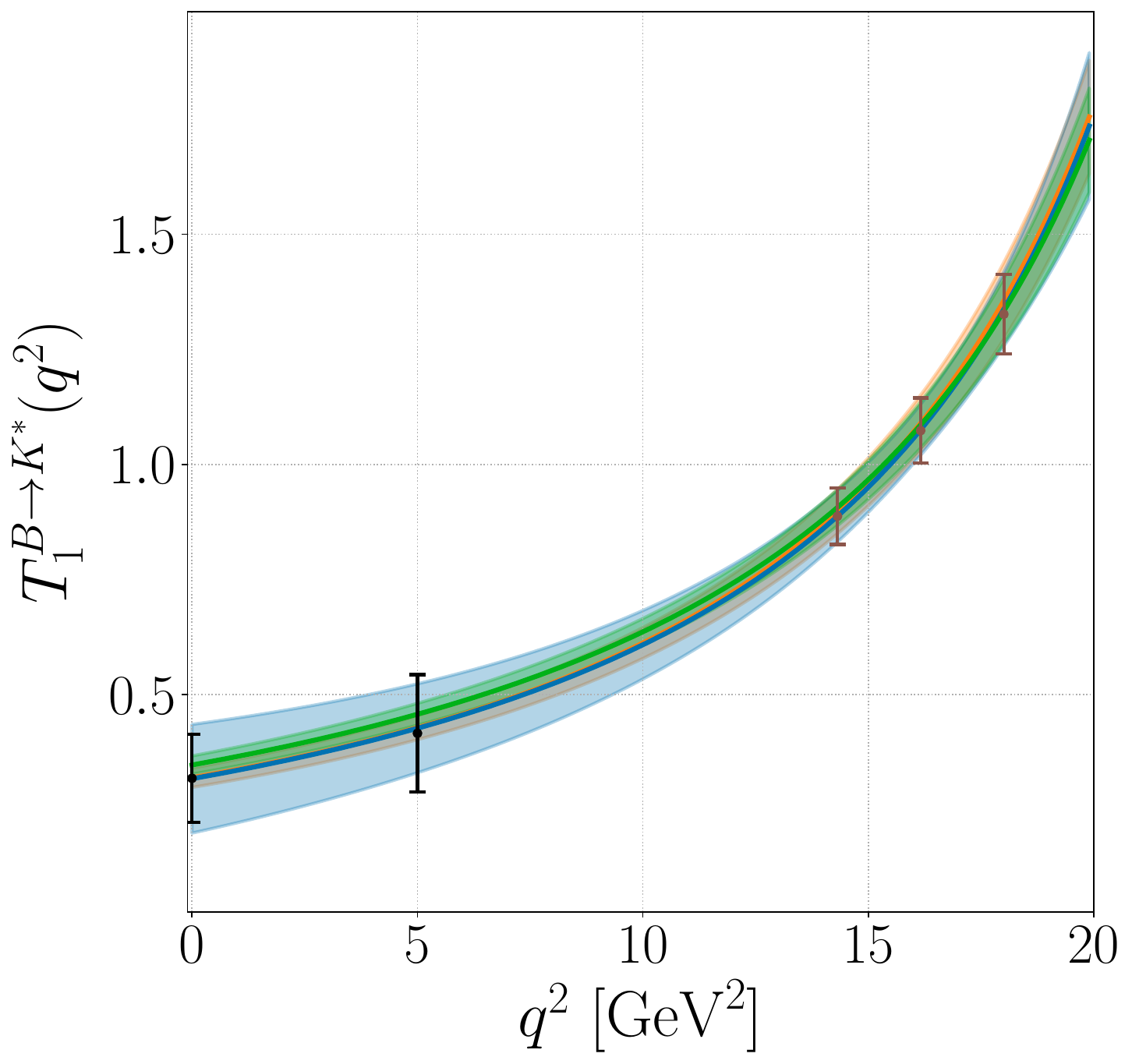}
    \hspace{0.008\linewidth}
    \includegraphics[width=0.32\linewidth]{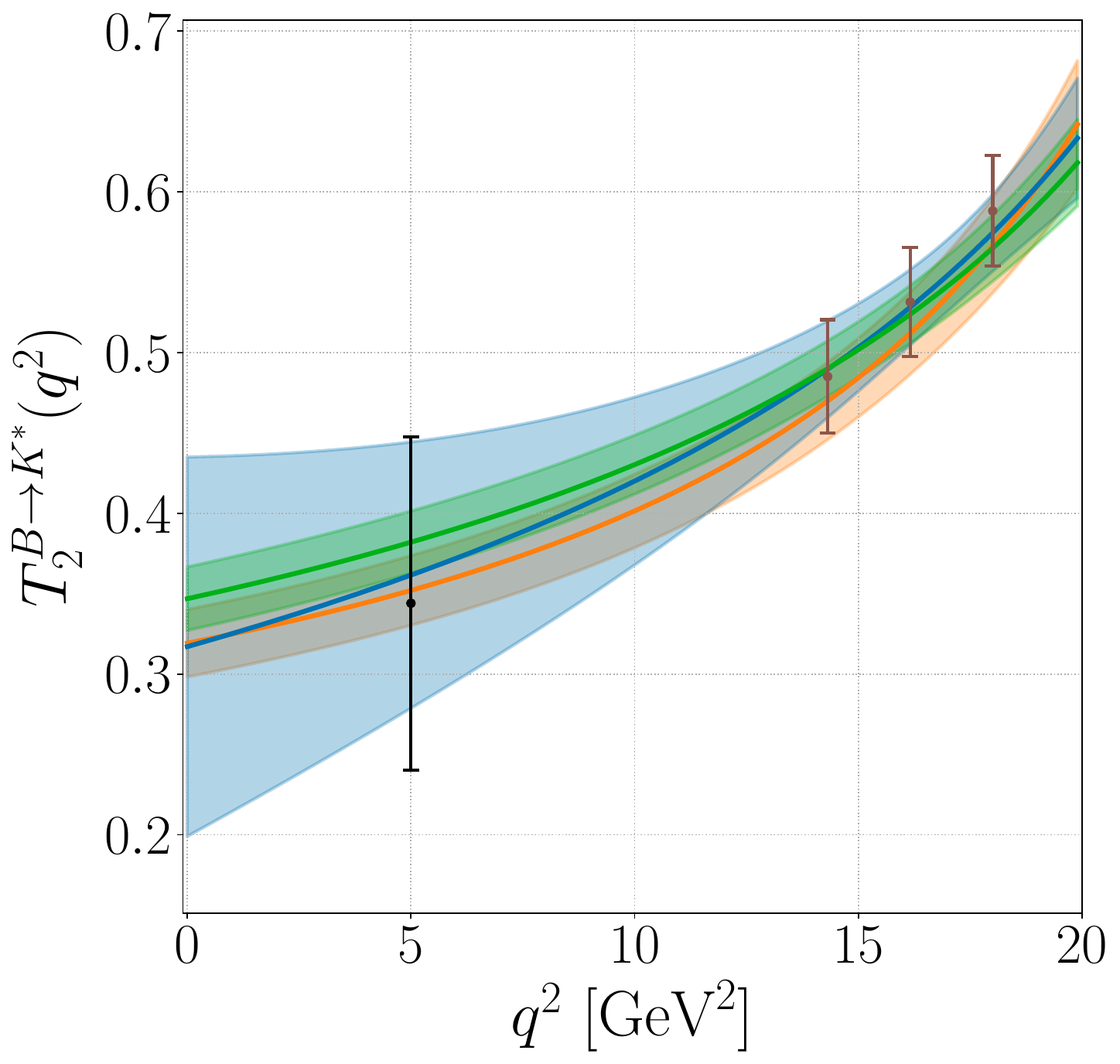}
    \hspace{0.008\linewidth}
    \includegraphics[width=0.32\linewidth]{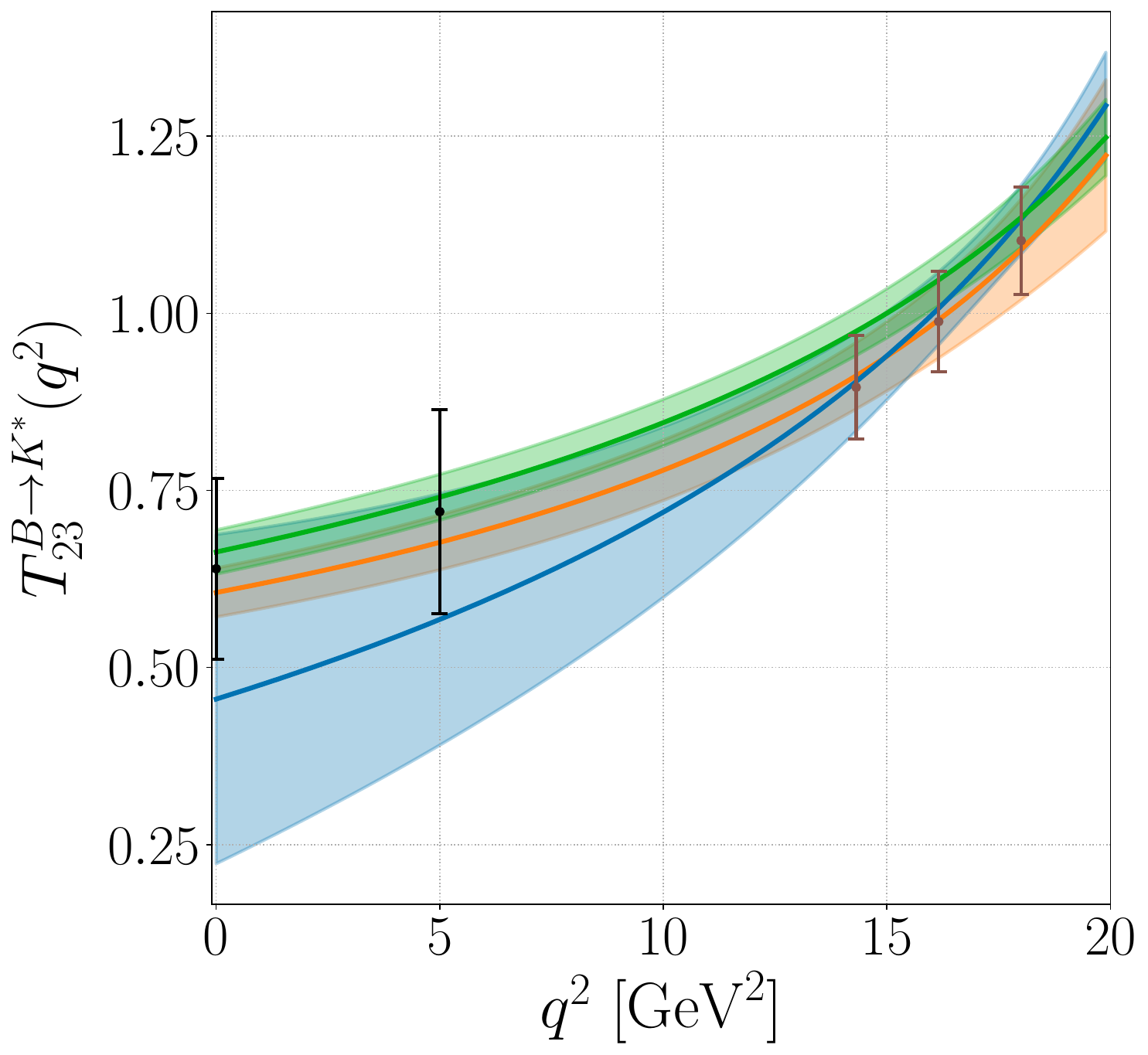}
    \includegraphics[width=0.32\linewidth]{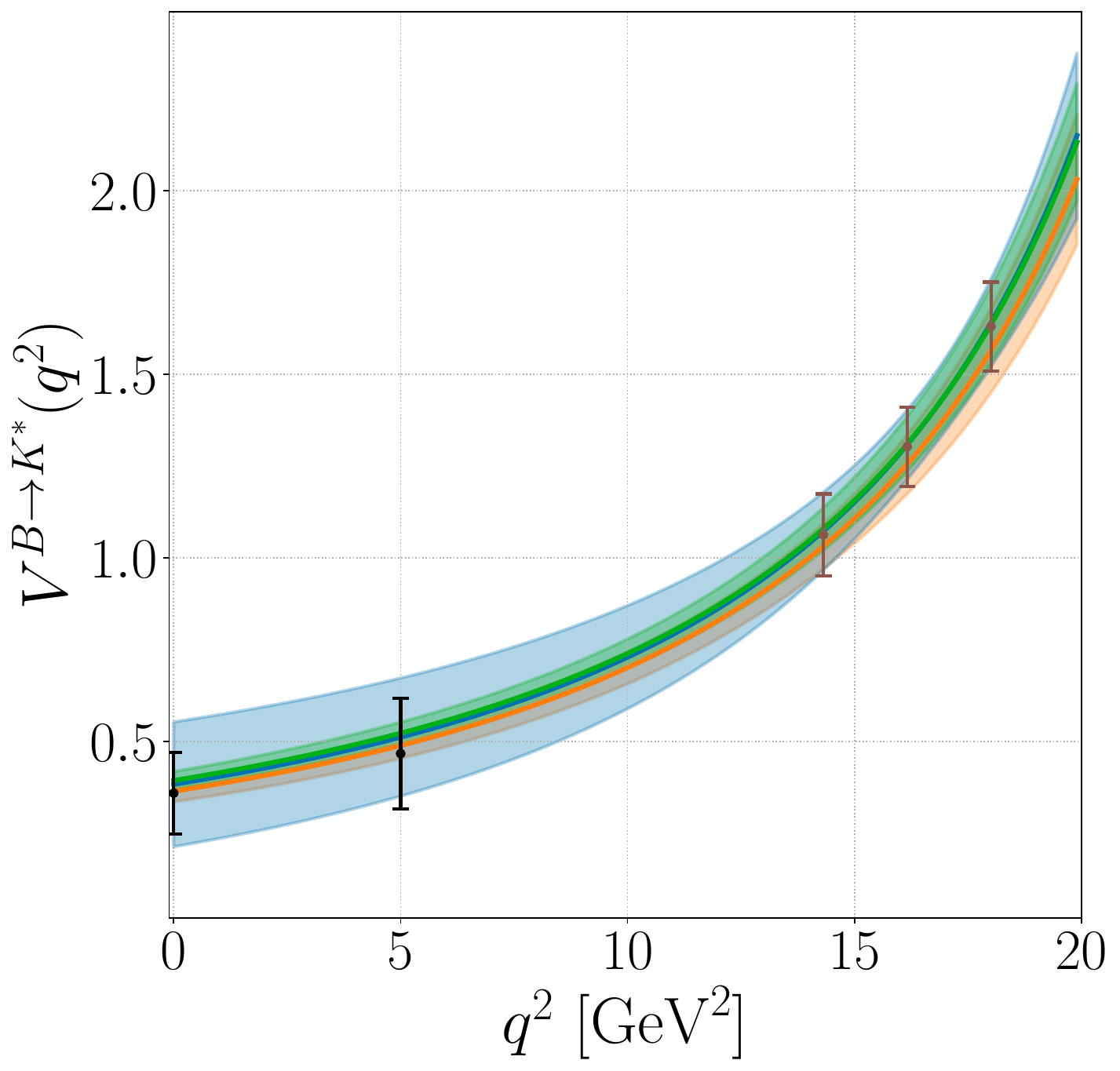}
    \caption{Comparison of the FFs entering the $B\to K^*$ transition. The blue and green bands describe the results obtained in this work without or with the inclusion of LCSR results in the BGL likelihood (see text for details), while the orange one reproduces the results from Ref.~\cite{Gubernari:2023puw}. For completeness, we report also the HLMW lattice data points~\cite{Horgan:2013hoa, Horgan:2015vla} in brown and the latest LCSR determinations as found on the EOS repository~\cite{Gubernari:2020eft,EOSAuthors:2021xpv} in black (the determinations at $q^2 = -15, -10, -5$ GeV$^2$, included in the fitting procedure, are not shown). The widths of the bands and the sizes of the error bars correspond to one standard deviation.}
    \label{fig:FF_BtoKstar}
\end{figure}

\begin{figure}[!tb!]
    \centering
    \includegraphics[width=0.32\linewidth]{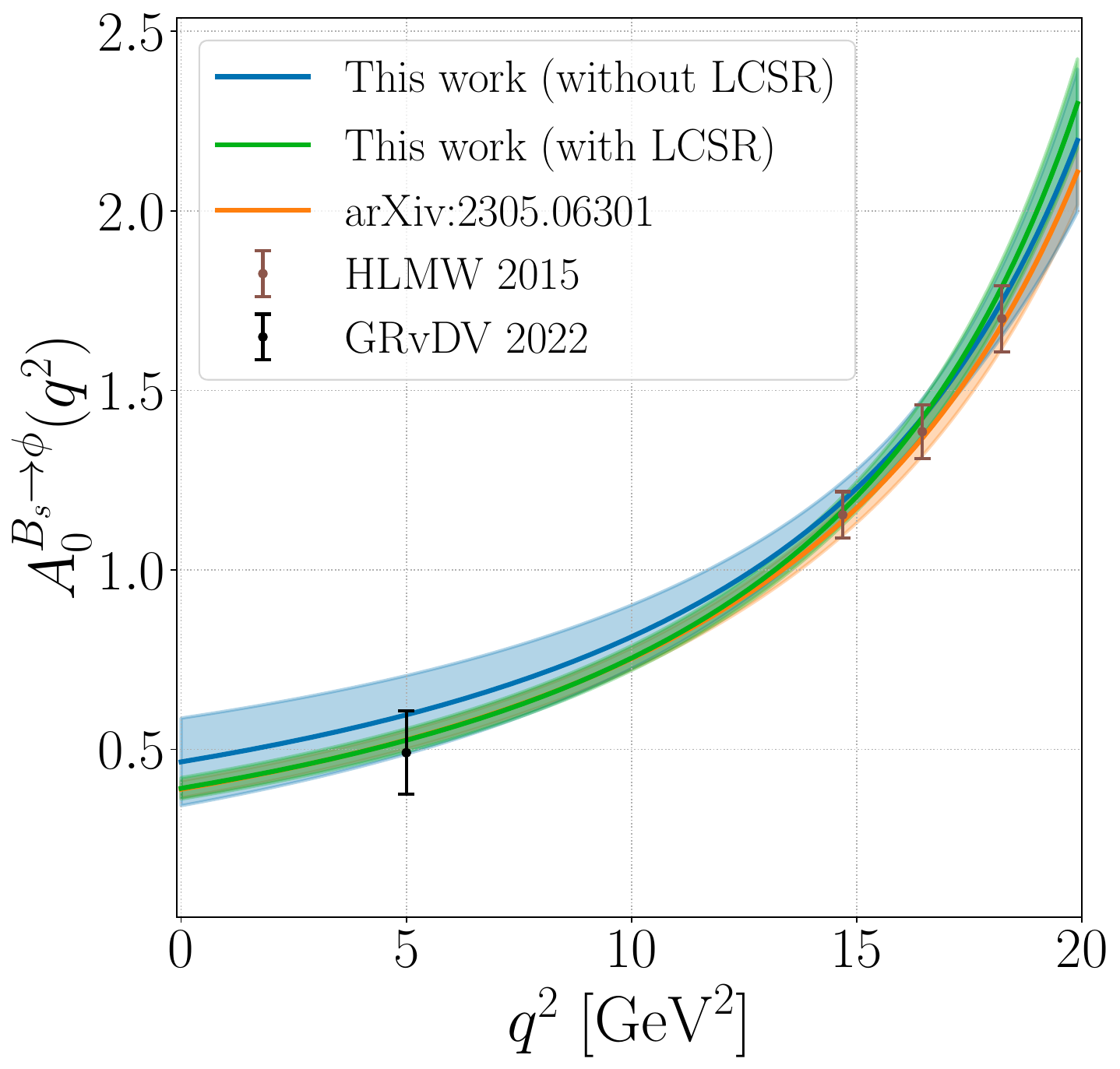}
    \hspace{0.008\linewidth}
    \includegraphics[width=0.32\linewidth]{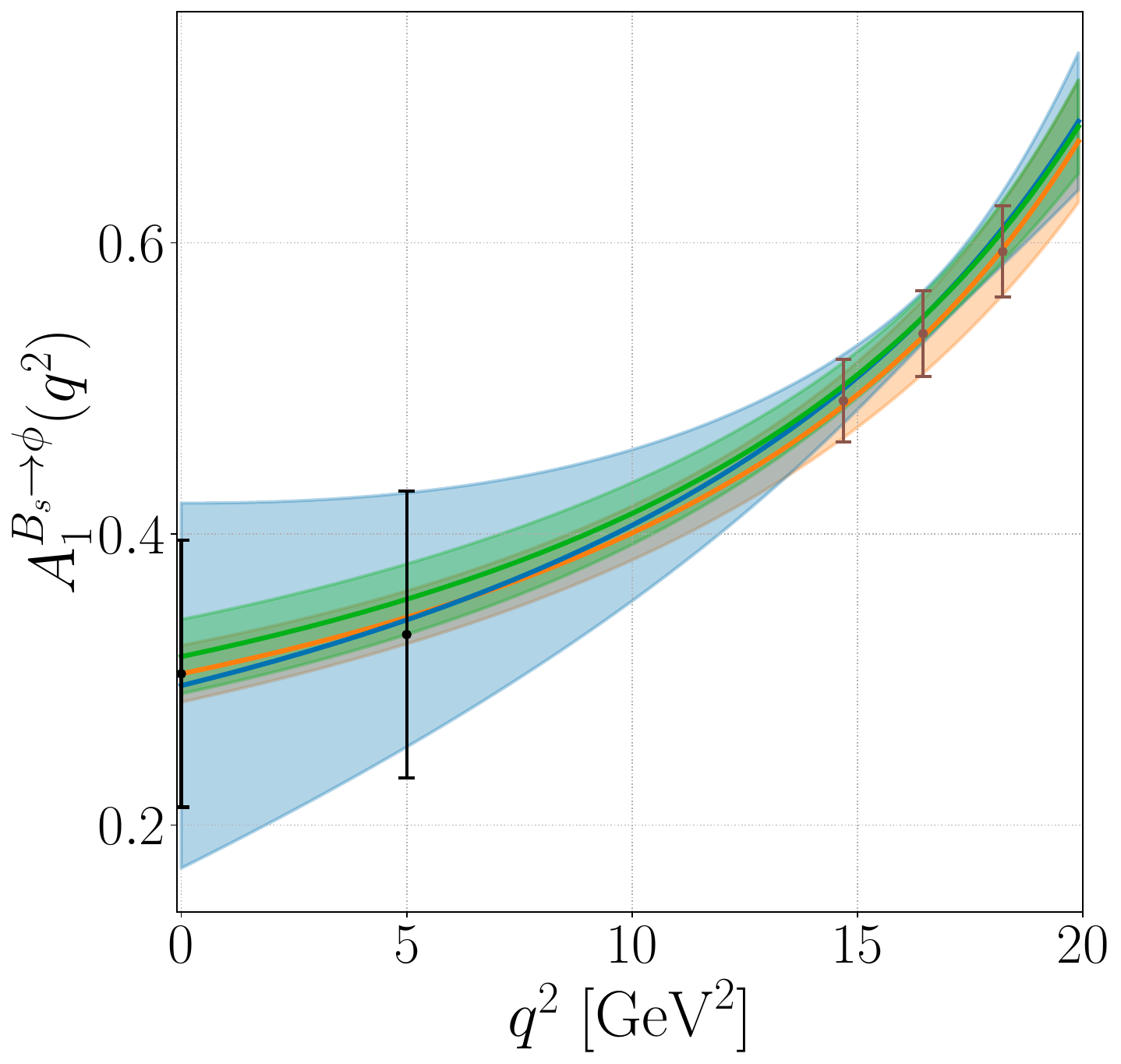}
    \hspace{0.008\linewidth}
    \includegraphics[width=0.32\linewidth]{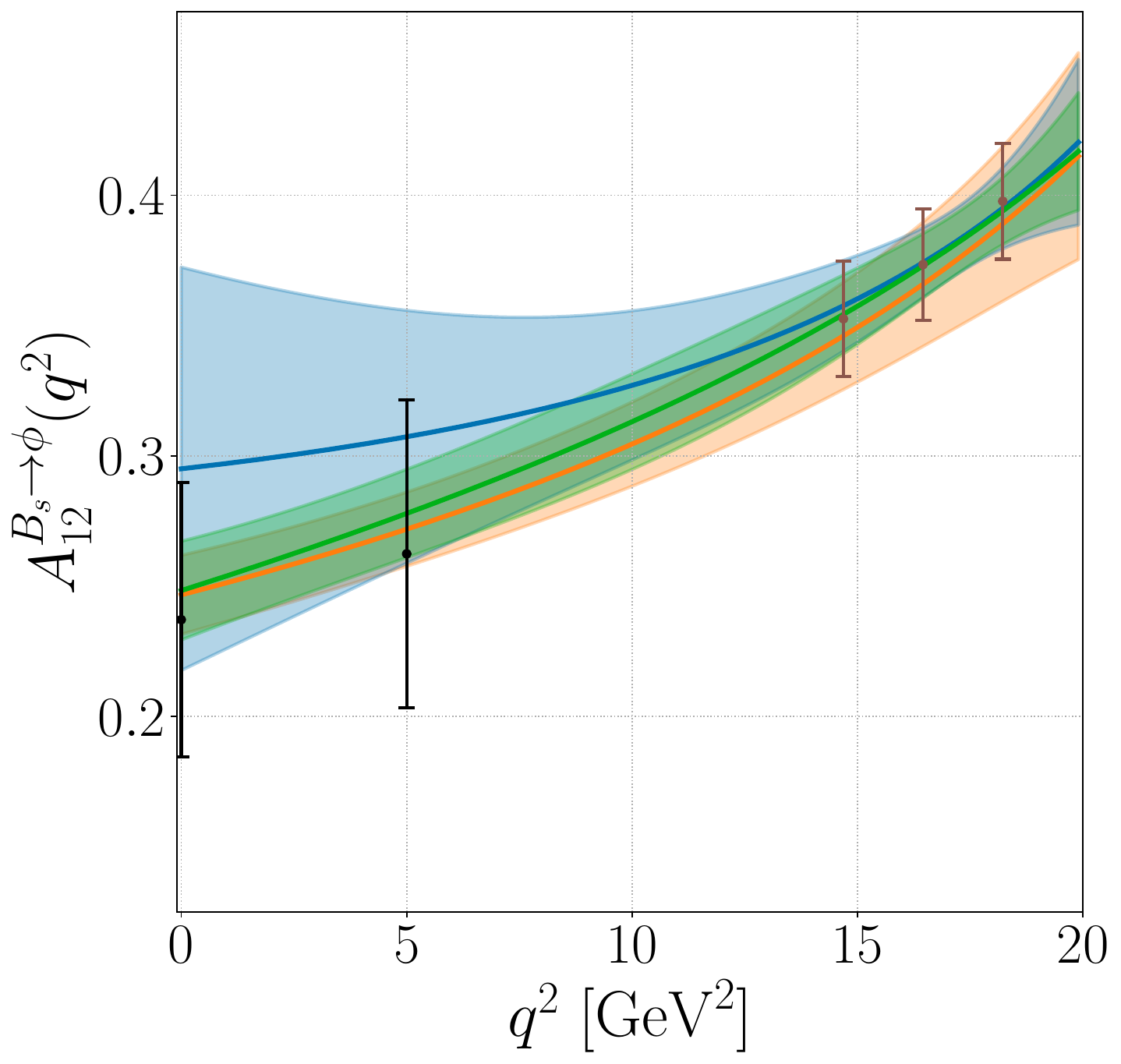}
    \includegraphics[width=0.32\linewidth]{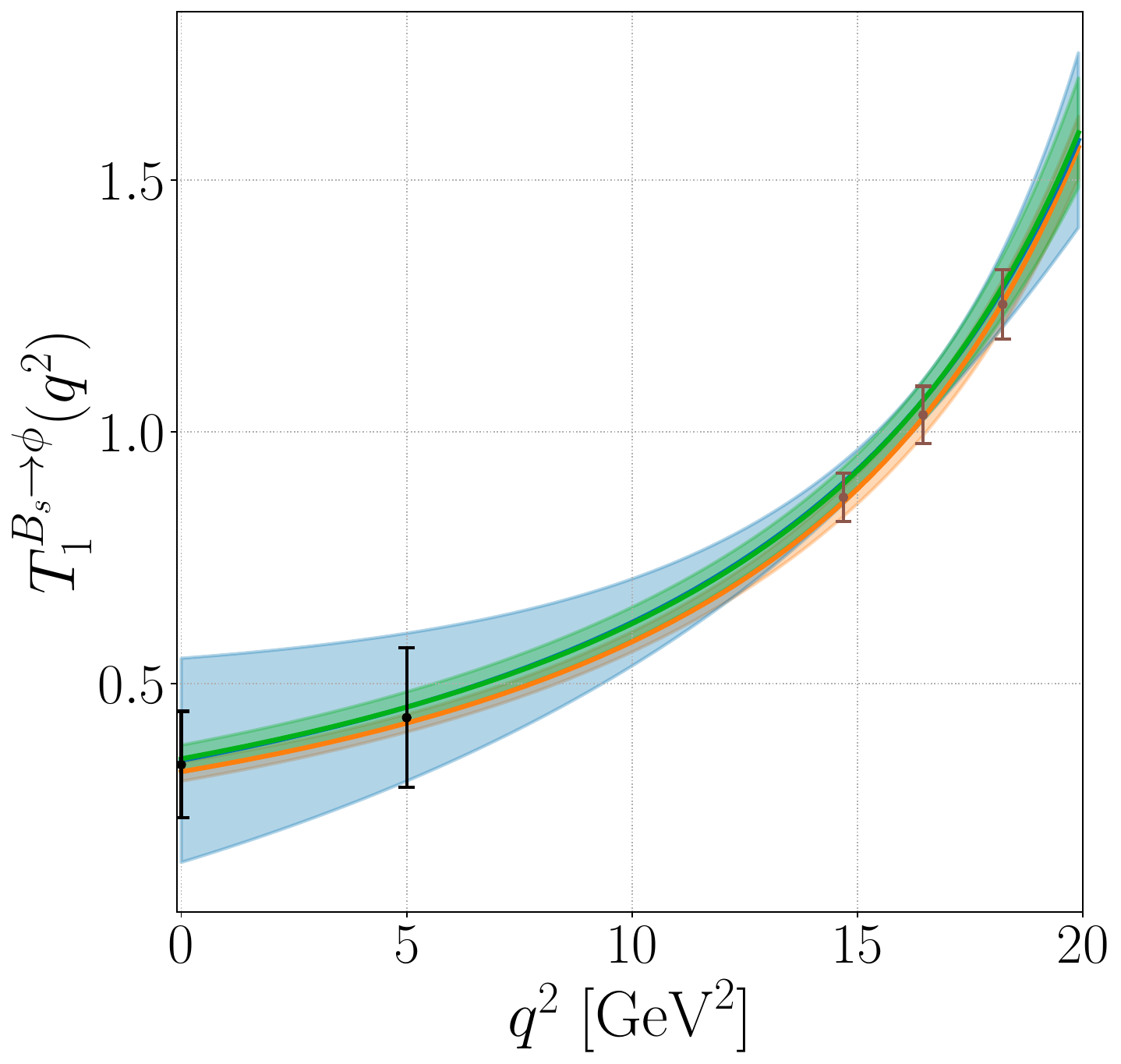}
    \hspace{0.008\linewidth}
    \includegraphics[width=0.32\linewidth]{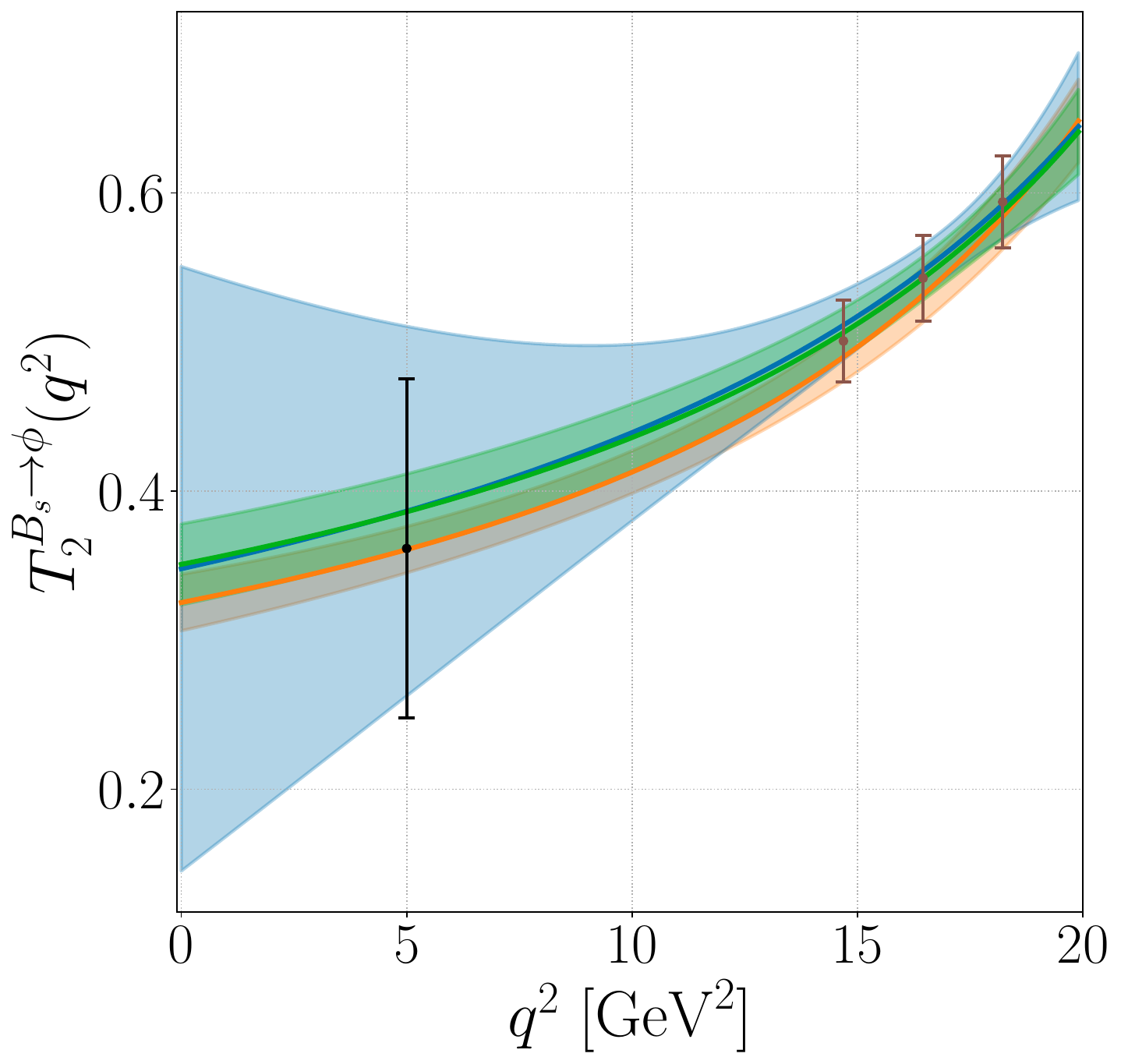}
    \hspace{0.008\linewidth}
    \includegraphics[width=0.32\linewidth]{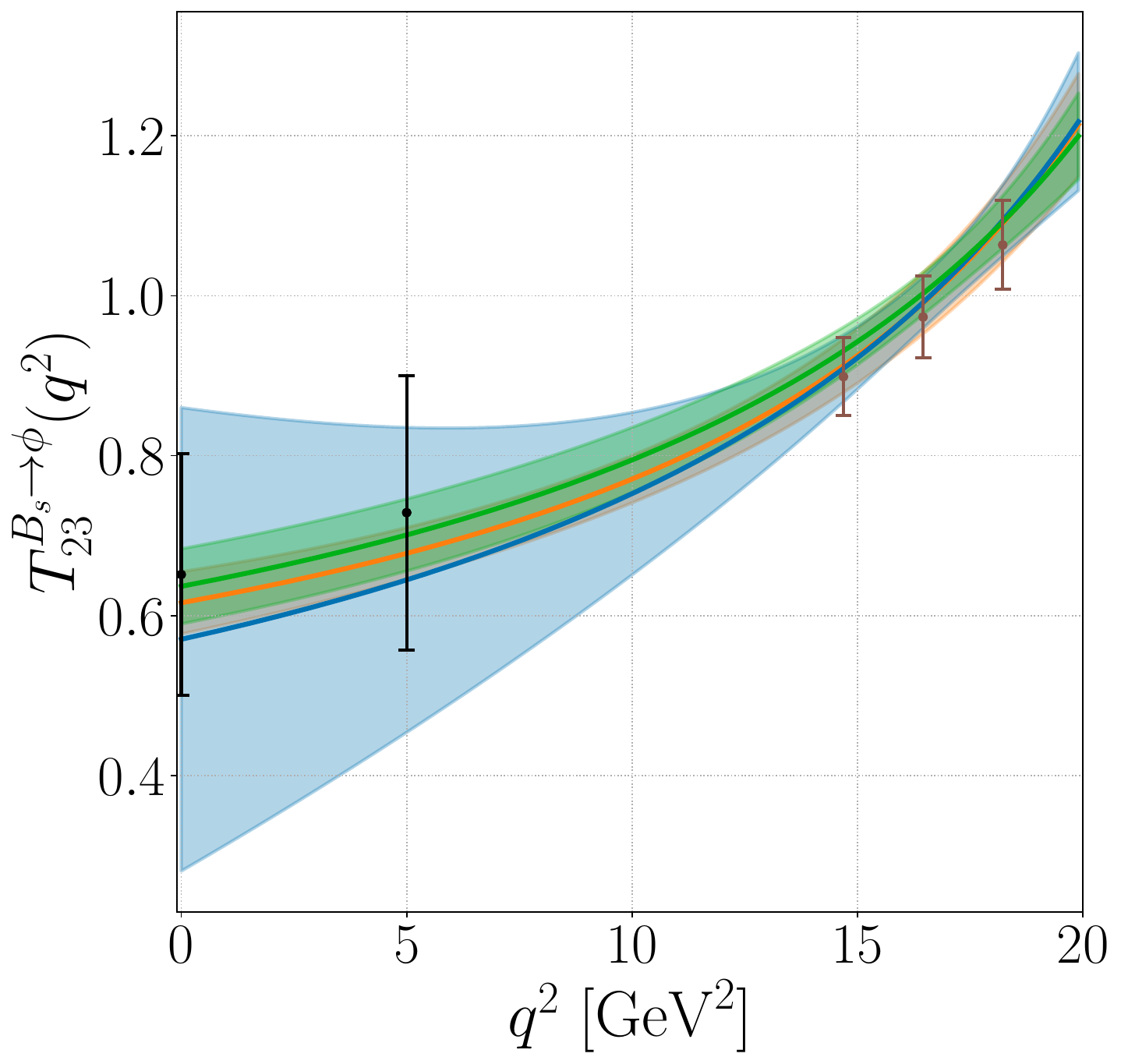}
    \includegraphics[width=0.32\linewidth]{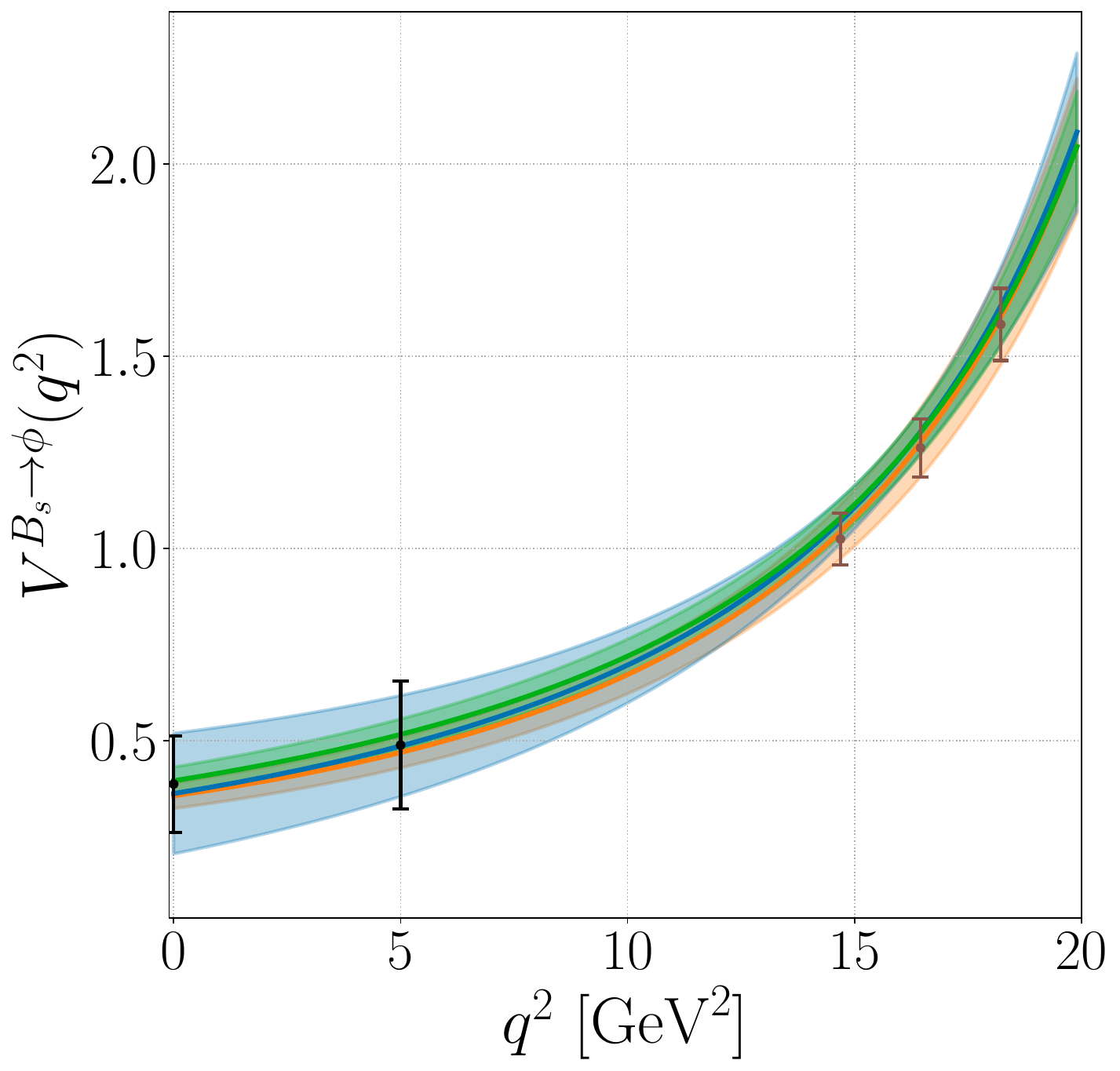}
    \caption{Same as Figure \ref{fig:FF_BtoKstar} for the $B_s\to \phi$ transition.}
    \label{fig:FF_BstoPhi}
\end{figure}

For what concerns the semileptonic $B \to K$ decay, the input lattice data have been obtained starting from the results of the studies in Refs.~\cite{Bailey:2015dka,Parrott:2022rgu}. To be more specific, four lattice points associated to Ref.~\cite{Parrott:2022rgu} have been considered at four reference values of the squared momentum transfer $q^2$, namely $q_i^2=\{0,\,7.6,\,15.3,\,22.9\}$ GeV$^2$ ($i=1,2,3,4$), covering the whole kinematic range. In addition to this, other four lattice inputs have been obtained from the results of the fits carried out in Ref.~\cite{Bailey:2015dka} at $q_j^2=\{17.0,\,18.9,\,20.9,\,22.8\}$ GeV$^2$ ($j=1,2,3,4$).\footnote{We did not include in our analysis the results of the lattice computations of the FFs performed in Ref.\,\cite{Bouchard:2013eph}, often referred to as HPQCD 2013 dataset. Indeed, as clear from Figure 27 of FLAG Review 2024\,\cite{FlavourLatticeAveragingGroupFLAG:2024oxs}, the HPQCD 2013 values of the tensor form factor $f_T^{B \to K}$ are in tension with the ones corresponding to the FNAL/MILC 2016 and the HPQCD 2022 datasets in the whole high-$q^2$ region.} We have thus used multivariate Gaussian distributions associated to these datasets in order to generate a sample
of events (of the order of $10^5$), each of which is composed by 8 data points for the $B \to K$ FFs (8 points for each FF).\footnote{Since the DM unitary filters become extremely selective when multiple lattice datasets are combined, the importance sampling procedure outlined in Ref.~\cite{Simula:2023ujs} has been used. Indeed, such a procedure allows to generate events for the FFs satisfying the unitary filters for any number of initial data points.} This sample has been analysed in the framework of the DM method, and the corresponding results are shown as blue bands in Fig.~\ref{fig:FF_BtoK}. As a comparison, in the same figure we report also as orange bands the results obtained in Ref.~\cite{Gubernari:2023puw}. The two sets of FFs bands are found to be in good agreement, albeit in the case of the $f_+^{B \to K}$ and $f_0^{B \to K}$ FFs we find small differences in the slope and, thus, in the values of the same FFs at $q^2=0$. Our understanding is that this effect is induced by the lattice inputs used for the FF $f_0^{B \to K}$. Indeed, quite strong correlations ($i.e.$ larger than 0.92) exist between the two HPQCD 2022 points at $q^2=\{0,\,7.6\}$ GeV$^2$ and between the two FNAL/MILC 2016 data at $q^2=\{17.0,\,18.9\}$ GeV$^2$. Once unitarity is imposed on both the HPQCD 2022 and FNAL/MILC 2016 datasets at the same time, the DM FF band of $f_0^{B \to K}$ at $q^2=0$ goes a bit below the HPQCD 2022 point located at the same value of the squared momentum transfer, and this effect is translated as well to the FF $f_+^{B \to K}$ by means of the kinematical constraint in Eq.~(\ref{eq:KC1}).

For what concerns, instead, the semileptonic $B \to K^*$ and $B_s \to \phi$ decays, the inputs for the DM analysis have been taken from the lattice study in Refs.~\cite{Horgan:2013hoa, Horgan:2015vla}. In this paper and in its addendum, the authors give the results of their fits of the lattice data extrapolated to the physical quark mass limit. Thus, we reconstructed the values of each of the FFs entering in semileptonic $B \to K^*$ and $B_s \to \phi$ decays at three reference values of the squared momentum transfer $q^2$. Since the authors reported explicitly the region at high-$q^2$ in which they performed their simulations on the lattice, we chose $q_i^2=\{14.3,\,16.2,\,18.0\}$ GeV$^2$ ($i=1,2,3$) as our $q^2$ reference values. The results of the application of the DM method to the sample datasets originated through appropriate multivariate Gaussian distributions are shown as blue bands in Figs.~\ref{fig:FF_BtoKstar}-\ref{fig:FF_BstoPhi} for $B \to K^* \ell\ell$ and $B_s \to \phi\ell\ell$, respectively. Also in this case we report as orange bands the results obtained in Ref.~\cite{Gubernari:2023puw}. 

Due to the absence of lattice results at low $q^2$ in the vectorial channels, the agreement between the blue and the orange bands is not as remarkable in Figs.~\ref{fig:FF_BtoKstar}-\ref{fig:FF_BstoPhi}, particularly concerning the tensorial FFs $T_1$, $T_2$ and $T_{23}$. For these two channels we therefore produced an additional set of results, where the DM analysis is extended in order to incorporate not only lattice results, but also LCSR ones. In particular, for the $B\to K^*$ channel we considered the LCSR inputs from Ref.~\cite{Gubernari:2018wyi}, provided at $q_i^2=\{-15,\,-10,\,-5,\,0,\,5\}$ GeV$^2$ ($i=1,2,3,4,5$)\footnote{Due to the kinematical constraints given at Eq.~\eqref{eq:KC2}, there is no input point at $q^2=0$ for the FFs $A_0$ and $T_2$.} and labelled as ``\texttt{GRvDV:2022A}'' on the EOS repository~\cite{EOSAuthors:2021xpv}; similarly, for the $B_s \to \phi$ channel we considered analogous LCSR inputs from Ref.~\cite{Gubernari:2020eft}, labelled ``\texttt{GvDV:2020A}'' on the EOS repository. These results are reported as green bands in Figs.~\ref{fig:FF_BtoKstar}-\ref{fig:FF_BstoPhi}, which show again a better level of compatibility with the results from Ref.~\cite{Gubernari:2023puw}. 

The analyses presented in the remainder of this paper are therefore performed using two
different approaches for the hadronic FFs.
The first, dubbed ``LQCD DM'', employs exclusively the FFs obtained through
the DM analysis based on lattice inputs only, corresponding to the
blue bands shown in Figs.~\ref{fig:FF_BtoK}--\ref{fig:FF_BstoPhi}.
The second, denoted as ``LQCD+LCSR DM'', incorporates additional constraints from
LCSR in the vector channels, and corresponds to using the DM results for
the $B\to K$ FFs (blue band in Fig.~\ref{fig:FF_BtoK}) together with the DM
analysis including LCSR inputs for the $B\to K^*$ and $B_s\to\phi$ FFs (green
bands in Figs.~\ref{fig:FF_BtoKstar} and~\ref{fig:FF_BstoPhi}).
Although this nomenclature is slightly imprecise, since LCSR inputs are included only for
the vector final states, it provides a convenient shorthand and will be used throughout
the paper for clarity.

In order to make the DM FFs results directly usable in phenomenological analyses,
we translate them into a set of parameters defined within the BGL
$z$-expansion framework~\cite{Boyd:1997kz}.
Concretely, for each FF $F_i\left(z(q^2)\right)$ entering the $B\to K$, $B\to K^*$ and
$B_s\to\phi$ transitions, we perform a BGL fit of the form
\begin{equation}
F_i\left(z(q^2)\right)
=
\frac{1}{{\phi_p}_i\left(z(q^2)\right)}\,
\sum_{n=0}^{2} a^i_n\, z(q^2)^n\,,
\end{equation}
where $z(q^2)$ is the conformal variable defined at Eq.~\eqref{eq:z} and
${\phi_p}_i(q^2)$ is the modified outer function defined at Eq.~\eqref{eq:poles} (with the unmodified ones given at Eq.~\eqref{eq:phif+}-\eqref{eq:phiT2}), 
chosen such that the unitarity bounds take a simple
quadratic form in terms of the coefficients $a^i_n$ corresponding 
to each specific spin-parity quantum channel.
The fits are performed independently for each transition and for each FF choice
(LQCD DM and LQCD+LCSR DM), fully accounting for correlations among the input data.

The resulting posterior distributions for the BGL coefficients, including their central
values, uncertainties and correlation matrices, constitute the hadronic input used in
all subsequent phenomenological analyses presented in this work. These results have been implemented in the \texttt{HEPfit} framework~\cite{deBlas:2019okz,HEPfit},
which has been used to perform all the numerical analyses and fits discussed in the
remainder of the paper.
For completeness and reproducibility, the numerical results of the BGL fits are reported
in Appendix~\ref{app:AppB}.

% ============================================================
\section{Updates on rare decays: Standard Model}
\label{sec:SM_updates}
% ============================================================

In the previous section we presented a new determination of the $B\to K^{(*)}$ and $B_s\to\phi$ FFs based on the DM approach, which exploits analyticity and unitarity constraints in order to provide a theoretically controlled description of the hadronic matrix elements over the full kinematic range. The resulting FFs constitute the main theoretical input for the phenomenological analysis of rare $b\to s$ transitions.

In this section we investigate the implications of these FF determinations for SM predictions of rare $B$ semileptonic decays.
We consider both $b\to s\nu\bar\nu$ transitions and $b\to s\ell^+\ell^-$ processes.
The former provide theoretically very clean observables, whose uncertainties are
dominated by the normalization and shape of the hadronic FFs, while the latter
also receive non-local contributions from charm loops and other hadronic effects.

% ------------------------------------------------------------
\subsection{\boldmath$b\to s\nu\bar\nu$ decays}
\label{subsec:SM_b2snunu}
% ------------------------------------------------------------

Rare flavour-changing neutral current decays mediated by the transition $b\to s\nu\bar\nu$ constitute one of the theoretically cleanest probes of short-distance physics in the SM. Unlike their charged-lepton counterparts, these modes are dominated by
a single semileptonic operator in the SM effective Hamiltonian.
As a consequence, the dominant irreducible theoretical uncertainty in the SM predictions for exclusive $B\to K^{(*)}\nu\bar\nu$ decays originates from hadronic FFs.

In this section we present updated SM predictions for the branching fractions of $B\to K\nu\bar\nu$ and $B\to K^*\nu\bar\nu$ decays, with particular emphasis on the impact of different FF determinations. Specifically, we compare predictions obtained using two different approaches for the FFs determination, as outlined in Sec.~\ref{sec:DM_FF}. This comparison allows us to isolate the effect of input from LCSR on the SM predictions for these rare invisible modes.

At scales $\mu \sim m_b$, the SM contribution to $b\to s\nu\bar\nu$ transitions is described by the effective Hamiltonian
\begin{equation}
\mathcal{H}_{\rm eff,\, SM}^{b\to s\nu\bar\nu}
=
-\frac{4G_F}{\sqrt{2}}\,\lambda_t\,
\frac{\alpha_e}{4\pi}\,
C_{L,\nu}^{\rm SM}\,
\left(\bar{s}\gamma_\mu P_L b\right)
\left(\bar{\nu}\gamma^\mu(1-\gamma_5)\nu\right)
\,+\,{\rm h.c.}\,,
\label{eq:Heff_b2snunu}
\end{equation}
where $P_L=(1-\gamma_5)/2$, $G_F$ and $\alpha_e$ are the Fermi and fine structure constant, and the size of $\lambda_t\equiv V_{tb}V_{ts}^*$ employing the latest result from the UTfit collaboration~\cite{UTfit:2022hsi,UTfit:2026qxi} is $|\lambda_t| \simeq 0.041$. In the SM only the left-handed Wilson coefficient $C_{L,\nu}^{\rm SM}$ is generated, and it is lepton flavour universal. The short-distance
coefficient can be expressed in terms of the loop function $X_t$ as $C_{L,\nu}^{\rm SM} = -X_t/\sin^2\theta_W$~\cite{Altmannshofer:2009ma}.
The function $X_t$ is known up to NLO corrections in QCD~\cite{Buchalla:1993bv,Buchalla:1998ba,Misiak:1999yg} and includes two-loop electroweak corrections~\cite{Brod:2010hi}, leading to $C_{L,\nu}^{\rm SM} = -6.32(7)$ in the SM. 

The exclusive decay amplitudes for $B\to K^{(*)}\nu\bar\nu$ are obtained by evaluating the
hadronic matrix elements of the quark current
$\bar{s}\gamma_\mu(1-\gamma_5)b$ between the $B$ meson and the $K$ or $K^*$ final state.
For the pseudoscalar channel $B\to K\nu\bar\nu$, the relevant hadronic input is provided by
the vector FF $f_+(q^2)$, while for the vector channel $B\to K^*\nu\bar\nu$ the
amplitude depends on the $A_1(q^2)$, $A_{12}(q^2)$ and $V(q^2)$ $B\to K^*$ FFs.

The differential decay rate for
$B\to K\nu\bar\nu$ can be written as
\begin{equation}
\frac{d\Gamma(B\to K\nu\bar\nu)}{dq^2}
=
\frac{G_F^2\,\alpha_e^2}{256\pi^5}\,
\frac{|\lambda_t|^2}{m_B^3}\,
|C_{L,\nu}^{\rm SM}|^2\,
\lambda_K^{3/2}(q^2)\,
|f_+(q^2)|^2\,,
\label{eq:dG_BtoK_nunu}
\end{equation}
where $\lambda_K(q^2)=\lambda(m_B^2,m_K^2,q^2)$ is the usual Källén function and $m_B$ is the mass of the $B$ meson.
For the vector final state $B\to K^*\nu\bar\nu$, the differential decay rate can be expressed
as
\begin{equation}
\frac{d\Gamma(B\to K^*\nu\bar\nu)}{dq^2}
=
\frac{G_F^2\,\alpha_e^2}{128\pi^5}\,
\frac{|\lambda_t|^2}{m_B^3}\,
|C_{L,\nu}^{\rm SM}|^2\,
q^2\,\lambda_{K^*}^{1/2}(q^2)\,(m_B+m_{K^*})^2 F(q^2)\,,
\label{eq:dG_BtoKst_nunu}
\end{equation}
where $\lambda_{K^*}(q^2)=\lambda(m_B^2,m_{K^*}^2,q^2)$, $m_{K*}$ is the mass of the $K^*$ resonance, and the FFs component is encoded in the term
\begin{equation}
F(q^2)=\left[
\left[A_1(q^2)\right]^2
+
\frac{32\,m_{K^*}^2\,m_B^2\,\left[A_{12}(q^2)\right]^2}{q^2\,(m_B+m_{K^*})^2}\,
+
\frac{\lambda_{K^*}(q^2)\, \left[V(q^2)\right]^2}{(m_B+m_{K^*})^4}
\right]\,.
\end{equation}

The phenomenologically relevant observables are obtained by integrating the above
expressions over the full kinematic range $0\le q^2\le (m_B-m_{K^{(*)}})^2$. 
Since neutrinos are not experimentally identified, the decay rates 
have already been implicitly summed over all light neutrino flavours,
$\nu_\ell=\nu_e,\nu_\mu,\nu_\tau$,
as dictated by lepton flavour universality in the SM.
As a consequence, the branching fractions receive an overall multiplicative
contribution from the three active neutrino species.
In the SM, all $B\to K^{(*)}\nu\bar\nu$ observables depend quadratically on the single
Wilson coefficient $C_L^\nu$, rendering these modes particularly sensitive to the
normalization and shape of the hadronic FFs, which therefore constitute the
dominant source of theoretical uncertainty.

\begin{table}[!t!]
    \centering
    \setlength{\tabcolsep}{10pt}
    \renewcommand{\arraystretch}{1.35}
    \begin{tabular}{|c|cc|}
    \hline
        \textbf{Channel } &
        \textbf{$\mathcal{B}$ w/ LQCD DM FFs} &
        \textbf{$\mathcal{B}$ w/ LQCD+LCSR DM FFs}\\
        \hline
        $B^+ \to K^+\nu\bar\nu$ & ($3.95 \pm 0.14$) $\times 10^{-6}$ & ($3.95 \pm 0.14$) $\times 10^{-6}$ \\
        $B^0 \to K^0\nu\bar\nu$ & ($3.71 \pm 0.13$) $\times 10^{-6}$ & ($3.71 \pm 0.13$) $\times 10^{-6}$ \\
        $B^+ \to K^{*+}\nu\bar\nu$ & ($9.9 \pm 1.8$) $\times 10^{-6}$ & ($9.5 \pm 0.8$) $\times 10^{-6}$ \\
        $B^0 \to K^{*0}\nu\bar\nu$ & ($9.3 \pm 1.7$) $\times 10^{-6}$ & ($8.9 \pm 0.7$) $\times 10^{-6}$ \\
        \hline
    \end{tabular}
    \caption{Updated SM predictions for the branching fractions of the rare invisible
    decays $B\to K^{(*)}\nu\bar\nu$.
    The first column corresponds to the results obtained employing FFs determined using the DM approach outlined in Sec.~\ref{sec:DM_FF} based solely on Lattice QCD, while the results of the second column are obtained including in the $B\to K^*$ FFs determination inputs from LCSR taken from Ref.~\cite{Gubernari:2023puw} as well.}
    \label{tab:btosnunu_SM}
\end{table}

Our updated SM predictions for the integrated branching fractions of
$B^+\to K^+\nu\bar\nu$, $B^0\to K^0\nu\bar\nu$,
$B^+\to K^{*+}\nu\bar\nu$,
and
$B^0\to K^{*0}\nu\bar\nu$
are summarized in Table~\ref{tab:btosnunu_SM}.
We report results obtained using the LQCD DM approach  
and compare them with predictions obtained when also LCSR results
are taken into account in the $B\to K^*$ decays, see Sec.~\ref{sec:DM_FF} for further details.
The quoted uncertainties correspond to one standard deviation and include the full
propagation of FF uncertainties and correlations, as well as parametric inputs.

Starting from the pseudoscalar channels $B^+\to K^+\nu\bar\nu$ and $B^0\to K^0\nu\bar\nu$, we observe that
the predictions based on the DM FFs are lower than the ones obtained in Refs.~\cite{Parrott:2022zte,Becirevic:2023aov}. This behaviour is consistent with the differences observed in the determination of the
$f_+(q^2)$ FF, as illustrated in the central panel of
Fig.~\ref{fig:FF_BtoK}, mildly increasing the excess to a level of $2.7\sigma$. 
The study of the vector channels $B^+\to K^{*+}\nu\bar\nu$ and $B^0\to K^{*0}\nu\bar\nu$
allows us instead to inspect the impact of LCSR inputs: indeed,
when LCSR constraints are included, the central values of the branching fractions are
shifted to slightly lower values and the associated uncertainties are significantly reduced.
This behaviour can be traced back to the tighter constraints 
over the $B\to K^*$ FFs
at low and intermediate values of $q^2$, which are not directly accessible in current
lattice simulations, as shown in Fig.~\ref{fig:FF_BtoKstar}.
As a result, the LQCD+LCSR DM predictions for the vectorial channels are more precise, while the LQCD DM one can be considered more conservative.

The updated SM predictions presented in this section provide an essential theoretical benchmark for ongoing and future searches for $B\to K^{(*)}\nu\bar\nu$ decays at Belle~II and hadron collider experiments. Moreover, improved control over hadronic uncertainties in these modes strengthens their complementarity with $b\to s\ell^+\ell^-$ decays in global analyses of potential NP effects.

% ------------------------------------------------------------
\subsection{\boldmath$b\to s\ell^+\ell^-$ transitions}
\label{subsec:SM_b2sll}
% ------------------------------------------------------------

The dynamics of the decays $B \to K^{(*)}\ell^+\ell^-$ and $B_s \to \phi\ell^+\ell^-$ can be conveniently described in terms of helicity amplitudes~\cite{Jager:2012uw,Gratrex:2015hna}. In the SM and at a renormalization scale of order the bottom-quark mass
$\mu\simeq m_b$, these amplitudes can be written schematically as
\bea\label{eq:Hamps}
H_V^{\lambda} &\propto&
\left[
C_9^{\rm SM}\,\widetilde{V}_{L\lambda}(q^2)
+
\frac{m_B^2}{q^2}
\left(
\frac{2m_b}{m_B}\,C_7^{\rm SM}\,\widetilde{T}_{L\lambda}(q^2)
-
16\pi^2\,h_{\lambda}(q^2)
\right)
\right]\,, \nonumber\\
H_A^{\lambda} &\propto&
C_{10}^{\rm SM}\,\widetilde{V}_{L\lambda}(q^2)\,, \qquad
H_P \propto
\frac{m_{\ell}\,m_b}{q^2}\,
C_{10}^{\rm SM}
\left(
\widetilde{S}_{L}(q^2)
-
\frac{m_s}{m_b}\,\widetilde{S}_{R}(q^2)
\right)\,,
\eea
where $\lambda = 0,\pm\,(0)$ denotes the helicity of the final state vector (pseudoscalar) meson and
$C_{7,9,10}^{\rm SM}$ are the Wilson coefficients (WC) of the
$|\Delta B| = 1$ weak effective Hamiltonian
normalized as in Ref.~\cite{Ciuchini:2019usw}
(see Refs.~\cite{Buchalla:1995vs,Grinstein:2015nya,Silvestrini:2019sey} for reviews).

The factorizable contributions to the helicity amplitudes are expressed, for vectorial $B\to K^*$ or $B_s\to\phi$ transitions, in terms of the 
FFs $\widetilde{V}_{L\lambda}^V$, $\widetilde{T}_{L\lambda}^V$, $\widetilde{S}_{L}^V$ and $\widetilde{S}_{R}^V$, and for pseudoscalar $B\to K$ ones, in terms on the FFs $\widetilde{V}_{L}^P$, $\widetilde{T}_{L}^P$, $\widetilde{S}_{L}^P$ and $\widetilde{S}_{R}^P$. The relations between these quantities and the ones introduced in Sec.~\ref{sec:DM_FF} can be found in Ref.~\cite{Jager:2012uw}.
Beyond factorization, additional contributions arise at loop level from the
insertion of four-quark operators that are neither CKM-suppressed nor multiplied by
small WCs.
The dominant effect is induced by the current–current operator
\bea
Q_2^c = (\bar{s}_L\gamma_\mu c_L)(\bar{c}_L\gamma^\mu b_L)\,,
\eea
which generates non-factorizable power corrections to the vector helicity amplitudes
$H_V^{\lambda}$ through the hadronic correlator $h_\lambda(q^2)$
\cite{Jager:2014rwa,Ciuchini:2015qxb,Chobanova:2017ghn}.
This correlator is defined in terms of the time-ordered product
\begin{equation}
\label{eq:hlambda}
\frac{\epsilon^*_\mu(\lambda)}{m_B^2}
\int d^4x\; e^{iq\cdot x}\,
\langle M \vert
\mathcal{T}\!\left\{
j^\mu_{\rm em}(x)\,
Q^{c}_{2}(0)
\right\}
\vert \bar B \rangle\,,
\end{equation}
where $j^\mu_{\rm em}(x)$ denotes the electromagnetic quark current and $M$ is the final 
state meson.

A method to compute the correlator in Eq.~\eqref{eq:hlambda} from first principles on the lattice has been recently proposed in Ref.~\cite{Frezzotti:2025hif}. However, a full-fledged calculation is not yet available, so in the following we use a phenomenological parametrization and determine the correlator from data. Alternative approaches rely on the estimate of a subset of the relevant contributions (namely those corresponding to using only the charm part of the electromagnetic current in Eq.~\eqref{eq:hlambda}) using LCSR at low $q^2$~\cite{Khodjamirian:2010vf,Khodjamirian:2012rm}, with the caveats discussed in 
Ref.~\cite{Melikhov:2022wct}, then using dispersion relations to obtain the correlator at larger values of $q^2$. Further developments include implementing the LCSR calculation at negative $q^2$ \cite{Bobeth:2017vxj}, refining the use of dispersion relations \cite{Gubernari:2020eft,Gubernari:2022hxn}, estimating in a model-dependent way some triangle diagrams related to charmed meson rescattering \cite{Isidori:2024lng,Isidori:2025dkp}, and taking anomalous thresholds into account \cite{Mutke:2024tww,Gopal:2024mgb,Hoferichter:2026jlh}. 

Extracting the correlator in Eq.~\eqref{eq:hlambda} from data is a conservative choice but comes at the price of giving up the possibility of identifying eventual lepton-universal NP contributions to $C_9$~\cite{Ciuchini:2015qxb,Ciuchini:2016weo,Ciuchini:2017mik,Ciuchini:2018anp,Ciuchini:2021smi,Ciuchini:2022wbq}, since the contribution of the correlator partly overlaps with $C_9$, as we discuss in detail below. 

In the following, we parametrize 
the correlator in Eq.~\eqref{eq:hlambda} for $B\to K^*$ and $B_s\to\phi$ transitions as follows~\cite{Ciuchini:2018anp}:
\begin{eqnarray} 
\label{eq:newhs}
h_-(q^2) & = & \frac{m_b}{8\pi^2 m_B} \widetilde T_{L -}^V(q^2) h_-^{(0)} + \frac{\widetilde V_{L -}^V(q^2)}{16\pi^2 m_B^2} h_-^{(1)} q^2 + {h_-^{(2)}} q^4+{\cal O}(q^6)\,, \nonumber\\
h_+(q^2) &=&  \frac{m_b}{8\pi^2 m_B} \widetilde T_{L +}^V(q^2) h_-^{(0)} + \frac{\widetilde V_{L +}^V(q^2)}{16\pi^2 m_B^2}  h_-^{(1)} q^2 + h_+^{(0)} + {h_+^{(1)}}q^2 + {h_+^{(2)}} q^4+{\cal O}(q^6)\,, \nonumber\\
h_0(q^2) &=& \frac{m_b}{8\pi^2 m_B} \widetilde T_{L 0}^V(q^2) h_-^{(0)} + \frac{\widetilde V_{L 0}^V(q^2)}{16\pi^2 m_B^2}  h_-^{(1)} q^2 + {h_0^{(0)}}\sqrt{q^2} + {h_0^{(1)}}(q^2)^\frac{3}{2}  +{\cal O}((q^2)^\frac{5}{2})\,,
\end{eqnarray}
which allows us to write the helicity amplitudes as:
\begin{eqnarray} 
H_V^{-} &\propto  
 & \ \ \frac{m_B^2}{q^2} \ \  \bigg[ \frac{2m_b}{m_B}\left(C_7^{\rm SM} - h_-^{(0)} \right) \widetilde T_{L -}^V  
  -  16\pi^2 {h_-^{(2)}}\, q^4 \bigg] + \left(C_9^{\rm SM} - h_-^{(1)}\right)\widetilde V_{L -}^V\,, \nonumber \\ 
H_V^{+} &\propto  
 & \ \ \frac{m_B^2}{q^2} \ \ \bigg[ \frac{2m_b}{m_B}\left(C_7^{\rm SM} - h_-^{(0)} \right) \widetilde T_{L +}^V  
  -  16\pi^2 \Big(h_{+}^{(0)} + {h_{+}^{(1)}}\, q^2 + {h_{+}^{(2)}}\, q^4\Big) \bigg] 
  + \left(C_9^{\rm SM} - h_-^{(1)}\right)\widetilde V_{L +}^V\,, \nonumber \\ 
H_V^{0} &\propto  
 & \ \ \frac{m_B^2}{q^2}\ \ \bigg[ \frac{2m_b}{m_B}\left(C_7^{\rm SM} - h_-^{(0)} \right) \widetilde T_{L 0}^V -16\pi^2 \sqrt{q^2} \Big({h_{0}^{(0)}}
  + {h_{0}^{(1)}}\, q^2 \Big) \bigg] + \left(C_9^{\rm SM} - h_-^{(1)}\right)\widetilde V_{L 0}^V\,.
\label{eq:HVs}
\end{eqnarray}
In a similar fashion, we parametrize the analogous correlator for $B\to K$ transitions in the following way
\begin{eqnarray} 
\label{eq:h_K}
h_K(q^2) &=& 
\frac{m_b}{8\pi^2 m_B} \widetilde T_{L}^P(q^2) h_-^{(0)} + \frac{\widetilde V_{L}^P(q^2)}{16\pi^2 m_B^2}  h_-^{(1)} q^2 + \frac{\sqrt{\lambda_K}}{2 m_B^3} \left(h_K^{(1)}\sqrt{q^2} + {h_K^{(2)}}(q^2)^\frac{3}{2}  +{\cal O}((q^2)^\frac{5}{2})\right)\,,
\end{eqnarray}
with $\lambda_K \equiv \lambda(m_B^2,m_K^2,q^2)$ the usual K\"allén function, which allows us to write its helicity amplitude as:
\begin{eqnarray} 
H_V^{P} &\propto
& \ \ \frac{m_B^2}{q^2}\ \ \bigg[ \frac{2m_b}{m_B}\left(C_7^{\rm SM} - h_-^{(0)} \right) \widetilde T_{L}^P -16\pi^2 \frac{\sqrt{\lambda_K} \sqrt{q^2}}{2 m_B^3} \Big(h_{K}^{(1)}
+ {h_{K}^{(2)}}\, q^2 \Big) \bigg] + \left(C_9^{\rm SM} - h_-^{(1)}\right)\widetilde V_{L}^P\,.
\label{eq:HV_K}
\end{eqnarray}
Although these parametrizations are primarily phenomenological in nature, they have the
advantage of making the interplay between hadronic effects and potential new physics
contributions fully transparent.
In particular, the real parts of the coefficients $h_-^{(0)}$ and $h_-^{(1)}$ enter the helicity amplitudes
in the same way as lepton-universal NP shifts in the real parts of the WCs
$C_7$ and $C_9$, respectively.\footnote{We have updated the parametrization of $B \to K$ matrix element to mimic possible lepton flavour universal $\Delta C_9$ also in this channel.}
As a result, theoretical assumptions on the allowed size of these hadronic parameters play
a crucial role in the extraction and interpretation of possible NP contributions to
$C_7$ and $C_9$ in global analyses. Notice that imaginary parts of $h_\lambda^{(i)}$ parameters as well as real parts of $h_-^{(2)}$, $h_+^{(0,1,2
)}$, $h_0^{(0,1)}$ and $h_K^{(1,2)}$ do not enter the helicity amplitudes in the same way as NP contributions to $C_7$ and $C_9$, so they can be identified as genuine hadronic contributions without any ambiguity. 

In the following, we employ two different approaches regarding 
the treatment of non-local hadronic effects:
\begin{itemize}
    \item \emph {Data Driven}: No strong theory bias is imposed on the size of
    non-factorizable contributions. The coefficients $h_\lambda^{(i)}$ are treated as free
    parameters with broad priors, allowed to be complex in general, and fitted to experimental
    data. This approach captures the
    possibility of significant charm-loop and rescattering effects beyond leading power. This is the approach we have employed in our previous studies, see Refs.~\cite{Ciuchini:2015qxb,Ciuchini:2016weo,Ciuchini:2017mik,Ciuchini:2018anp,Ciuchini:2019usw,Ciuchini:2020gvn,Ciuchini:2021smi,Ciuchini:2022wbq}.
    \item \emph {Model Dependent}: Power corrections are assumed to be
    parametrically suppressed and dominated by leading contributions that mimic shifts in
    short-distance coefficients. The correlator in eq.~\eqref{eq:hlambda} is assumed to be well described by the approach of Refs.~\cite{Khodjamirian:2010vf,Khodjamirian:2012rm,Bobeth:2017vxj,Chrzaszcz:2018yza,Gubernari:2020eft,Gubernari:2022hxn}, yielding a subleading effect to the hadronic effects computable in QCD factorization. 
\end{itemize}

In this Section, we focus on updating the SM fit for the determination of the non-local hadronic parameters in the \emph{Data Driven} approach (further studies 
performed with the \emph{Model Dependent} approach will be relegated 
to Sec.~\ref{sec:NP_updates}). Results in the \emph{Data Driven} approach are summarized in Appendix~\ref{app:AppC}, where we report the 68\% and 95\% Highest Posterior Density Intervals (HPDI) for the hadronic parameters $h_\lambda^{(i)}$ together with their correlations.

\begin{figure}[!t!]
    \centering
    \includegraphics[width=0.45\linewidth]{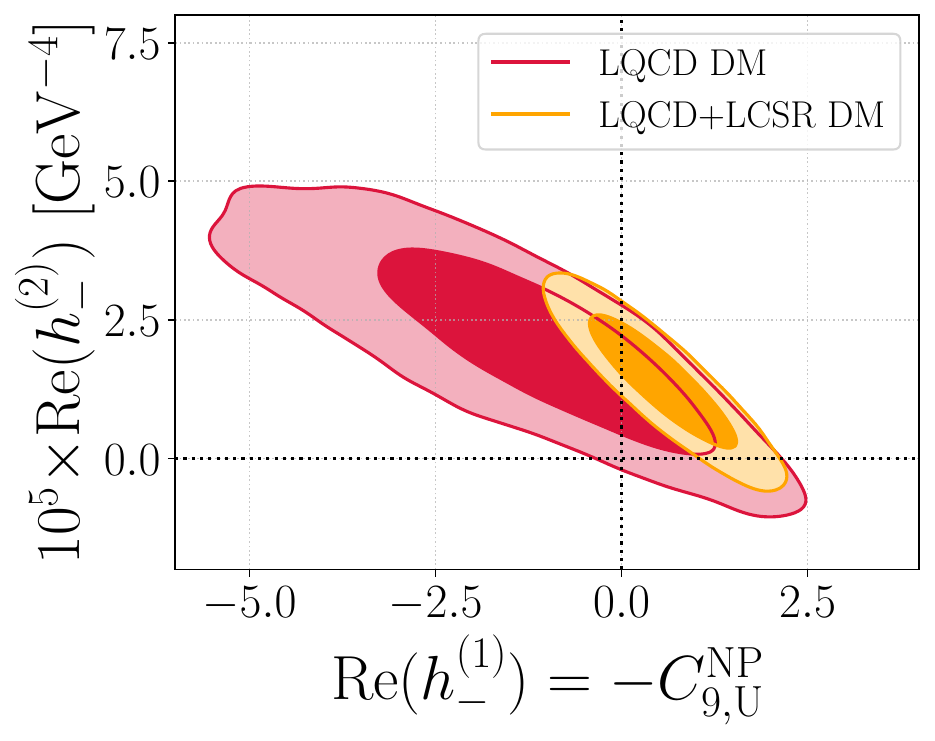}
    \hfill
    \includegraphics[width=0.465\linewidth]{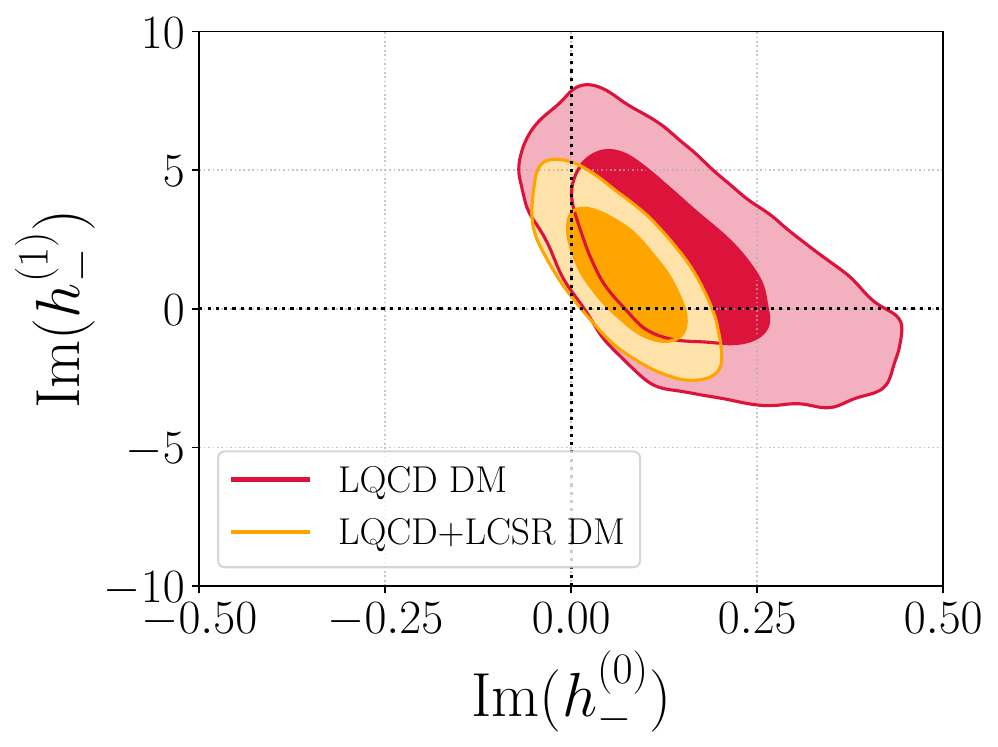}
    \caption{Left Panel: Joint posterior p.d.f. for Re($h^{(1)}_-$) and Re($h^{(2)}_-$) in the SM fit in the \emph{Data Driven} scenario. The results obtained employing FF based on Lattice results only are shown in red, while the ones obtained when also LCSR input are included in the FF determination are shown in orange. Darker (lighter) regions correspond to 68\% (95\%) probability. Notice that according to our hadronic parameterization given in Eqs.~\eqref{eq:hlambda} and~\eqref{eq:h_K}, Re($h^{(1)}_-$) can be reinterpreted as a lepton universal NP contribution, $C_{9,{\rm U}}^{\rm NP}$. Right Panel: Joint posterior p.d.f. for Im($h^{(0)}_-$) and Im($h^{(1)}_-$) in the SM fit in the \emph{Data Driven} scenario.}
    \label{fig:hm1_hm2}
\end{figure}

Two key results of our analysis are displayed in Fig.~\ref{fig:hm1_hm2}:
\begin{itemize}
\item On the left we report the two-dimensional p.d.f.~for Re($h^{(2)}_-$) versus Re($h^{(1)}_-$), which is degenerate with flavour-universal NP contributions to $C_9$. Using LCSR information in the local FFs the two-dimensional 95\% probability contour does not include the origin, pointing to a non-zero combination of Re($h^{(1)}_-$) and Re($h^{(2)}_-$), compatible with a $C_{9,\rm U}^{\rm NP}\neq 0$. In the more conservative LQCD DM approach, however, the origin is included in the 95\% probability region. Furthermore, the 68\% probability contour requires Re($h^{(2)}_-$)~$\neq 0$, hinting at non-negligible hadronic effects rather than lepton-flavour universal NP. 
\item A second interesting two-dimensional p.d.f.~for Im($h^{(1)}_-$) versus Im($h^{(0)}_-$) is reported in the right panel. As recently pointed out in Ref.~\cite{Altmannshofer:2026cwk}, the updated angular analysis of $B^0 \to K^{*0} \mu^{+}\mu^{-}$ by the LHCb collaboration \cite{LHCb:2025mqb} hints at a non-vanishing $S_7$, which requires a CP-conserving imaginary part coming necessarily from the hadronic matrix element. Notice that even if Re($h^{(2)}_-$)~$=0$ and Im($h^{(0)}_-$)~$=0$, then data hint at Re($h^{(1)}_-$)$\,\sim\,$Im($h^{(1)}_-$)~$\neq 0$, disfavouring the interpretation in terms of NP only. We observe also in this case that the allowed region is much larger when the LQCD DM approach is used, but the requirement of non-vanishing imaginary parts is robust at 95\% probability. 
\end{itemize}

The robustness of these conclusions has been tested by varying the binning of the
experimental data entering the fit. We found that the joint posterior distributions
obtained using different binning choices are fully compatible within uncertainties.
This indicates that, at the current level of experimental precision, the use of finer binning schemes -- despite increasing the number of data points -- does not lead to a significant improvement in the determination of the hadronic parameters, due to the correspondingly larger uncertainties.

Finally, it is worth mentioning that a $q^2$ dependence is found also for the hadronic parameters entering the $B\to K$ transition,\footnote{Due to branching fractions being the only experimental results currently available in this channel, we refrain from fitting for hadronic imaginary parts, which would be redundant.} with the coefficient ${\rm Re}\,(h_K^{(2)})$ being 
different from 0 at $\sim3\sigma$, see Tab.~\ref{tab:hlambda}. This $q^2$ dependence is induced by data in the [$0.1, 0.98$]~GeV$^2$ bin; the presence of sizeable long-distance effects in this bin could also be expected from the presence of light resonances, which are visible in the $q^2$ spectrum of $B \to K^{*} \mu^+\mu^-$ (see Fig.~5 of Ref.~\cite{LHCb:2024onj}).

Overall, these results confirm that a careful treatment of hadronic uncertainties -- both in form factors and in non-local matrix elements -- plays a crucial role in the interpretation of $b\to s\ell^+\ell^-$ observables. They also highlight the importance of controlling hadronic contributions before claiming  the presence of short-distance NP effects.

% ============================================================
\section{Updates on rare decays: New Physics}
\label{sec:NP_updates}
% ============================================================

We now turn to the interpretation of rare $b\to s$ decays in the presence of possible physics beyond the Standard Model, with the caveats discussed in the previous section. In the following we analyse potential NP effects in the two complementary classes of processes discussed above. First, we consider $b\to s\nu\bar\nu$ transitions, whose theoretical cleanliness makes them particularly suitable for probing modifications of the short-distance Wilson
coefficients associated with left- and right-handed quark currents.
We then study $b\to s\ell^+\ell^-$ decays, where the interpretation of the data depends more strongly on the treatment of non-local hadronic effects.
In this case we perform global fits under two alternative assumptions regarding the size of power corrections — the \emph{Data Driven} and \emph{Model Dependent} 
approaches detailed in Sec.~\ref{subsec:SM_b2sll} — in order to assess how theoretical assumptions influence the inferred preference for NP scenarios.

% ------------------------------------------------------------
\subsection{Rare invisible decays: $b\to s\nu\bar\nu$}
\label{subsec:NP_b2snunu}
% ------------------------------------------------------------

In the presence of physics beyond the Standard Model, the short-distance structure
of $b\to s\nu\bar\nu$ transitions can be modified by additional contributions to the
WCs multiplying left- and right-handed quark currents. The most general
low-energy effective Hamiltonian relevant for $b\to s\nu\bar\nu$ decays can be written as
\begin{equation}
\mathcal{H}_{\rm eff,\, NP}^{b\to s\nu\bar\nu}
=
-\frac{4G_F}{\sqrt{2}}\,V_{tb}V_{ts}^*\,
\frac{\alpha_e}{4\pi}\,
\sum_{i,j=e,\mu,\tau}
\left[
C_L^{\nu_i\nu_j}\,
\left(\bar{s}\gamma^\mu P_L b\right)
\left(\bar{\nu}_i\gamma_\mu(1-\gamma_5)\nu_j\right)
+
C_R^{\nu_i\nu_j}\,
\left(\bar{s}\gamma^\mu P_R b\right)
\left(\bar{\nu}_i\gamma_\mu(1-\gamma_5)\nu_j\right)
\right]\,,
\label{eq:Heff_b2snunu_NP}
\end{equation}
with NP effects that can be
encoded in flavour-dependent shifts of the WCs,
\begin{equation}
C_{L,R}^{\nu_i\nu_j}
=
C_{L,\nu}^{\rm SM}\,\delta_{ij}
+
\delta C_{L,R}^{\nu_i\nu_j}\,,
\end{equation}
where $i,j$ label neutrino flavours.
In this framework, the relative NP correction to the branching fractions of
$B\to K^{(*)}\nu\bar\nu$ decays can be written as~\cite{Buras:2014fpa,Allwicher:2023xba}
\begin{equation}
\delta\mathcal{B}_{K^{(*)}}^{\nu\bar\nu}
=
\sum_i
\frac{2\,\mathrm{Re}\!\left[C_{L,\nu}^{\rm SM}\left(\delta C_L^{\nu_i\nu_i}
+
\delta C_R^{\nu_i\nu_i}\right)\right]}
{3\,|C_{L,\nu}^{\rm SM}|^2}
+
\sum_{i,j}
\frac{\left|\delta C_L^{\nu_i\nu_j}
+
\delta C_R^{\nu_i\nu_j}\right|^2}
{3\,|C_{L,\nu}^{\rm SM}|^2}
-
\eta_{K^{(*)}}
\sum_{i,j}
\frac{\mathrm{Re}\!\left[
\delta C_R^{\nu_i\nu_j}
\left(
C_{L,\nu}^{\rm SM}\,\delta_{ij}
+
\delta C_L^{\nu_i\nu_j}
\right)\right]}
{3\,|C_{L,\nu}^{\rm SM}|^2}\,,
\label{eq:deltaB_nunu_general}
\end{equation}
where $\eta_K=0$ for $B\to K\nu\bar\nu$ and % $\eta_{K^*}=3.33(7)$ 
$\eta_{K^*}\sim \mathcal{O}(1)$ for $B\to K^*\nu\bar\nu$.
The first two terms correspond to purely left-handed and purely right-handed contributions, while the last term accounts for the interference between left- and right-handed quark currents, which is present only in the vector final state.

\begin{table*}[!t!]
\centering
\setlength{\tabcolsep}{10pt}
\renewcommand{\arraystretch}{1.35}
\begin{tabular}{|c|c|cc|}
\hline
\textbf{Wilson coefficient} & \textbf{Form Factors} &
\textbf{68\% HPDI} & \textbf{95\% HPDI} \\
\hline
\multicolumn{4}{|c|}{\textbf{Left-handed NP only: $\delta C_L^{\nu}\neq 0$, $\delta C_R^{\nu}=0$}} \\
\hline
\multirow{2}{*}{$\delta C_L^{\nu}$}
& LQCD DM & $[-4.22, 1.81] \cup [10.90, 16.49]$ & $[-5.33, 5.84] \cup [6.36, 17.65]$ \\
& LQCD+LCSR DM & $[-3.71, 1.64] \cup [10.88, 16.66]$ & $[-4.80, 5.59] \cup [6.75, 17.63]$ \\
\hline
\multicolumn{4}{|c|}{\textbf{Right-handed NP only: $\delta C_R^{\nu}\neq 0$, $\delta C_L^{\nu}=0$}} \\
\hline
\multirow{2}{*}{$\delta C_R^{\nu}$}
& LQCD DM & $[-9.21, -4.59]$ & $[-11.11, -1.21]$ \\
& LQCD+LCSR DM & $[-9.24, -4.72]$ & $[-11.07, -1.53]$ \\
\hline
\multicolumn{4}{|c|}{\textbf{General NP: $\delta C_L^{\nu}\neq 0$, $\delta C_R^{\nu}\neq 0$}} \\
\hline
\multirow{2}{*}{$\delta C_L^{\nu}+\delta C_R^{\nu}$}
& LQCD DM & $[-9.40, -4.33] \cup [17.10, 22.35]$ & $[-11.67, -0.04] \cup [12.99, 24.53]$  \\
& LQCD+LCSR DM & $[-9.45, -4.28] \cup [17.07, 22.15]$ & $[-11.69, -0.06] \cup [12.77, 24.39]$ \\
\hline
\multirow{2}{*}{$\delta C_L^{\nu}-\delta C_R^{\nu}$}
& LQCD DM & $[1.75, 10.79]$ & $[-1.94, 14.48]$ \\
& LQCD+LCSR DM & $[1.92, 11.05]$ & $[-1.88, 14.48]$ \\
\hline
\end{tabular}
\caption{68\% and 95\% HPDI of the posterior distributions of the NP WCs
$\delta C_L^{\nu}$ and $\delta C_R^{\nu}$ obtained from fits to
$B\to K^{(*)}\nu\bar\nu$ observables under different assumptions on the chiral structure
of NP contributions.
Results are shown for both the LQCD DM FFs and the LQCD+LCSR DM FFs.
All fits assume flavour-diagonal and flavour-universal NP.}
\label{tab:btosnunu_NP}
\end{table*}

In this work we restrict ourselves to the flavour-diagonal and flavour-universal NP
scenario, in which NP effects do not induce lepton-flavour violation and are identical
for the three light neutrino flavours. In this limit the WCs satisfy 
$\delta C_{L,R}^{\nu_i\nu_j} = \delta_{ij}\,\delta C_{L,R}^{\nu}$, 
and Eq.~(\ref{eq:deltaB_nunu_general}) simplifies to 
\begin{equation}
\delta\mathcal{B}_{K^{(*)}}^{\nu\bar\nu}
=
\frac{2\,\mathrm{Re}\!\left[C_L^{\rm SM}
\left(\delta C_L^{\nu}+\delta C_R^{\nu}\right)\right]}
{|C_L^{\rm SM}|^2}
+
\frac{\left|\delta C_L^{\nu}+\delta C_R^{\nu}\right|^2}
{|C_L^{\rm SM}|^2}
-
\eta_{K^{(*)}}
\frac{\mathrm{Re}\!\left[
\delta C_R^{\nu}
\left(C_L^{\rm SM}+\delta C_L^{\nu}\right)\right]}
{|C_L^{\rm SM}|^2}\,.
\label{eq:deltaB_nunu_LFU}
\end{equation}
The full branching fractions in the presence of NP are then obtained as
\begin{equation}
\mathcal{B}(B\to K^{(*)}\nu\bar\nu)
=
\mathcal{B}(B\to K^{(*)}\nu\bar\nu)_{\rm SM}
\left(1+\delta\mathcal{B}_{K^{(*)}}^{\nu\bar\nu}\right)\,.
\end{equation}
In the SM limit, $\delta C_L^{\nu}=\delta C_R^{\nu}=0$, one recovers the predictions
discussed in the previous subsection. A notable feature of the NP parametrization in
Eq.~(\ref{eq:deltaB_nunu_LFU}) is the different sensitivity of pseudoscalar and vector
final states to right-handed quark currents.
For $B\to K\nu\bar\nu$ decays, corresponding to $\eta_K=0$, NP effects enter only through
the combination $\delta C_L^\nu+\delta C_R^\nu$, and the decay rate is
insensitive to the chirality of the quark current.
In contrast, for $B\to K^*\nu\bar\nu$ decays the presence of $\eta_{K^*}\neq 0$ gives rise
to an additional interference term between left- and right-handed contributions.
As a consequence, vector final states provide enhanced sensitivity to right-handed
currents and allow, in principle, to disentangle different NP chiral structures.
This complementarity between $B\to K\nu\bar\nu$ and $B\to K^*\nu\bar\nu$ decays plays an
important role in constraining NP scenarios that modify the chiral structure of
$b\to s\nu\bar\nu$ transitions~\cite{Bause:2023mfe,Allwicher:2023xba}.

The impact of this complementarity is further sharpened by the current experimental
situation. For the $B\to K^*\nu\bar\nu$ and $B^0\to K^0 \nu \bar \nu$ modes, only upper limits are available at present \cite{Belle:2013tnz,Belle:2017oht}:
\begin{eqnarray}
\mathcal{B}(B^0\to K^{0}\nu\bar\nu)
\;&<&\; 2.6 \cdot 10^{-5}\qquad (90\%~{\rm C.L.})\,,\nonumber \\
\mathcal{B}(B^0\to K^{*0}\nu\bar\nu)
\;&<&\; 1.8 \cdot 10^{-5}\qquad (90\%~{\rm C.L.})\,,\nonumber \\
\mathcal{B}(B^+\to K^{*+}\nu\bar\nu)
\;&<&\; 4.0 \cdot 10^{-5}\qquad (90\%~{\rm C.L.})\,,
\label{eq:exp_b2sstar_nunu}
\end{eqnarray}
thereby constraining the allowed size of NP contributions, in particular those involving right-handed currents that are more efficiently probed by vector final states.
In contrast, for the pseudoscalar channel $B^+\to K^+\nu\bar\nu$ we have at disposal the Belle~II experimental measurement~\cite{Belle-II:2023esi}:
\be
\mathcal{B}(B^+\to K^{+}\nu\bar\nu)=(2.3\pm0.7)\times10^{-5}\,,
\ee
above the SM prediction of Table~\ref{tab:btosnunu_SM} at the $2.7\,\sigma$ level. 

\begin{figure}[!t!]
    \centering
    \includegraphics[width=0.45\linewidth]{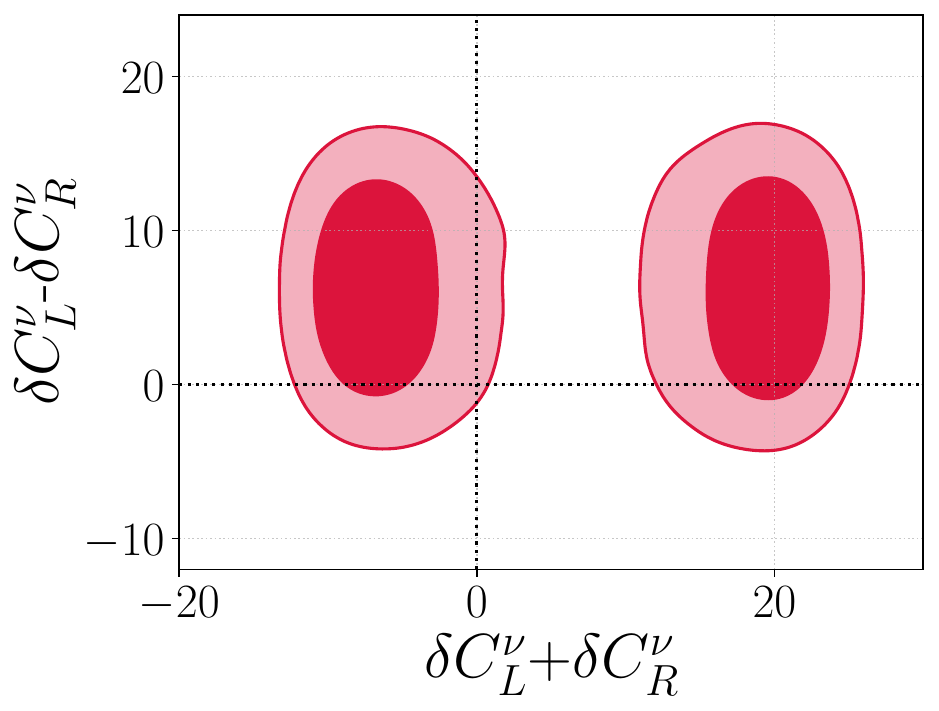}
    \hfill
    \includegraphics[width=0.45\linewidth]{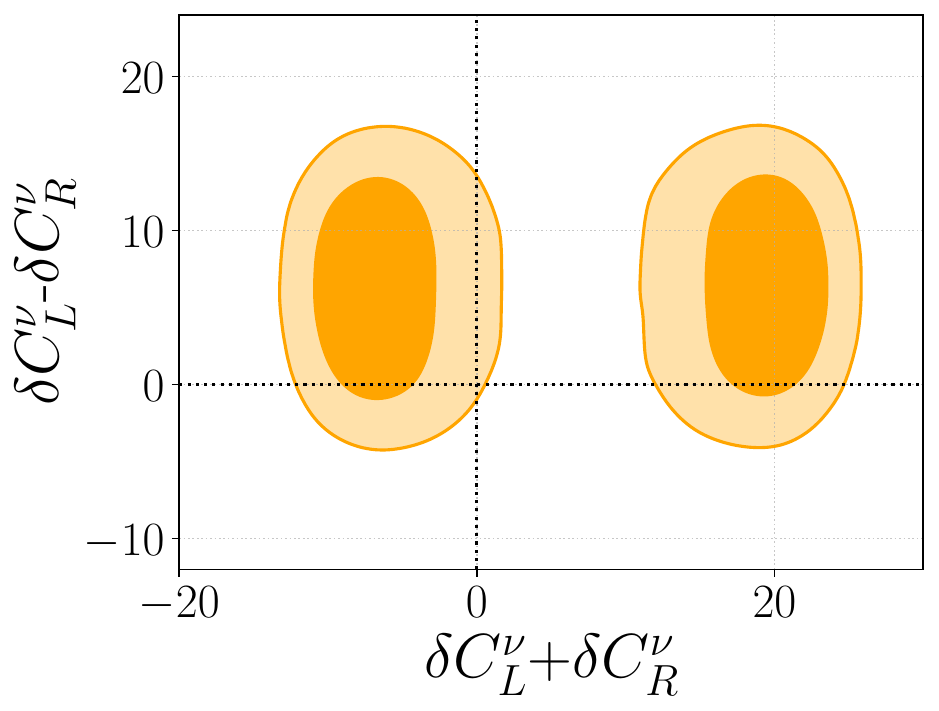}
    \caption{Two-dimensional marginalized posterior distribution in the
    $(\delta C_L^{\nu}+\delta C_R^{\nu},\,\delta C_L^{\nu}-\delta C_R^{\nu})$ plane obtained in the general NP scenario.
    Results are shown for the LQCD DM FFs (left) and for the LQCD+LCSR DM FFs
    (right). Darker (lighter) regions correspond to the 68\% (95\%) probability contours.}
    \label{fig:b2snunu_corr}
\end{figure}

In our numerical analysis, the parameters $\delta C_L^{\nu}$ and $\delta C_R^{\nu}$ are
treated as real and flavour-universal NP coefficients and are constrained using the
combined experimental information on $B\to K^{(*)}\nu\bar\nu$ decays. In order to quantify the impact of NP contributions to $b\to s\nu\bar\nu$ transitions, we consider three benchmark scenarios, defined by different assumptions on the chiral structure of the effective interaction. Specifically, we perform separate fits assuming:
(i) a purely left-handed NP contribution, $\delta C_L^{\nu}\neq 0$ and $\delta C_R^{\nu}=0$;
(ii) a purely right-handed NP contribution, $\delta C_R^{\nu}\neq 0$ and $\delta C_L^{\nu}=0$;
and (iii) a general scenario in which both $\delta C_L^{\nu}$ and $\delta C_R^{\nu}$ are
allowed to vary simultaneously.
In all cases, we work under the assumption of lepton flavour-diagonal and universal NP, as discussed above. For each scenario, we perform fits using both sets of hadronic inputs, namely the LQCD DM FFs and the LQCD+LCSR DM FFs, and we report the corresponding posterior in terms of 68\% and 95\% HPDI. Our results are summarized in Table~\ref{tab:btosnunu_NP}. 

In the scenario with purely left-handed NP contributions, the current data allow for two disconnected regions in $\delta C_L^{\nu}$, reflecting the quadratic dependence of the branching fractions on the WCs. These two solutions, when expressed in terms of the total coefficient $C_L^\nu = C_{L,\nu}^{\rm SM}+\delta C_L^\nu$, correspond to equal magnitude but opposite signs for $C_L^\nu$.
In the case of purely right-handed NP contributions, constraints on $\delta C_R^{\nu}$ are driven predominantly by the $B\to K^*\nu\bar\nu$ modes, which are uniquely sensitive to right-handed quark currents through the interference term proportional to $\eta_{K^*}\neq 0$ in Eq.~\eqref{eq:deltaB_nunu_LFU}. Contrarily to the purely left-handed scenario, the interplay among the different modes leads to the presence of a single solution in this scenario.

In the NP scenario in which both $\delta C_L^{\nu}$ and $\delta C_R^{\nu}$ are allowed to vary simultaneously, correlations between the two WCs play an important role in determining the allowed parameter space. For this reason we also present the two-dimensional marginalized posterior distribution in the $(\delta C_L^{\nu},\,\delta C_R^{\nu})$ plane. This correlation plot illustrates the interplay between left- and right-handed NP contributions and highlights degeneracies arising from the current experimental constraints on $B\to K^{(*)}\nu\bar\nu$ decays.
The corresponding contours at 68\% and 95\% probability are shown in
Fig.~\ref{fig:b2snunu_corr}, for both choices of FFs inputs. All our results are qualitatively in agreements with the early finding of Refs.~\cite{Bause:2023mfe,Allwicher:2023xba,Belle-II:2025lfq}, taking at the same time into account the combined effect of all the upper limits of \eqref{eq:exp_b2sstar_nunu}, including the one for the pseudoscalar neutral mode.

\begin{figure}[!t!]
    \centering
    \includegraphics[width=0.45\linewidth]{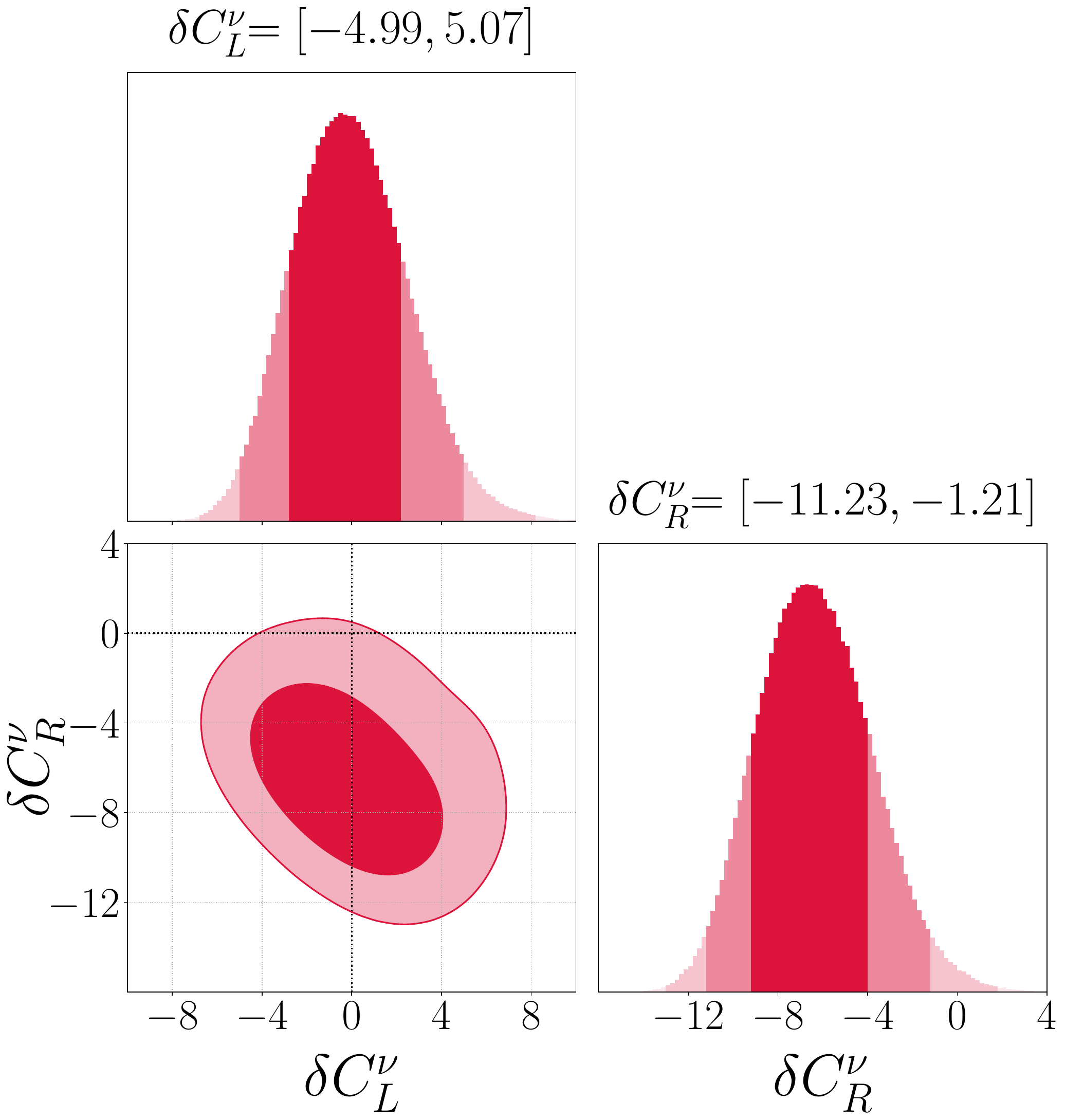}
    \hfill
    \includegraphics[width=0.45\linewidth]{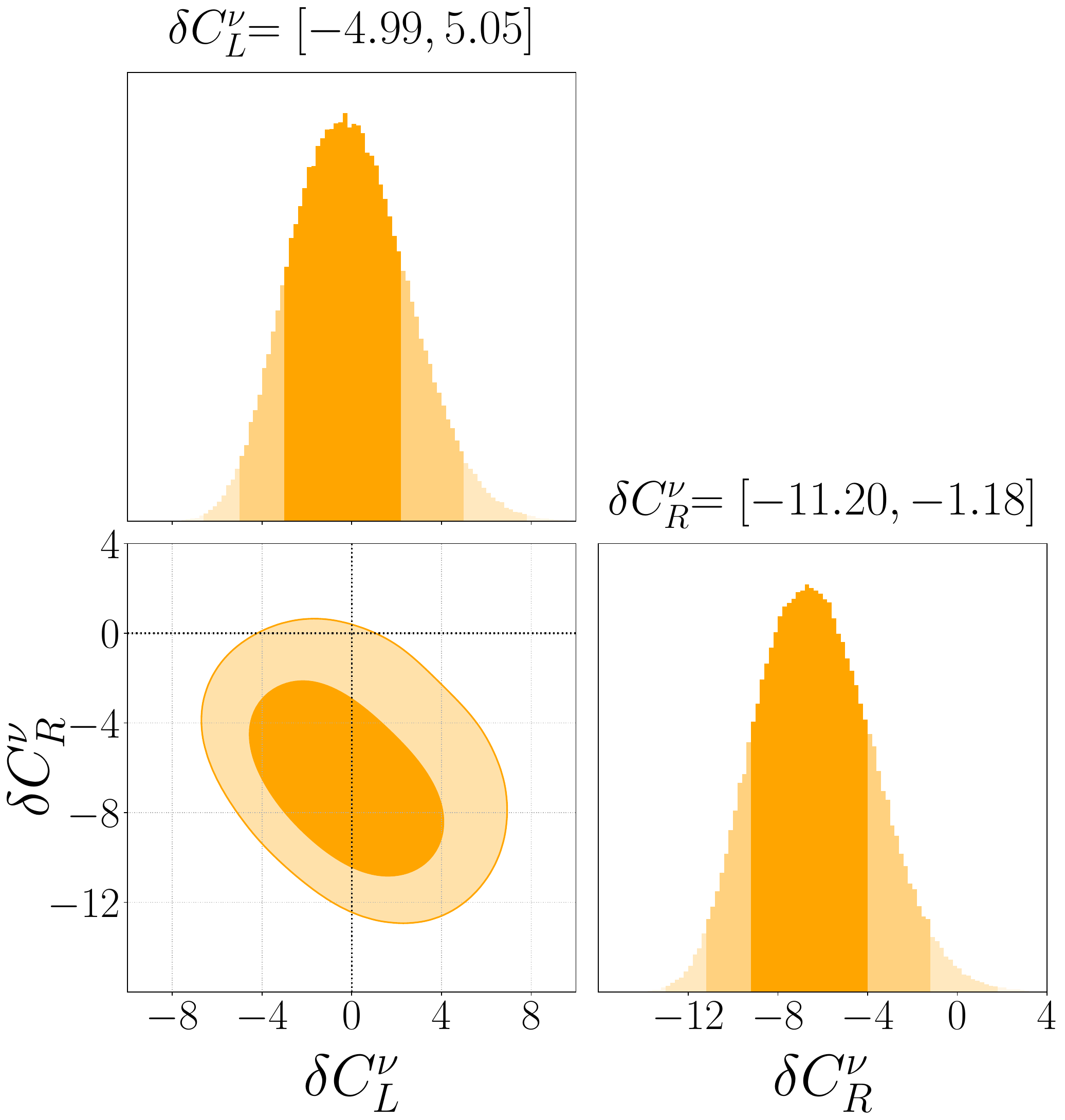}
    \caption{Two- and one-dimensional marginalized posterior distribution in the
    $(\delta C_L^{\nu},\,\delta C_R^{\nu})$ plane obtained in the general NP scenario.
    Results are shown for the LQCD DM FFs (left) and for the LQCD+LCSR DM FFs
    (right). Darker (lighter) regions correspond to the 68\% (95\%) probability contours.}
    \label{fig:b2snunu_corr_tri}
\end{figure}

Notice that in the more general NP scenario where both $\delta C_L^{\nu}$ and $\delta C_R^{\nu}$ are varied simultaneously, strong correlations between the two WCs emerge. Since the $B\to K\nu\bar\nu$ decay depends only on the combination $\delta C_L^{\nu}+\delta C_R^{\nu}$, this mode alone would constrain the parameter space to two vertical bands in the $(\delta C_L^{\nu},\,\delta C_R^{\nu})$ plane, corresponding to
the quadratic dependence of the BR on this sum.
However, the inclusion of constraints from $B\to K^*\nu\bar\nu$, which are sensitive also to the orthogonal combination $\delta C_L^{\nu}-\delta C_R^{\nu}$ through interference effects, restricts these bands to finite regions. As a result, the allowed parameter space is reduced to two disconnected, stripe-like regions, as shown in Fig.~\ref{fig:b2snunu_corr}, clearly illustrating the complementary
role of pseudoscalar and vector final states in constraining the chiral structure of $b\to s\nu\bar\nu$ transitions. 
It is eventually interesting to further inspect the region closer to the SM solution in the original (un-rotated) basis of WCs. The resulting triangle plots are shown in Fig.~\ref{fig:b2snunu_corr_tri}. As illustrated in this figure, the constraints on $\delta C_L^{\nu}$ become compatible with zero within uncertainties, while a sizeable negative contribution to $\delta C_R^{\nu}$ is preferred. In particular, the posterior distribution of $\delta C_R^{\nu}$ excludes the SM value at more than $2\sigma$, a behaviour that is consistently observed for both the LQCD DM and the LQCD+LCSR DM form-factor determinations.

Overall, the comparison between the LQCD DM and LQCD+LCSR DM results shows that the inclusion of LCSR inputs has only a limited impact on the inferred NP parameter space, particularly in the general scenario.
This indicates that the present constraints on NP in $b\to s\nu\bar\nu$ transitions are primarily driven by the experimental situation, with hadronic uncertainties playing a subleading role at the current level of precision.

% ------------------------------------------------------------
\subsection{Rare visible decays: $b\to s\ell^+\ell^-$}
\label{subsec:NP_b2sll}
% ------------------------------------------------------------

In the study of NP effects to $b\to s\ell^+\ell^-$ transitions, we focus here 
on the scenario where deviations from the SM values of the WCs are allowed only
for left-handed quark currents, namely considering non-vanishing values for
$C_{9,e}^{\rm NP}$, $C_{9,\mu}^{\rm NP}$, $C_{10,e}^{\rm NP}$ and 
$C_{10,\mu}^{\rm NP}$. In particular, we performed fits adopting either a
\emph{Data Driven} or a \emph{Model Dependent} description of power
corrections, as defined in Sec.~\ref{subsec:SM_b2sll}, and using either 
the LQCD DM or LQCD+LCSR DM FFs determinations, as detailed in Sec.~\ref{sec:DM_FF}. The fit results where NP
effects are considered in right-handed quark currents as well are reported
in Appendix~\ref{app:AppD}.

\begin{table*}[!t!]
\centering
\setlength{\tabcolsep}{10pt}
\renewcommand{\arraystretch}{1.35}
\begin{tabular}{|c|c|cc|}
\hline
\textbf{Wilson coefficient} & \textbf{Form Factors} & \textbf{68\% HPDI} & \textbf{95\% HPDI} \\
\hline

\multirow{2}{*}{$C_{9,-}^{\rm NP}$}
& LQCD DM & $[-0.75, 0.07]$ & $[-1.28, 0.44]$ \\
& LQCD+LCSR DM & $[-0.57, 0.01]$ & $[-0.88, 0.31]$ \\
\hline

\multirow{2}{*}{$C_{10,e}^{\rm NP}$}
& LQCD DM & $[-0.56, 0.02]$ & $[-0.79, 0.36]$ \\
& LQCD+LCSR DM & $[-0.20, 0.32]$ & $[-0.45, 0.59]$ \\
\hline

\multirow{2}{*}{$C_{10,\mu}^{\rm NP}$}
& LQCD DM & $[-0.08, 0.29]$ & $[-0.27, 0.47]$ \\
& LQCD+LCSR DM & $[0.23, 0.56]$ & $[0.07, 0.74]$ \\
\hline

\end{tabular}
\caption{68\% and 95\% HPDI of the posterior distribution of the WCs $C_{9,-}^{\rm NP}$, $C_{10,e}^{\rm NP}$ and $C_{10,\mu}^{\rm NP}$ obtained from a fit employing the \emph{Data Driven} approach concerning non-local hadronic effects (see text for details). These results have been obtained employing either the FFs based on Lattice results only (LQCD DM), or the ones obtained when also LCSR input are included in the FF determination (LQCD+LCSR DM). Results concerning the $C_{9,+}^{\rm NP}$ WCs are not reported since this coefficient is flatly distributed.}
\label{tab:4WC_FDD}
\end{table*}

\begin{figure}[!t!]
    \centering
    \includegraphics[width=0.75\linewidth]{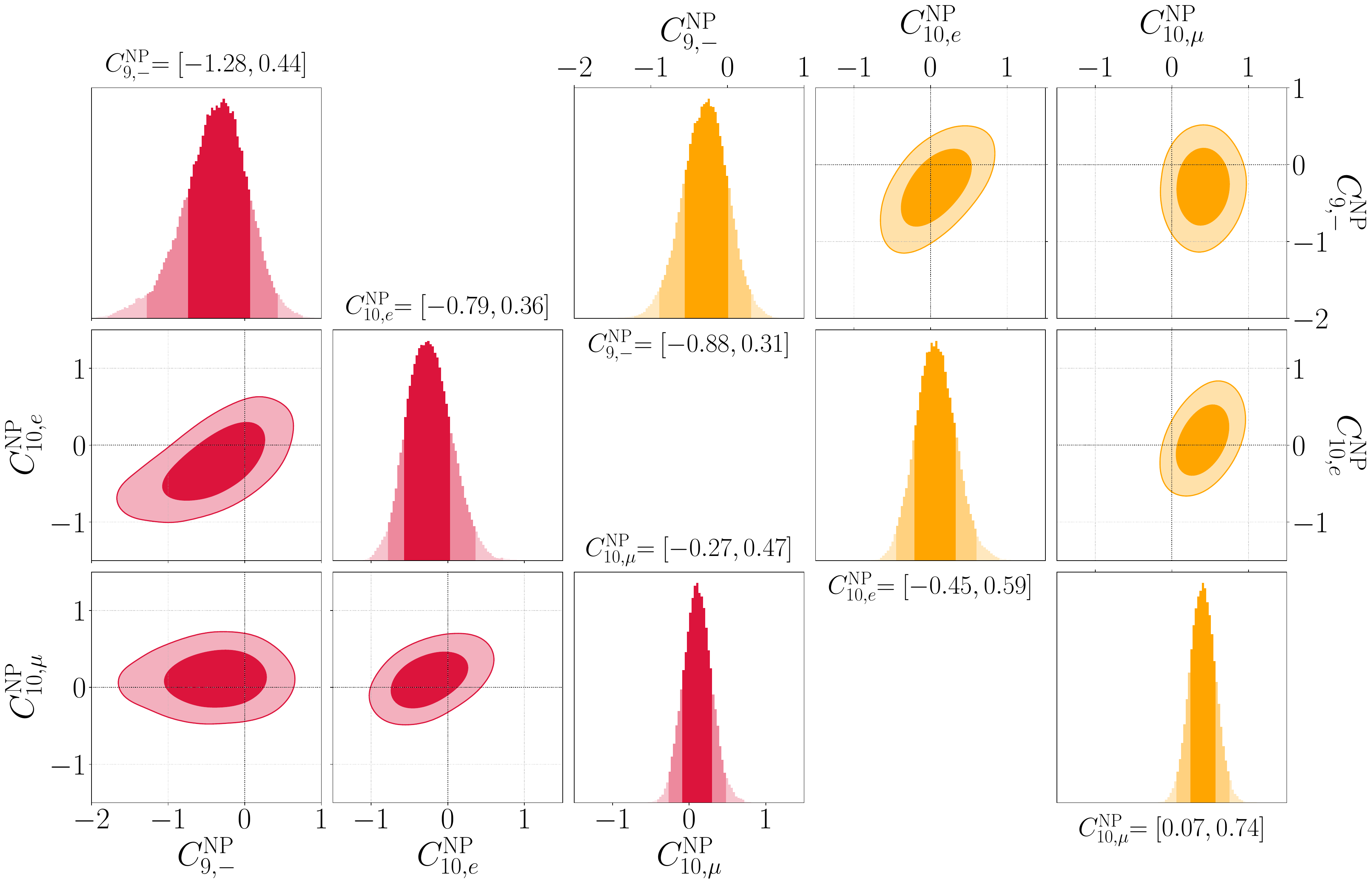}
    \caption{Two- and one-dimensional marginalized joint p.d.f. for the set of  WCs $C_{9,-}^{\rm NP}$, $C_{10,e}^{\rm NP}$ and $C_{10,\mu}^{\rm NP}$ obtained from a fit employing the \emph{Data Driven} approach concerning non-local hadronic effects (see text for details). Results concerning the $C_{9,+}^{\rm NP}$ WCs are not reported since this coefficient is flatly distributed. The results obtained employing LQCD DM FFs are shown in red, while the ones obtained with LQCD+LCSR DM FFs are shown in orange. For both scenarios, we show the 68\% and 95\% probability regions and we quote 95\% probability regions numbers.}
    \label{fig:4D_FDD}
\end{figure}

\begin{table*}[!t!]
\centering
\setlength{\tabcolsep}{10pt}
\renewcommand{\arraystretch}{1.35}
\begin{tabular}{|c|c|cc|}
\hline
\textbf{Wilson coefficient} & \textbf{Form Factors} & \textbf{68\% HPDI} & \textbf{95\% HPDI} \\
\hline

\multirow{2}{*}{$C_{9,e}^{\rm NP}$}
& LQCD DM & $[-1.74, -1.05]$ & $[-2.08, -0.70]$ \\
& LQCD+LCSR DM & $[-1.67, -1.04]$ & $[-1.99, -0.72]$ \\
\hline

\multirow{2}{*}{$C_{10,e}^{\rm NP}$}
& LQCD DM & $[-0.43, -0.01]$ & $[-0.62, 0.22]$ \\
& LQCD+LCSR DM & $[-0.39, 0.00]$ & $[-0.55, 0.24]$ \\
\hline

\multirow{2}{*}{$C_{9,\mu}^{\rm NP}$}
& LQCD DM & $[-1.24, -0.93]$ & $[-1.41, -0.78]$\\
& LQCD+LCSR DM & $[-1.21, -0.95]$ & $[-1.34, -0.82]$ \\
\hline

\multirow{2}{*}{$C_{10,\mu}^{\rm NP}$}
& LQCD DM & $[-0.04, 0.20]$ & $[-0.16, 0.31]$ \\
& LQCD+LCSR DM & $[-0.01, 0.20]$ & $[-0.11, 0.30]$ \\
\hline

\end{tabular}
\caption{68\% and 95\% HPDI of the posterior distribution of the WCs $C_{9,e}^{\rm NP}$, $C_{10,e}^{\rm NP}$, $C_{9,\mu}^{\rm NP}$ and $C_{10,\mu}^{\rm NP}$ obtained from a fit employing the \emph{Model Dependent} approach concerning non-local hadronic effects (see text for details). These results have been obtained employing either the FFs based on Lattice results only (LQCD DM), or the ones obtained when also LCSR input are included in the FF determination (LQCD+LCSR DM). }
\label{tab:4WC_PMD}
\end{table*}

\begin{figure}[!t!]
    \centering
    \includegraphics[width=0.85\linewidth]{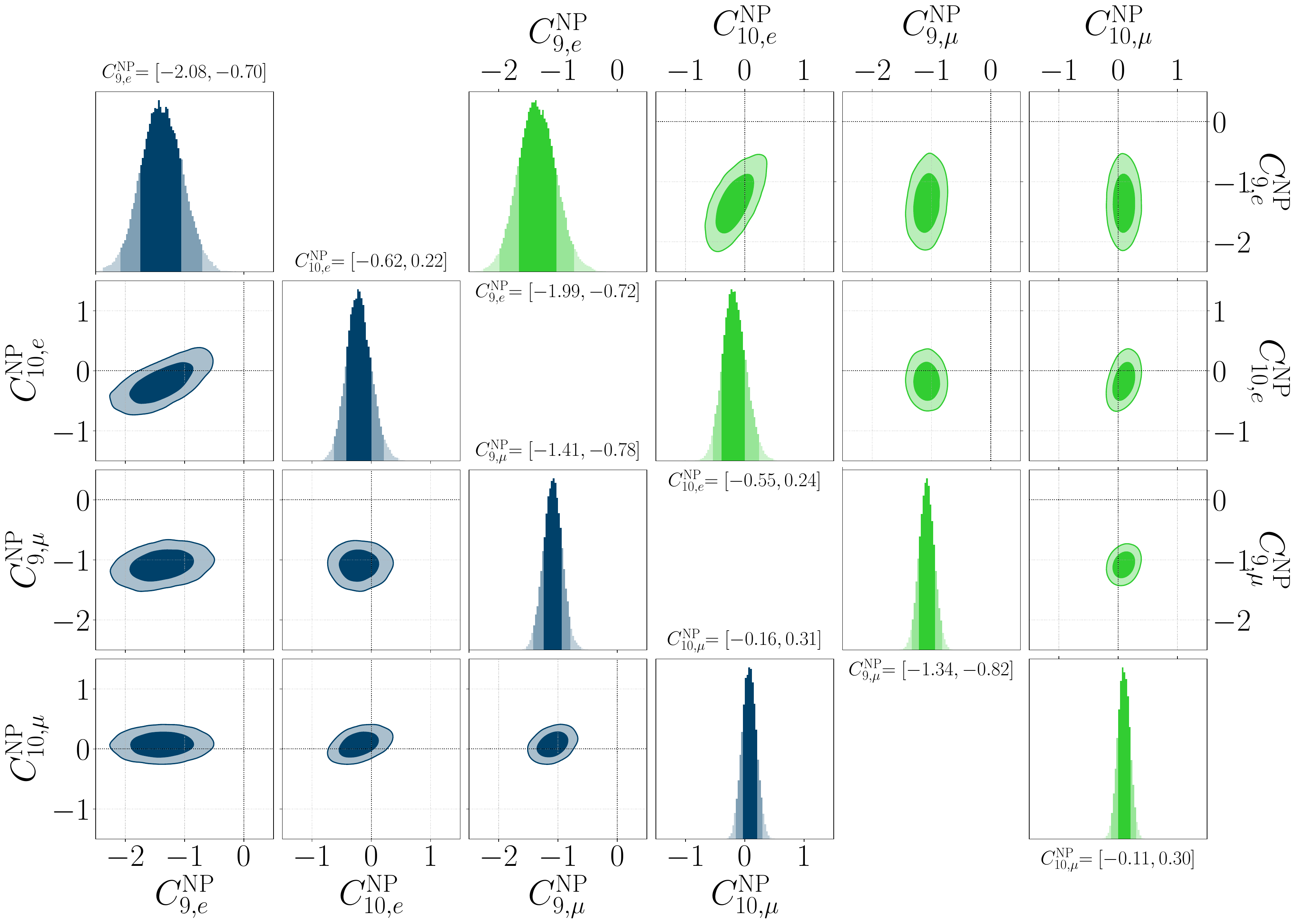}
    \caption{Two- and one-dimensional marginalized joint p.d.f. for the set of  4 WCs $C^{\rm NP}_{i,j}$ with $i=9,10$ and $j=e,\mu$ obtained from a fit employing the \emph{Model Dependent} approach concerning non-local hadronic effects (see text for details). The results obtained employing LQCD DM FFs are shown in blue, while the ones obtained with LQCD+LCSR DM FFs are shown in green. For both scenarios, we show the 68\% and 95\% probability regions and we quote 95\% probability regions numbers.}
    \label{fig:4D_PMD}
\end{figure}

Before discussing the results of our fits, it is useful to recall that, in the 
\emph{Data Driven} approach, the parameter ${\rm Re}\,(h_-^{(1)} )$ is fully 
degenerate with a LFU shift to $C_9$, see Eqs.~\eqref{eq:HVs} and~\eqref{eq:HV_K}. In this context it is therefore useful to
introduce the following notation:
\begin{equation}
C_{9,\pm}^{\rm NP} \equiv C_{9,e}^{\rm NP} \pm C_{9,\mu}^{\rm NP}\,.    
\end{equation}
Indeed, in this parametrization the $C_{9,+}^{\rm NP}$ coefficient is proportional 
to a LFU contribution, and therefore degenerate to ${\rm Re}\,(h_-^{(1)} )$, while
the $C_{9,-}^{\rm NP}$ one encodes genuine LFUV effects. Hence, it is possible to
constrain in a fit only the latter WC, since the former one will give origin to a flat
direction in the $(C_{9,+}^{\rm NP},{\rm Re}\,(h_-^{(1)} ))$ plane. For this reason,
we will report in the \emph{Data Driven} approach below 
only the results relative to the $C_{9,-}^{\rm NP}$ coefficient when discussing
vectorial currents.

We are now ready to discuss our fit results.
The WCs posteriors in the \emph{Data Driven} case are reported
in Table~\ref{tab:4WC_FDD}, while the corresponding marginalized
distributions are shown in Fig.~\ref{fig:4D_FDD}.
Meaningful constraints are obtained for $C_{9,-}^{\rm NP}$,
$C_{10,e}^{\rm NP}$ and $C_{10,\mu}^{\rm NP}$, which are found to be compatible with the SM
within uncertainties.
The choice on whether to employ LQCD DM or LCQD+LCSR DM FFs leads to only mild quantitative changes, indicating that FF uncertainties are subdominant in this fit.

Turning to the \emph{Model Dependent} scenario, whose results are summarized in
Table~\ref{tab:4WC_PMD} and Fig.~\ref{fig:4D_PMD}, a qualitatively different picture
emerges.
In this case, the data favour sizeable negative values of both $C_{9,e}^{\rm NP}$ and
$C_{9,\mu}^{\rm NP}$, while the corresponding $C_{10}$ coefficients remain more weakly
constrained and compatible with the SM.
This pattern is in line with previous global analyses of $b\to s\ell^+\ell^-$ transitions
and reflects the more restrictive treatment of non-factorizable hadronic power
corrections adopted in this framework.
By limiting the freedom of hadronic effects to mimic short-distance contributions, the
\emph{Model Dependent} approach forces potential tensions between data and SM predictions to be
absorbed predominantly into the WCs, resulting in an apparent preference
for NP contributions in $C_{9,e}^{\rm NP}$ and $C_{9,\mu}^{\rm NP}$. Once again,
the FFs choice influences quantitatively the bounds set on these WCs.

These findings are corroborated by studying the goodness-of-fit for each scenarios.
Let us introduce the Information Criterion (IC) approach as defined in Ref.~\cite{IC}. 
In this framework, the information criterion for a given model $\mathcal{M}$ 
is defined as ${\rm IC}_\mathcal{M} \equiv -2 \overline{\log \mathcal{L}} \, + \, 4 \sigma^{2}_{\log \mathcal{L}}$, 
where the first and second terms are the mean and variance of the 
log-likelihood posterior distribution. This definition incorporates both 
the goodness of fit (through the first term) and a penalty for complexity 
(through the second one), thereby providing a balanced
measure of model performance. In this framework, models with smaller 
values of the information criterion are preferred~\cite{BayesFactors}. It is therefore
useful to compare the result of a NP fit in a given scenario to the corresponding
SM one.

Within the \emph{Data Driven} approach the NP fits all show larger IC compared
to the SM case, while the  situation is the opposite in the \emph{Model Dependent} approach,
independently from the FFs employed.
This behaviour reflects the fact that the large flexibility of the hadronic 
parametrization typical of the former approach
allows the data to be well described within the SM without requiring additional
short-distance contributions; conversely, the NP scenario is strongly
preferred over the SM one in the \emph{Model Dependent} approach, in which the SM is  incapable to reproduce experimental data.

Overall, these results illustrate the crucial role played by the treatment of hadronic
uncertainties in global analyses of rare $b\to s\ell^+\ell^-$ decays.
While the \emph{Model Dependent} approach tends to favour NP contributions in $C_{9,e}^{\rm NP}$ and $C_{9,\mu}^{\rm NP}$, the \emph{Data Driven}
analyses show that a substantial part of the observed tensions can be accommodated by
non-local hadronic effects.
This underlines the importance of improving theoretical control over power corrections, as
well as of future experimental measurements with increased precision, in order to achieve
a more definitive interpretation of the flavour anomalies.

\section{Conclusions and Outlook}
\label{sec:conclusions}

In this work we have presented a systematic and deliberately conservative
reassessment of rare $b \to s \ell^+\ell^-$ and $b \to s \nu\bar\nu$ transitions, combining recent theoretical developments with the latest experimental results from LHCb, CMS, Belle and Belle~II. The cornerstone of our analysis is a new determination of the $B\to K^{(*)}$ and $B_s\to\phi$ FFs, obtained by applying the Dispersive Matrix (DM) framework to lattice QCD (LQCD) inputs. Grounded in analyticity and unitarity, this approach extrapolates first-principle lattice results to the full kinematic range, providing at the same time a reliable estimate of the associated uncertainties. To gauge the impact of the additional theoretical information usually employed in the literature, we have contrasted this LQCD DM setup with a LCSR+LQCD DM approach, in which light-cone sum-rule inputs supplement the lattice data in the channels and $q^2$ regions where the latter are currently unavailable.

LQCD DM and LCSR+LQCD DM approaches differ markedly in their uncertainty
budget: discarding LCSR information substantially enlarges the FF errors in the
large-recoil region, precisely where the tensions with data are most pronounced.
This inflation is phenomenologically consequential because, once power corrections such as charming penguins are properly included, FF uncertainties no longer cancel in the optimized angular observables. The comparison between the two setups therefore does more than quantify the weight carried by low-$q^2$ inputs in the vector channels: it exposes how sensitively the room left for SM penguin matrix elements and/or possible NP effects depends on the assumptions made about the local FFs.

A second, logically independent choice concerns the treatment of the 
hadronic dynamics itself. Following what was developed in a series of previous works, we have compared a \emph{Data Driven} framework, in which charm-loop and other power-correction effects are parametrized with minimal theoretical bias and constrained directly by the data, with a \emph{Model Dependent} framework, in which additional theoretical assumptions actively restrict the size and structure of these contributions. We find that the inferred preference for New Physics (NP) hinges critically on this choice and, through the mechanism just described, is further modulated by the FF approach adopted.

Within the \emph{Data Driven} approach, the genuinely novel input comes from the latest high-precision updates to the $B \to K^* \mu^+ \mu^-$ angular observables by LHCb and CMS, and in particular from those angular observables that are intrinsically sensitive to strong phases. These provide an unprecedented probe of the imaginary parts of the decay amplitudes \cite{Altmannshofer:2026cwk}, pointing more clearly than ever to sizeable hadronic contributions and thereby steering the global fit toward nonfactorizable hadronic dynamics rather than short-distance NP (see Fig.~\ref{fig:hm1_hm2}). This conclusion is corroborated by an independent and longer-standing indication, namely the $q^2$ dependence of the $B^+ \to K^+ \mu^+ \mu^-$ differential branching fraction in the low-$q^2$ region: its lowest bin sits below the short-distance expectation, and the resulting $q^2$ profile is naturally reproduced by a $q^2$-dependent hadronic amplitude interfering with the short-distance one, rather than by a constant shift in $C_9$. The two pieces of evidence are complementary -- one probing the absorptive parts of the hadronic contributions, the other their $q^2$ shape -- and together they are difficult to mimic with a $q^2$-independent short distance shift. Notice that a sizeable negative shift in $C_9$ -- in line with previous global analyses of the so-called $P_5^{\prime}$ anomaly -- survives only within the more restrictive \emph{Model Dependent} framework. We have stressed that this short-distance interpretation now carries a steep price: it cannot reproduce the strong-phase-sensitive observables, most notably $S_7$, and it is not the preferred option even in the analysis of the $B^+ \to K^+ \mu^+ \mu^-$ differential branching fraction. Finally, our NP analysis of $b \to s \ell^+\ell^-$ updates what we have shown in Ref.~\cite{Ciuchini:2022wbq}.

The $b \to s \nu\bar\nu$ modes, being free from charming-penguin contamination and dependent solely on local FFs, provide complementary information. Using the LQCD DM form factors, we have provided updated SM predictions for the $B\to K^{(*)}\nu\bar\nu$ branching fractions, collected in
Table~\ref{tab:btosnunu_SM}. In line with the pattern observed in the
charged-lepton modes, the pseudoscalar channels -- fully covered by lattice data -- are essentially insensitive to the inclusion of LCSR inputs, whereas in the vector channels the purely LQCD DM determination carries a markedly larger uncertainty, by roughly a factor of two. For $B^+\to K^+\nu\bar\nu$ we provided the updated prediction $\mathrm{BR}(B^+ \to K^+\nu\bar\nu)_\mathrm{SM} = (3.95 \pm 0.14) \times 10^{-6}$, which is the reference SM prediction to which experimental results should be compared.
We have then explored NP scenarios involving both left- and right-handed quark currents within the Weak Effective Theory, presenting the marginalized posteriors for $(\delta C_L^{\nu},\,\delta C_R^{\nu})$ in Figs.~\ref{fig:b2snunu_corr} and~\ref{fig:b2snunu_corr_tri}. Current data still leave broad allowed regions in parameter space, with the pseudoscalar and vector channels offering complementary sensitivity to the chiral structure of possible NP.

In conclusion, the $b \to s \nu\bar\nu$ modes are theoretically clean: $B \to K\nu\bar\nu$ is fully controlled by lattice form factors and is not limited by hadronic assumptions, so here the decisive step is experimental, and a measurement of $B \to K^{*}\nu\bar\nu$ -- complementary in its chiral sensitivity -- would be especially valuable. The $b \to s \ell^+\ell^-$ modes, by contrast, are limited by theory: not only by the penguin matrix elements, but also by the local form factors themselves, which in the vector channels still rest largely on LCSR estimates rather than on  first-principle lattice computations. Waiting for future first-principles estimates of penguin matrix elements from lattice QCD along the lines suggested in Ref.~\cite{Frezzotti:2025hif}, which will finally allow to disentangle possible NP effects in $C_9$, we pointed out that current data hint at sizeable penguin matrix elements. It will be interesting to see if future measurements with increased accuracy confirm this pattern, calling for a conservative treatment of hadronic uncertainties following our \emph{Data Driven} scenario.

\begin{acknowledgments}
The authors wish to thank Andrea Mauri and Mark Smith for fruitful discussions concerning the latests results from the LHCb collaboration, and Nico Gubernari for clarifications on the usage of their light cone sum rules results and for interesting discussions.
The work of MF was supported by the Cluster of Excellence \textit{PRISMA}$^{++}$ (EXC 2118/2, Project ID 390831469).
The work of JS received funding from the European Research Council (ERC) under the European Union’s Horizon 2022 Research and Innovation Program (ERC Advanced Grant agreement No.101097780, EFT4jets). Views and opinions expressed are those of the authors and do not reflect those of the European Union or the ERC Executive Agency. Neither the European Union nor the granting authority can be held responsible for them. JS also thanks INFN - Sezione di Roma for hospitality, where part of this work was carried out.
LV is supported by the Italian Ministry of University and Research (MUR)
and by the European Union’s NextGenerationEU program under the Young Researchers 2024
SoE Action, research project ‘SHYNE’, ID: SOE\_20240000025.
\end{acknowledgments}

\appendix

\section{Definitions of Hadronic Form Factors}
\label{app:AppA}

For the exclusive decay of a $B$-meson into a pseudoscalar one $P$ due to the quark transition $b \to q$ the relevant hadronic matrix elements are~\cite{Bharucha:2010im}
\bea
    \label{eq:BtoP_vector}
     \langle P(p_P) | V^\mu | B(p_B) \rangle & = & \left[ P^\mu - \frac{m_B^2 - m_P^2}{q^2} q^\mu \right] f_+(q^2) \color{black} +
    \frac{m_B^2 - m_P^2}{q^2} q^\mu f_0(q^2) \color{black}  ~ , ~~~ \\[2mm]
    \label{eq:BtoP_vectorT}
     \langle P(p_P) | T^{\mu \nu} q_\nu | B(p_B) \rangle & = &  \frac{i}{m_B + m_P}  \left[ q^2 P^\mu - (m_B^2 - m_P^2) q^\mu \right] 
    f_T(q^2) \color{black} ~ , ~
\eea
where $V^\mu = \bar{b} \gamma^\mu q$ and $T^{\mu \nu} = \bar{b} \sigma^{\mu \nu} q$ are respectively the quark (weak) vector and vector-tensor currents, $P \equiv p_B + p_P$ and $q \equiv p_B - p_P$ is the 4-momentum transfer.
In order to avoid a kinematical singularity, the following kinematical constraint holds at $q^2 = 0$:
\be
    \label{eq:KC1}
    f_+(0) = f_0(0) \,.
\ee
All the three FFs are dimensionless and have definite spin-parity. 

In the case of an exclusive decay of a $B_{(s)}$-meson into a vector one the relevant hadronic matrix elements are~\cite{Bharucha:2010im, Boyd:1997kz}
\bea
    \label{eq:BtoV_vector}
     \langle V(p_V, \epsilon) | V^\mu | B_{(s)}(p_{B_{(s)}}) \rangle & = & i \epsilon^{\mu \nu \rho \sigma} \epsilon_\nu^* P_\rho q_\sigma \frac{1}{m_{B_{(s)}} + m_V} 
    V(q^2) \color{black} ~ , ~ \\[2mm]
    \label{eq:BtoV_axial}
     \langle V(p_V, \epsilon) | A^\mu | B_{(s)}(p_{B_{(s)}}) \rangle & = & - (\epsilon^* \cdot q) \frac{2m_V}{q^2} q^\mu A_0(q^2) \color{black} \nonumber \\
    & - &  \left[ \epsilon^{* \mu} - (\epsilon^* \cdot q) \frac{q^\mu}{q^2} \right] (m_{B_{(s)}} + m_V)
    A_1(q^2) \color{black} \nonumber \\ 
    & + & (\epsilon^* \cdot q) \frac{m_{B_{(s)}} - m_V}{q^2} \left( \frac{q^2}{m_{B_{(s)}}^2 - m_V^2} P^\mu - q^\mu \right) 
    A_2(q^2) \color{black} ~ , ~
\eea
where $A^\mu = \bar{b} \gamma^\mu \gamma_5 q$ is the quark (weak) axial current and
\bea
    \label{eq:BtoV_vectorT}
     \langle V(p_V, \epsilon) | T^{\mu \nu} q_\nu | B_{(s)}(p_{B_{(s)}}) \rangle & = & i \epsilon^{\mu \nu \rho \sigma} \epsilon_\nu^* P_\rho q_\sigma 
    T_1(q^2) \color{black} ~ , ~ \\[2mm]
    \label{eq:BtoV_axialT}
     \langle V(p_V, \epsilon) | A^{\mu \nu} q_\nu | B_{(s)}(p_{B_{(s)}}) \rangle & = & \left[ \epsilon^{* \mu} (m_{B_{(s)}}^2 - m_V^2) - (\epsilon^* \cdot q) P^\mu \right] 
    T_2(q^2) \color{black} \nonumber \\ 
    & + & (\epsilon^* \cdot q) \left( q^\mu - \frac{q^2}{m_{B_{(s)}}^2 - m_V^2} P^\mu \right) 
    T_3(q^2) \color{black} ~ , ~
\eea
where $A^{\mu \nu} = \bar{b} \sigma^{\mu \nu} \gamma_5 q$ is the quark (weak) axial-tensor current, $P \equiv p_B + p_V$ and $q \equiv p_B - p_V$. Only the FFs $A_2(q^2)$ and $T_3(q^2)$ do not have definite spin-parity. The relations of the (dimensionless) FFs $V$, $A_{0,1,2}$ and $T_{1,2,3}$ with those having definite spin-parity, as defined in Refs.~\cite{Bharucha:2010im, Boyd:1997kz}, are
\bea
\label{eq:gf}
g(q^2) &\equiv& \frac{2}{m_{B_{(s)}} + m_V} V(q^2) \, , \qquad
f(q^2) \equiv (m_{B_{(s)}} + m_V) A_1(q^2) \, , \\[2mm]
\label{eq:F12}
\mathcal{F}_1(q^2) &\equiv& \frac{m_{B_{(s)}}+m_V}{m_V}
\left[ \frac{m_{B_{(s)}}^2 - m_V^2 - q^2}{2} A_1(q^2)
- \frac{\lambda_V(q^2)}{2(m_{B_{(s)}}+m_V)^2} A_2(q^2) \right] \, , \qquad
\mathcal{F}_2(q^2) \equiv 2 A_0(q^2) \, , \\[2mm]
\label{eq:T12}
T_1(q^2) &\equiv& T_1(q^2) \, , \qquad
T_2(q^2) \equiv T_2(q^2) \, , \\[2mm]
\label{eq:T0}
T_0(q^2) &\equiv& \frac{1}{2m_V (m_{B_{(s)}}+m_V)}
\left[ (m_{B_{(s)}}^2 + 3 m_V^2 - q^2) T_2(q^2)
- \frac{\lambda_V(q^2)}{m_{B_{(s)}}^2 - m_V^2} T_3(q^2) \right] \, .
\eea
It is then convenient to introduce the following combinations of axial and tensor form
factors:
\bea
A_{12}(q^2)
&=&
\frac{
\bigl(m_{B_{(s)}}+m_V\bigr)^2
\bigl(m_{B_{(s)}}^2-m_V^2-q^2\bigr)\,A_1(q^2)
-
\lambda_V(q^2)\,A_2(q^2)
}
{
16\,m_{B_{(s)}}\,m_V^2\,
\bigl(m_{B_{(s)}}+m_V\bigr)
}\,,
\label{eq:A12_def}
\\
T_{23}(q^2)
&=&
\frac{
\bigl(m_{B_{(s)}}^2-m_V^2\bigr)
\bigl(m_{B_{(s)}}^2+3m_V^2-q^2\bigr)\,T_2(q^2)
-
\lambda_V(q^2)\,T_3(q^2)
}
{
8\,m_{B_{(s)}}\,m_V^2\,
\bigl(m_{B_{(s)}}-m_V\bigr)
}\,.
\label{eq:T23_def}
\eea
In the above expressions $\lambda_V(q^2) \equiv ((m_{B_{(s)}}+m_V)^2 - q^2)((m_{B_{(s)}}-m_V)^2-q^2)$, with $t_{\pm}$ defined just below Eq.\,(\ref{eq:z}). Two kinematical constraints hold at $q^2 = 0$, namely
\be
    \label{eq:KC2}
    \mathcal{F}_2(0) = \frac{2}{m_{B_{(s)}}^2 - m_V^2} \mathcal{F}_1(0)\,, \qquad T_1(0) = T_2(0) ~ , ~
\ee
while at $q^2 = t_-$ other two kinematical constraints apply, $i.e.$
\be
     \label{eq:KC3}
     \mathcal{F}_1(t_-) = (m_{B_{(s)}} - m_V) f(t_-)\,, \qquad T_0(t_-) = T_2(t_-) ~ . ~
\ee

The kinematic functions $\phi$, which enter directly in the matrix (\ref{eq:Delta2}), they depend on the particular FF one is looking at. By following the procedure outlined in Refs.\,\cite{Boyd:1997kz, Bharucha:2010im}, one finds (see also Ref.\,\cite{Gubernari:2023puw}) that
%\footnote{Analogous results can be found in Section 2 of Ref.\,\cite{Gubernari:2023puw}.}
\bea
    \label{eq:phif+}
   \sqrt{\chi^{1-}}  \phi^{f_+}(z) & = & \frac{16 r_P^2}{m_B} \sqrt{\frac{2 n_I}{3 \pi}} \frac{(1 + z)^2 \sqrt{1 - z}}{\left[ (1 + r_P)(1 - z) + 2 \sqrt{r_P}(1 + z) \right]^5} ~ , ~ \\[2mm]
    \label{eq:phif0}
    \sqrt{\chi^{0+}}  \phi^{f_0}(z) & = & 2 r_P (1 - r_P^2) \sqrt{\frac{2 n_I}{ \pi}} \frac{(1-z^2) \sqrt{1 - z}}{\left[ (1 + r_P)(1 - z) + 2 \sqrt{r_P}(1 + z) \right]^4} ~ , ~ \\[2mm]
    \label{eq:phifT}
    \sqrt{\chi^{TT}} \phi^{f_T}(z) & = & \frac{16 r_P^2}{m_B (1 + r_P)} \sqrt{\frac{2 n_I}{3\pi}} \frac{(1 + z)^{2} (1 - z)^{-1/2}}{\left[ (1 + r_P)(1 - z) + 2 \sqrt{r_P}(1 + z) \right]^4} \qquad ~ 
\eea 
with $r_P \equiv m_P / m_B$, and
\bea
    \label{eq:phig}
    \sqrt{\chi^{1-}} \phi^g(z) & = & 16 r_V^2 \sqrt{\frac{n_I}{3\pi}} \frac{(1 + z)^{2} (1 - z)^{-1/2}}{\left[ (1 + r_V)(1 - z) + 2 \sqrt{r_V}(1 + z) \right]^4} ~ , ~ \\[2mm]
    \label{eq:phif}
    \sqrt{\chi^{1+}} \phi^f(z) & = & 4 \frac{r_V}{m_{B_{(s)}}^2} \sqrt{\frac{n_I}{3\pi}} \, \frac{(1 + z)(1 - z)^{3/2}}{\left[ (1 + r_V)(1 - z) + 2 \sqrt{r_V}(1 + z) \right]^4} ~ , ~ \\[2mm]
    \label{eq:phiF1}
    \sqrt{\chi^{1+}} \phi^{\mathcal{F}_1}(z) & = & 2 \frac{r_V}{m_{B_{(s)}}^3} \sqrt{\frac{2n_I}{3\pi}} \frac{(1 + z)(1 - z)^{5/2}}{\left[ (1 + r_V)(1 - z) + 2 \sqrt{r_V}(1 + z) \right]^5} ~ , ~ \\[2mm]
    \label{eq:phiF2}
    \sqrt{\chi^{0-}} \phi^{\mathcal{F}_2}(z) & = &  8 r_V^2 \sqrt{\frac{2 n_I}{\pi}} \frac{(1 + z)^{2} (1 - z)^{-1/2}}{\left[ (1 + r_V)(1 - z) + 2 \sqrt{r_V}(1 + z) \right]^4} ~ , \\[2mm]
    \label{eq:phiT0}
    \sqrt{\chi^{BB}} \phi^{T_0}(z) & = & 2 \frac{r_V (1 + r_V)}{m_{B_{(s)}}} \sqrt{\frac{2 n_I}{3\pi}} \, \frac{(1 + z)(1 - z)^{3/2}}{\left[ (1 + r_V)(1 - z) + 2 \sqrt{r_V}(1 + z) \right]^4}  ~ , \qquad ~ \\[2mm]
    \label{eq:phiT1}
   \sqrt{\chi^{TT}}  \phi^{T_1}(z) & = & \frac{32 r_V^2}{m_{B_{(s)}}} \sqrt{\frac{n_I}{3 \pi}} \frac{(1 + z)^2 \sqrt{1 - z}}{\left[ (1 + r_V)(1 - z) + 2 \sqrt{r_V}(1 + z) \right]^5} ~ , ~ \\[2mm]
    \label{eq:phiT2}
    \sqrt{\chi^{BB}} \phi^{T_2}(z) & = & 4 \frac{r_V (1 - r_V^2)}{m_{B_{(s)}}} \sqrt{\frac{n_I}{3\pi}} \frac{(1 + z)(1 - z)^{5/2}}{\left[ (1 + r_V)(1 - z) + 2 \sqrt{r_V}(1 + z) \right]^5} ~   
\eea
with $r_V \equiv m_V/ m_{B_{(s)}}$. Note that in this paper the Clebsch-Gordan factor $n_I$ has been taken equal to 2 for $B \to K^{(*)}$ decays and equal to 1 for $B_s \to \phi$ one. 

As outlined in Section \ref{sec:DM_FF}, the above expressions of the kinematic functions have to be modified according to the prescription in Eq.(\ref{eq:poles}) in presence of sub-threshold poles, namely particles whose masses lie below $t_+$. The masses of such states are shown in Table \ref{tab:poles}. Since in our study the definition of $t_+$ varies channel by channel (as outlined below Eq.(\ref{eq:z})), the number of poles to be included in Eq.(\ref{eq:poles}) thus varies according to the transition and to the spin-parity quantum channel one is looking at.
\begin{itemize}
\item $B \to K$ FFs: one has to consider the first $B_s^*$ state for the FFs $f_+$ and $f_T$, while no pole is present for $f_0$;
\item $B \to K^*$ FFs: both the poles in the spin-parity quantum channels $J^P = 0^-,\, 1^-$, $i.e.$ for the FFs $\mathcal{F}_2$ and $g, T_1$ respectively, have to be considered, while only the first pole is relevant for $J^P = 1^+$, namely for the FFs $f,\,\mathcal{F}_1,\,T_0$ and $T_2$;
\item $B_s \to \phi$ FFs: one has to consider both poles for each of the seven FFs, $i.e.$ two poles for each spin-parity quantum channel.
\end{itemize}
It is worth stressing that there is no $B_{s0}^*$ state (belonging to $J^P = 0^+$) with a mass below the threshold $m_B+m_K$\,\cite{Godfrey:2016nwn}.

\begin{table}[!t!]
\renewcommand{\arraystretch}{1.35}
\begin{center}
\begin{tabular}{|c|c|c|c|}
\hline
&$B_s$ $(J^P = 0^-)$ & $B_s^*$ $(J^P = 1^-)$ & $B_{s1}$ $(J^P = 1^+)$\\ \hline
First Pole & 5.367 & 5.415 & 5.829\\
Second Pole & 5.984 & 6.012 & 6.295\\\hline
\end{tabular}
\end{center}
\caption{Values of the different pole masses (in GeV) of the $B_s^{(*)}$ and $B_{s1}$ states used in our study. These numbers have been taken from Refs.\,\cite{Godfrey:2016nwn, ParticleDataGroup:2024cfk}.}
\label{tab:poles}
\renewcommand{\arraystretch}{1.0}
\end{table}

Let us also highlight here that the values of the susceptibilities $\chi^{0^{\pm}}$, $\chi^{1^{\mp}}$, $\chi^{TT}$ and $\chi^{BB}$ may, in principle, be computed on lattice QCD. This has been done, for instance, for $b \to c$ quark transitions in Refs.\,\cite{Martinelli:2021frl, Melis:2024wpb, Harrison:2024iad} or for $b \to u$ ones in Ref.\,\cite{Martinelli:2022tte}. However, at present no lattice computation has been carried out for $b \to s$ quark transitions. Thus, in this work we have taken the results of the perturbative computations of the susceptibilites performed in Refs.\,\cite{Bharucha:2010im, Generet:2025hsv, Gubernari:2026sqc}\footnote{Using the results of Ref.\,\cite{Grigo:2012ji} for $\chi^{0^{\pm}},\,\chi^{1^{\mp}}$ does not produce any appreciable change in the shapes of the FFs shown in Section\,\ref{sec:DM_FF}.}, where the tensor ones are given in the $\overline{\rm{MS}}$ scheme at a renormalization scale $\mu = \overline{m}_b(\overline{m}_b)  = 4.2$ GeV. Then, we have subtracted the ground state contributions through the expressions
\begin{equation}
\chi_{\rm (gs)}^{TT} \simeq \chi_{\rm (gs)}^{1^-} = \frac{f_{1^-}^2}{m_{1^-}^4},\qquad\chi_{\rm (gs)}^{BB} \simeq \chi_{\rm (gs)}^{1^+} = \frac{f_{1^+}^2}{m_{1^+}^4},\qquad\chi_{\rm (gs)}^{0^-} = \frac{f_{0^-}^2}{m_{0^-}^2}.
\end{equation}
Here $m_{0^-,\,1^-\,1^+}$ are the ground-state masses of the first poles in Table \ref{tab:poles}, which are associated to different spin-parity quantum channels $J^P$, while $f_{0^-,\,1^-\,1^+}$ are their (leptonic) decay constants. The approximate equality $\chi_{\rm (gs)}^{TT} \simeq \chi_{\rm (gs)}^{1^-}$ ($\chi_{\rm (gs)}^{BB} \simeq \chi_{\rm (gs)}^{1^+}$) derives from the fact that the decay constants of the $B_s^*$ pole ($B_{s1}$ pole) associated to the vector (axial-vector) and tensor (axial-tensor) currents are numerically quite similar to each other, as $e.g.$ outlined in Ref.\,\cite{Pullin:2021ebn}. The values of the masses $m_{0^-,\,1^-\,1^+}$ can be thus read from the first row of Table \ref{tab:poles}, while the ones of the corresponding decay constants $f_{0^-,\,1^-\,1^+}$ coincide with the averages computed in Appendix A of Ref.\,\cite{Guadagnoli:2023zym}. The subtracted values of the susceptibilities eventually read
\begin{eqnarray}
\chi^{0^{+}} &=& 0.0142 \,, \qquad
\chi^{0^{-}} = 0.0139 \,, \qquad
\chi^{1^{-}} = 0.00062\,{\rm GeV}^{-2} \,, \qquad
\chi^{1^{+}} = 0.00054\,{\rm GeV}^{-2} \nonumber\\
&&\hspace{2.1cm}
\chi^{TT} = 0.00037\,{\rm GeV}^{-2} \,, \qquad
\chi^{BB} = 0.00029\,{\rm GeV}^{-2}\, .
\end{eqnarray}

\section{BGL fit results for the Form Factors}
\label{app:AppB}
In this Appendix we give the results obtained for the BGL fits to the FFs obtained employing the DM approach. The results concerning the $B\to K$ FFs are reported in Table \ref{tab:BGL_BtoK}. For the $B\to K^*$ case, we give the results obtained applying the LQCD DM approach to $f$, $g$, $F_1$ and $F_2$ FFs in Table \ref{tab:BGL_BtoKst_1}, and the ones relative to $T_1$, $T_2$ and $T_0$ FFs in Table \ref{tab:BGL_BtoKst_2}. The results concerning the full set of FFs obtained with the LQCD+LCSR DM approach are given in Table \ref{tab:BGL_BtoKst_3}. Analogous results for the $B_s \to \phi$ FFs can be found in Tables \ref{tab:BGL_BstoPhi_1}-\ref{tab:BGL_BstoPhi_3}. Notice that, due the kinematical constraints shown at Eqs.~\eqref{eq:KC1},~\eqref{eq:KC2} and~\eqref{eq:KC3}, the parameters $a_0^{f_0}$, $a_0^{F_1}$, $a_0^{F_2}$, $a_0^{T_2}$ and $a_0^{T_3}$ are not independent quantities but instead linear combinations of the other BGL parameters, and therefore do not appear in the tables.

\begin{table}[!ht!]
\centering
\renewcommand{\arraystretch}{0.9}
\begin{tabular}{lrrrrrrrr}
\toprule
 & $a_0^{f_+}$ & $a_1^{f_+}$ & $a_2^{f_+}$ &
   $a_1^{f_0}$ & $a_2^{f_0}$ &
   $a_0^{f_T}$ & $a_1^{f_T}$ & $a_2^{f_T}$ \\
\midrule
Mean
& 0.0304 & -0.0313 & -0.073 & -0.451 & 0.28 & 0.0620 & -0.012 & -0.22 \\
Std. Dev.
& 0.0005 & \phantom{-}0.0084 & \phantom{-}0.027 & \phantom{-}0.035 & 0.11 & 0.0014 & \phantom{-}0.025 & \phantom{-}0.10 \\
\midrule
\multicolumn{9}{c}{Correlation Matrix} \\
\midrule
$a_0^{f_+}$ &  1.00 & -0.70 &  0.53 & -0.02 &  0.03 &  0.01 &  0.00 & -0.01 \\
$a_1^{f_+}$ &       &  1.00 & -0.97 &  0.08 & -0.13 &  0.00 &  0.00 &  0.00 \\
$a_2^{f_+}$ &       &       &  1.00 & -0.13 &  0.21 &  0.00 &  0.00 &  0.00 \\
$a_1^{f_0}$ &       &       &       &  1.00 & -0.96 &  0.01 &  0.02 & -0.02 \\
$a_2^{f_0}$ &       &       &       &       &  1.00 & -0.01 & -0.01 &  0.02 \\
$a_0^{f_T}$ &       &       &       &       &       &  1.00 & -0.62 &  0.39 \\
$a_1^{f_T}$ &       &       &       &       &       &       &  1.00 & -0.93 \\
$a_2^{f_T}$ &       &       &       &       &       &       &       &  1.00 \\
\bottomrule
\end{tabular}
\caption{\footnotesize Posterior summaries for the BGL fit to the $B\to K$ FFs obtained employing the LQCD DM approach. The first row gives the posterior means, the second row the standard deviations, and the lower block shows the correlation matrix.}
\label{tab:BGL_BtoK}
\end{table}

\FloatBarrier

\begin{table}[p]
\centering
\renewcommand{\arraystretch}{0.8}
\begin{tabular}{lrrrrrrrrrr}
\toprule
 & $a_0^{f}$ & $a_1^{f}$ & $a_2^{f}$ &
   $a_0^{g}$ & $a_1^{g}$ & $a_2^{g}$ &
   $a_1^{F_1}$ & $a_2^{F_1}$ &
   $a_1^{F_2}$ & $a_2^{F_2}$ \\
\midrule
Mean
& 0.0418 & -0.066 & 0.00
& 0.0326 & -0.042 & -0.01
& -0.019 & -0.03
& -0.068 & -0.02 \\
Std. Dev.
& 0.0023 & \phantom{-}0.063 & 0.23
& 0.0030 & \phantom{-}0.073 & 0.23
& \phantom{-}0.021 & \phantom{-}0.11
& \phantom{-}0.076 & \phantom{-}0.24 \\
\midrule
\multicolumn{11}{c}{Correlation Matrix} \\
\midrule
$a_0^{f}$   & 1.00 & -0.60 &  0.03 & -0.01 &  0.03 & -0.04 & -0.51 &  0.19 & -0.18 & -0.06 \\
$a_1^{f}$   &      &  1.00 & -0.18 &  0.00 &  0.04 &  0.03 &  0.31 & -0.13 &  0.07 &  0.07 \\
$a_2^{f}$   &      &       &  1.00 &  0.02 & -0.04 & -0.05 & -0.16 &  0.22 &  0.06 &  0.02 \\
$a_0^{g}$   &      &       &       &  1.00 & -0.68 &  0.14 & -0.01 & -0.02 & -0.02 & -0.05 \\
$a_1^{g}$   &      &       &       &       &  1.00 & -0.27 &  0.00 &  0.03 &  0.01 &  0.06 \\
$a_2^{g}$   &      &       &       &       &       &  1.00 & -0.01 &  0.03 & -0.02 &  0.03 \\
$a_1^{F_1}$ &      &       &       &       &       &       &  1.00 & -0.65 &  0.39 &  0.06 \\
$a_2^{F_1}$ &      &       &       &       &       &       &       &  1.00 &  0.23 &  0.24 \\
$a_1^{F_2}$ &      &       &       &       &       &       &       &       &  1.00 & -0.28 \\
$a_2^{F_2}$ &      &       &       &       &       &       &       &       &       &  1.00 \\
\bottomrule
\end{tabular}
\caption{\footnotesize Posterior summaries for the BGL fit to the $B\to K^*$ FFs $f$, $g$, $F_1$ and $F_2$ obtained employing the LQCD DM approach. The first row gives the posterior means, the second row the standard deviations, and the lower block shows the correlation matrix.}
\label{tab:BGL_BtoKst_1}
\end{table}

\begin{table}[p]
\centering
\renewcommand{\arraystretch}{0.8}
\begin{tabular}{lrrrrrrr}
\toprule
 & $a_0^{T_1}$ & $a_1^{T_1}$ & $a_2^{T_1}$
 & $a_1^{T_2}$ & $a_2^{T_2}$
 & $a_1^{T_0}$ & $a_2^{T_0}$ \\
\midrule
Mean
& 0.0202 & -0.034 & -0.02
& -0.052 & 0.00
& -0.112 & 0.01 \\
Std. Dev.
& 0.0016 & \phantom{-}0.040 & \phantom{-}0.24
& \phantom{-}0.034 & 0.21
& \phantom{-}0.059 & 0.27 \\
\midrule
\multicolumn{8}{c}{Correlation Matrix} \\
\midrule
$a_0^{T_1}$ & 1.00 & -0.77 &  0.44 & -0.01 & -0.10 &  0.01 &  0.01 \\
$a_1^{T_1}$ &      &  1.00 & -0.70 &  0.14 & -0.03 & -0.03 & -0.02 \\
$a_2^{T_1}$ &      &       &  1.00 &  0.08 &  0.48 &  0.07 &  0.07 \\
$a_1^{T_2}$ &      &       &       &  1.00 & -0.55 &  0.39 & -0.08 \\
$a_2^{T_2}$ &      &       &       &       &  1.00 & -0.19 &  0.13 \\
$a_1^{T_3}$ &      &       &       &       &       &  1.00 & -0.24 \\
$a_2^{T_3}$ &      &       &       &       &       &       &  1.00 \\
\bottomrule
\end{tabular}
\caption{\footnotesize Posterior summaries for the BGL fit to the $B\to K^*$ FFs $T_1$, $T_2$ and $T_0$ obtained employing the LQCD DM approach. The first row gives the posterior means, the second row the standard deviations, and the lower block shows the correlation matrix.}
\label{tab:BGL_BtoKst_2}
\end{table}

\begin{table}[p]
\centering
\renewcommand{\arraystretch}{0.8}
\resizebox{\textwidth}{!}{%
\begin{tabular}{lrrrrrrrrrrrrrrrrr}
\toprule
 & $a_0^{f}$ & $a_1^{f}$ & $a_2^{f}$ & $a_0^{g}$ & $a_1^{g}$ & $a_2^{g}$ & $a_1^{F_1}$ & $a_2^{F_1}$ & $a_1^{F_2}$ & $a_2^{F_2}$ & $a_0^{T_1}$ & $a_1^{T_1}$ & $a_2^{T_1}$ & $a_1^{T_2}$ & $a_2^{T_2}$ & $a_1^{T_0}$ & $a_2^{T_0}$ \\
\midrule
Mean
& 0.0411 & -0.073 & 0.035 & 0.0324 & -0.035 & -0.016 & -0.015 & -0.030 & -0.056 & -0.022 & 0.0200 & -0.022 & -0.043 & -0.044 & 0.011 & -0.050 & -0.023 \\
Std.\ Dev.
& 0.0019 & \phantom{-}0.022 & 0.065 & 0.0023 & \phantom{-}0.025 & 0.075 & \phantom{-}0.011 & \phantom{-}0.057 & \phantom{-}0.030 & 0.085 & 0.0012 & \phantom{-}0.017 & 0.053 & \phantom{-}0.015 & \phantom{-}0.051 & \phantom{-}0.022 & \phantom{-}0.076 \\
\midrule
\multicolumn{18}{c}{Correlation Matrix} \\
\midrule
$a_0^{f}$   & 1.00 &-0.66 & 0.46 & 0.00 & 0.04 & 0.00 &-0.38 & 0.11 & 0.03 &-0.04 &-0.01 & 0.06 &-0.04 & 0.13 &-0.11 & 0.10 &-0.06 \\
$a_1^{f}$   &      & 1.00 &-0.90 & 0.10 & 0.14 &-0.10 & 0.43 &-0.17 & 0.16 &-0.08 & 0.00 & 0.21 &-0.17 & 0.16 &-0.08 & 0.06 & 0.00 \\
$a_2^{f}$   &      &      & 1.00 &-0.08 &-0.07 & 0.04 &-0.51 & 0.36 &-0.11 & 0.08 &-0.01 &-0.14 & 0.12 &-0.09 & 0.05 &-0.03 &-0.01 \\
$a_0^{g}$   &      &      &      & 1.00 &-0.69 & 0.47 & 0.02 & 0.01 & 0.09 &-0.03 &-0.03 & 0.08 &-0.05 & 0.05 &-0.02 & 0.04 &-0.02 \\
$a_1^{g}$   &      &      &      &      & 1.00 &-0.91 &-0.04 & 0.07 & 0.00 &-0.04 & 0.04 & 0.15 &-0.13 & 0.20 &-0.14 & 0.10 &-0.03 \\
$a_2^{g}$   &      &      &      &      &      & 1.00 & 0.02 &-0.04 & 0.01 & 0.04 &-0.04 &-0.10 & 0.09 &-0.14 & 0.10 &-0.07 & 0.02 \\
$a_1^{F_1}$ &      &      &      &      &      &      & 1.00 &-0.91 & 0.11 &-0.08 & 0.08 &-0.10 & 0.09 &-0.11 & 0.08 &-0.08 & 0.07 \\
$a_2^{F_1}$ &      &      &      &      &      &      &      & 1.00 & 0.02 & 0.04 &-0.07 & 0.13 &-0.12 & 0.12 &-0.08 & 0.11 &-0.08 \\
$a_1^{F_2}$ &      &      &      &      &      &      &      &      & 1.00 &-0.78 & 0.01 & 0.06 &-0.04 & 0.07 &-0.03 & 0.06 & 0.00 \\
$a_2^{F_2}$ &      &      &      &      &      &      &      &      &      & 1.00 &-0.05 &-0.01 & 0.00 &-0.04 & 0.01 &-0.02 &-0.02 \\
$a_0^{T_1}$ &      &      &      &      &      &      &      &      &      &      & 1.00 &-0.73 & 0.55 &-0.10 & 0.05 &-0.09 & 0.05 \\
$a_1^{T_1}$ &      &      &      &      &      &      &      &      &      &      &      & 1.00 &-0.94 & 0.47 &-0.39 & 0.17 &-0.06 \\
$a_2^{T_1}$ &      &      &      &      &      &      &      &      &      &      &      &      & 1.00 &-0.44 & 0.40 &-0.10 & 0.01 \\
$a_1^{T_2}$ &      &      &      &      &      &      &      &      &      &      &      &      &      & 1.00 &-0.93 & 0.49 &-0.25 \\
$a_2^{T_2}$ &      &      &      &      &      &      &      &      &      &      &      &      &      &      & 1.00 &-0.33 & 0.17 \\
$a_1^{T_0}$ &      &      &      &      &      &      &      &      &      &      &      &      &      &      &      & 1.00 &-0.91 \\
$a_2^{T_0}$ &      &      &      &      &      &      &      &      &      &      &      &      &      &      &      &      & 1.00 \\
\bottomrule
\end{tabular}
}
\caption{\footnotesize Posterior summaries for the BGL fit to all the $B\to K^*$ FFs obtained employing the LQCD+LCSR DM approach. The first row gives the posterior means, the second row the standard deviations, and the lower block shows the correlation matrix.}
\label{tab:BGL_BtoKst_3}
\end{table}

\begin{table}[p]
\centering
\renewcommand{\arraystretch}{0.8}
\begin{tabular}{lrrrrrrrrrr}
\toprule
 & $a_0^{f}$ & $a_1^{f}$ & $a_2^{f}$ & $a_0^{g}$ & $a_1^{g}$ & $a_2^{g}$ & $a_1^{F_1}$ & $a_2^{F_1}$ & $a_1^{F_2}$ & $a_2^{F_2}$ \\
\midrule
Mean
& 0.0148 & -0.022 & 0.00 & 
0.0139 & -0.014 & -0.01 & 0.0012 & 
0.001 & 0.009 & -0.01 \\
Std.\ Dev.
& 0.0007 & \phantom{-}0.028 & \phantom{-}0.25 & 
0.0010 & \phantom{-}0.033 & 0.25 & \phantom{-}0.0098 & 
\phantom{-}0.085 & \phantom{-}0.046 & \phantom{-}0.25 \\
\midrule
\multicolumn{11}{c}{Correlation Matrix} \\
\midrule
$a_0^{f}$   & 1.00 & -0.77 &  0.27 &  0.08 & -0.09 &  0.05 & -0.72 &  0.47 &  0.01 &  0.02 \\
$a_1^{f}$   &      &  1.00 & -0.55 & -0.05 &  0.04 &  0.01 &  0.59 & -0.40 & -0.01 & -0.04 \\
$a_2^{f}$   &      &       &  1.00 &  0.01 &  0.02 & -0.05 & -0.27 &  0.22 &  0.01 &  0.07 \\
$a_0^{g}$   &      &       &       &  1.00 & -0.80 &  0.26 & -0.08 &  0.04 &  0.00 &  0.00 \\
$a_1^{g}$   &      &       &       &       &  1.00 & -0.54 &  0.07 & -0.06 & -0.03 & -0.01 \\
$a_2^{g}$   &      &       &       &       &       &  1.00 & -0.02 &  0.02 &  0.02 & -0.01 \\
$a_1^{F_1}$ &      &       &       &       &       &       &  1.00 & -0.83 & -0.15 & -0.12 \\
$a_2^{F_1}$ &      &       &       &       &       &       &       &  1.00 &  0.40 &  0.37 \\
$a_1^{F_2}$ &      &       &       &       &       &       &       &       &  1.00 & -0.47 \\
$a_2^{F_2}$ &      &       &       &       &       &       &       &       &       &  1.00 \\
\bottomrule
\end{tabular}
\caption{\footnotesize Posterior summaries for the BGL fit to the $B_s\to \phi$ FFs $f$, $g$, $F_1$ and $F_2$ obtained employing the LQCD DM approach. The first row gives the posterior means, the second row the standard deviations, and the lower block shows the correlation matrix.}
\label{tab:BGL_BstoPhi_1}
\end{table}

\begin{table}[p]
\centering
\renewcommand{\arraystretch}{0.8}
\begin{tabular}{lrrrrrrr}
\toprule
 & $a_0^{T_1}$ & $a_1^{T_1}$ & $a_2^{T_1}$ & $\alpha_1^{T_2}$ & $\alpha_2^{T_2}$ & $\alpha_1^{T_0}$ & $\alpha_2^{T_0}$ \\
\midrule
Mean
& 0.0081 & 0.001 & -0.04 & 
-0.007 & -0.01 & 
-0.016 & 0.01 \\
Std.\ Dev.
& 0.0006 & \phantom{-}0.025 & \phantom{-}0.27 & 
\phantom{-}0.019 & 0.23 & 
\phantom{-}0.028 & 0.32 \\
\midrule
\multicolumn{8}{c}{Correlation Matrix} \\
\midrule
$a_0^{T_1}$        & 1.00 & -0.82 &  0.44 &  0.00 &  0.04 & -0.04 &  0.04 \\
$a_1^{T_1}$        &      &  1.00 & -0.77 &  0.16 & -0.34 &  0.06 & -0.05 \\
$a_2^{T_1}$        &      &       &  1.00 & -0.40 &  0.80 & -0.10 &  0.08 \\
$\alpha_1^{T_2}$   &      &       &       &  1.00 & -0.79 &  0.58 & -0.22 \\
$\alpha_2^{T_2}$   &      &       &       &       &  1.00 & -0.33 &  0.16 \\
$\alpha_1^{T_0}$   &      &       &       &       &       &  1.00 & -0.65 \\
$\alpha_2^{T_0}$   &      &       &       &       &       &       &  1.00 \\
\bottomrule
\end{tabular}
\caption{\footnotesize Posterior summaries for the BGL fit to the $B_s\to \phi$ FFs $T_1$, $T_2$ and $T_0$ obtained employing the LQCD DM approach. The first row gives the posterior means, the second row the standard deviations, and the lower block shows the correlation matrix.}
\label{tab:BGL_BstoPhi_2}
\end{table}

\begin{table}[p]
\centering
\renewcommand{\arraystretch}{0.8}
\resizebox{\textwidth}{!}{%
\begin{tabular}{lrrrrrrrrrrrrrrrrr}
\toprule
 & $a_0^{f}$ & $a_1^{f}$ & $a_2^{f}$ & $a_0^{g}$ & $a_1^{g}$ & $a_2^{g}$ & 
   $a_1^{F_1}$ & $a_2^{F_1}$ & $a_1^{F_2}$ & $a_2^{F_2}$ &
   $a_0^{T_1}$ & $a_1^{T_1}$ & $a_2^{T_1}$ &
   $a_1^{T_2}$ & $a_2^{T_2}$ & $a_1^{T_0}$ & $a_2^{T_0}$ \\
\midrule
Mean
& 0.0147 & -0.021 & 0.027 & 0.0137 & -0.008 & 0.007
& 0.0023 & -0.034 & -0.018 & -0.021
& 0.0081 & -0.002 & -0.027
& -0.0074 &  -0.005 & -0.007 & 0.005 \\
Std.\ Dev.
& 0.0006 & \phantom{-}0.015 & 0.060 & 0.0008 & \phantom{-}0.019 & 0.073
& 0.0069 & \phantom{-}0.042 & \phantom{-}0.022 & \phantom{-}0.085
& 0.0004 & \phantom{-}0.011 & 0.049
& \phantom{-}0.0085 & 0.038 & \phantom{-}0.016 & 0.068 \\
\midrule
\multicolumn{18}{c}{Correlation Matrix} \\
\midrule
$a_0^{f}$   & 1.00 & -0.69 & 0.56 & 0.03 & -0.09 & 0.03 & -0.59 & 0.43 & -0.09 & 0.00 & 0.05 & -0.09 & 0.02 & -0.09 & 0.00 & -0.08 & 0.03 \\
$a_1^{f}$   &      & 1.00 & -0.90 & -0.11 & 0.23 & -0.09 & 0.51 & -0.34 & 0.19 & -0.02 & -0.10 & 0.18 & -0.03 & 0.18 & 0.00 & 0.17 & -0.05 \\
$a_2^{f}$   &      &      & 1.00 & 0.07 & -0.11 & 0.06 & -0.54 & 0.46 & -0.04 & -0.01 & 0.05 & -0.08 & 0.03 & -0.05 & -0.03 & -0.10 & 0.05 \\
$a_0^{g}$   &      &      &      & 1.00 & -0.67 & 0.47 & -0.09 & 0.06 & -0.01 & -0.05 & 0.04 & -0.06 & 0.02 & -0.05 & 0.00 & 0.01 & -0.04 \\
$a_1^{g}$   &      &      &      &      & 1.00 & -0.89 & 0.09 & -0.03 & 0.07 & 0.03 & -0.09 & 0.16 & -0.04 & 0.13 & 0.00 & 0.03 & 0.06 \\
$a_2^{g}$   &      &      &      &      &      & 1.00 & -0.06 & 0.05 & -0.01 & -0.01 & 0.05 & -0.06 & 0.03 & -0.03 & -0.01 & 0.02 & -0.05 \\
$a_1^{F_1}$ &      &      &      &      &      &      & 1.00 & -0.95 & 0.09 & -0.03 & -0.05 & 0.06 & -0.05 & 0.01 & 0.01 & 0.03 & -0.01 \\
$a_2^{F_1}$ &      &      &      &      &      &      &      & 1.00 & 0.11 & -0.03 & 0.01 & -0.01 & 0.04 & 0.04 & -0.02 & 0.02 & 0.00 \\
$a_1^{F_2}$ &      &      &      &      &      &      &      &      & 1.00 & -0.84 & -0.08 & 0.06 & 0.02 & 0.07 & 0.00 & 0.08 & -0.01 \\
$a_2^{F_2}$ &      &      &      &      &      &      &      &      &      & 1.00 & 0.00 & 0.03 & -0.05 & 0.03 & -0.02 & 0.00 & 0.00 \\
$a_0^{T_1}$ &      &      &      &      &      &      &      &      &      &      & 1.00 & -0.75 & 0.60 & -0.23 & 0.11 & -0.19 & 0.13 \\
$a_1^{T_1}$ &      &      &      &      &      &      &      &      &      &      &      & 1.00 & -0.92 & 0.45 & -0.24 & 0.22 & -0.12 \\
$a_2^{T_1}$ &      &      &      &      &      &      &      &      &      &      &      &      & 1.00 & -0.28 & 0.17 & -0.12 & 0.09 \\
$a_1^{T_2}$ &      &      &      &      &      &      &      &      &      &      &      &      &      & 1.00 & -0.90 & 0.47 & -0.31 \\
$a_2^{T_2}$ &      &      &      &      &      &      &      &      &      &      &      &      &      &      & 1.00 & -0.33 & 0.26 \\
$a_1^{T_0}$ &      &      &      &      &      &      &      &      &      &      &      &      &      &      &      & 1.00 & -0.90 \\
$a_2^{T_0}$ &      &      &      &      &      &      &      &      &      &      &      &      &      &      &      &      & 1.00 \\
\bottomrule
\end{tabular}
}
\caption{\footnotesize Posterior summaries for the BGL fit to all the $B_s\to \phi$ FFs obtained employing the LQCD+LCSR DM approach. The first row gives the posterior means, the second row the standard deviations, and the lower block shows the correlation matrix.}
\label{tab:BGL_BstoPhi_3}
\end{table}

\FloatBarrier

\section{Fit results for hadronic parameters}
\label{app:AppC}

In this Appendix we report additional details on the results of the SM fits for the
non-local hadronic parameters $h_\lambda^{(i)}$ discussed in
Sec.~\ref{subsec:SM_b2sll}, obtained within the \emph{Data Driven} approach. Table~\ref{tab:hlambda} summarizes the 68\% and 95\% HPDI of the
posterior distributions for these parameters, obtained using either the LQCD DM or the
LQCD+LCSR DM form factors. The corresponding correlation matrices among the hadronic
parameters are given in Table~\ref{tab:corr_dm} for the LQCD DM case and in
Table~\ref{tab:corr_dm_lcsr} for the LQCD+LCSR DM case.

\begin{table*}[!hp!]
\centering
\setlength{\tabcolsep}{10pt}
\renewcommand{\arraystretch}{1.33}
\begin{tabular}{|c|c|cc|}
\hline
\textbf{Hadronic parameter} & \textbf{Form Factors} & \textbf{68\% HPDI} & \textbf{95\% HPDI} \\
\hline
\multirow{2}{*}{$ {\rm Re} \, (h_0^{(0)} ) \times 10^4 $ [GeV$^{-1}$]}
& LQCD DM & $[-0.11, 6.92]$ & $[-2.94, 11.81]$ \\
& LQCD+LCSR DM & $[-0.94, 2.63]$ & $[-2.54, 4.71]$ \\
\hline

\multirow{2}{*}{$ {\rm Im} \, (h_0^{(0)} ) \times 10^4 $ [GeV$^{-1}$]}
& LQCD DM & {\color{red} $[-8.81, -0.20]$} & $[-12.78, 4.31]$ \\
& LQCD+LCSR DM & $[-6.07, 0.16]$ & $[-9.36, 3.13]$\\
\hline

\multirow{2}{*}{$ {\rm Re} \, (h_+^{(0)} )\times 10^4 $}
& LQCD DM & {\color{red} $[-0.96, -0.06]$} & $[-1.45, 0.41]$ \\
& LQCD+LCSR DM & {\color{red} $[-0.95, -0.01]$} & $[-1.43, 0.45]$ \\
\hline

\multirow{2}{*}{$ {\rm Im} \, (h_+^{(0)} )\times 10^4 $}
& LQCD DM  & $[-0.73, 0.23]$ & $[-1.21, 0.74]$ \\
& LQCD+LCSR DM  & $[-0.73, 0.26]$ & $[-1.20, 0.76]$ \\
\hline

\multirow{2}{*}{$ {\rm Re} \, (h_-^{(0)} )\equiv - {\rm Re} \, (\Delta C_{7})$}
& LQCD DM & $[-0.08, 0.21]$ & $[-0.12, 0.57]$ \\
& LQCD+LCSR DM & $[-0.03, 0.01]$ & $[-0.06, 0.03]$\\
\hline

\multirow{2}{*}{$ {\rm Im} \, (h_-^{(0)} ) $}
& LQCD DM & {\color{red} $[0.03, 0.20]$} & $[-0.03, 0.36]$ \\
& LQCD+LCSR DM & {\color{red} $[0.02, 0.12]$} & $[-0.03, 0.17]$ \\
\hline

\multirow{2}{*}{$ {\rm Re} \, (h_0^{(1)} )\times 10^5 $[GeV$^{-3}$]}
& LQCD DM & {\color{red} $[3.12, 11.77]$} & $[-0.24, 16.90]$ \\
& LQCD+LCSR DM & $[-0.37, 4.75]$ & $[-2.76, 7.71]$ \\
\hline

\multirow{2}{*}{$ {\rm Im} \, (h_0^{(1)} )\times 10^5 $[GeV$^{-3}$]}
& LQCD DM & $[-9.24, 3.07]$ & $[-13.89, 9.69]$ \\
& LQCD+LCSR DM & $[-5.66, 5.02]$ & $[-10.59, 9.63]$ \\
\hline

\multirow{2}{*}{$ {\rm Re} \, (h_+^{(1)} )\times 10^4 $[GeV$^{-2}$]}
& LQCD DM & {\color{red} $[0.65, 1.99]$} & $[-0.04, 2.68]$ \\
& LQCD+LCSR DM & $[-0.08, 1.17]$ & $[-0.69, 1.80]$ \\
\hline

\multirow{2}{*}{$ {\rm Im} \, (h_+^{(1)} )\times 10^4 $[GeV$^{-2}$]}
& LQCD DM & {\color{red} $[-2.54, -0.63]$} & $[-3.43, 0.29]$ \\
& LQCD+LCSR DM & {\color{red} $[-1.69, -0.12]$} & $[-2.42, 0.68]$ \\
\hline

\multirow{2}{*}{$ {\rm Re} \, (h_-^{(1)} )\equiv - {\rm Re} \, ( \Delta C_{9}) $}
& LQCD DM & $[-2.44, 0.53]$ & $[-4.63, 1.82]$ \\
& LQCD+LCSR DM & $[-0.12, 1.15]$ & $[-0.74, 1.82]$ \\
\hline

\multirow{2}{*}{$ {\rm Im} \, (h_-^{(1)} ) $}
& LQCD DM & $[-0.63, 3.83]$ & $[-2.69, 6.32]$ \\
& LQCD+LCSR DM & $[-0.40, 2.72]$ & $[-1.89, 4.38]$ \\
\hline

\multirow{2}{*}{$ {\rm Re} \, (h_+^{(2)} )\times 10^5 $[GeV$^{-4}$]}
& LQCD DM & $[-1.21, 1.14]$ & $[-2.41, 2.32]$ \\
& LCQD+LCSR DM & $[-1.43, 0.81]$ & $[-2.55, 1.95]$ \\
\hline

\multirow{2}{*}{$ {\rm Im} \, (h_+^{(2)} )\times 10^5 $[GeV$^{-4}$]}
& LQCD DM & $[-0.01, 3.84]$ & $[-1.85, 5.11]$ \\
& LQCD+LCSR DM & $[-0.50, 2.53]$ & $[-1.93, 3.96]$ \\
\hline

\multirow{2}{*}{$ {\rm Re} \, (h_-^{(2)} )\times 10^5 $[GeV$^{-4}$]}
& LQCD DM & {\color{red} $[0.90, 3.23]$} & $[-0.31, 4.49]$ \\
& LQCD+LCSR DM & {\color{red} $[0.59, 2.15]$} & $[-0.21, 2.92]$ \\
\hline

\multirow{2}{*}{$ {\rm Im} \, (h_-^{(2)} )\times 10^5 $[GeV$^{-4}$]}
& LQCD DM & {\color{red} $[-5.62, -0.56]$} & $[-7.48, 2.31]$ \\
& LQCD+LCSR DM & $[-3.58, 0.43]$ & $[-5.45, 2.32]$ \\
\hline

\multirow{2}{*}{$ {\rm Re} \, (h_K^{(1)} )\times 10^3 $}
& LQCD DM & $[-0.10, 5.40]$ & $[-2.61, 9.08]$ \\
& LCQD+LCSR DM & $[-1.64, 1.15]$ & $[-3.08, 2.50]$ \\
\hline

\multirow{2}{*}{$ {\rm Re} \, (h_K^{(2)} )\times 10^4 $[GeV$^{-2}$]}
& LQCD DM & {\color{red} $[5.97, 11.44]$} & {\color{red} $[3.36, 15.01]$} \\
& LQCD+LCSR DM & {\color{red} $[4.47, 7.99]$} & {\color{red} $[2.82, 9.89]$} \\
\hline
\end{tabular}
\caption{68\% and 95\% HPDI of the posterior distribution of the hadronic parameters $h_\lambda^{(i)}$. The \textcolor{red}{red} colour highlights ranges not including 0. Genuine hadronic effects encoded in ${\rm Re} \, (h_K^{(2)} )$ are found to be non-vanishing at the 2$\sigma$ level independently from the employed FFs.}
\label{tab:hlambda}
\end{table*}

\FloatBarrier

\begin{table*}[!t]
\centering
\renewcommand{\arraystretch}{1.1}
\setlength{\tabcolsep}{2.5pt}
\resizebox{\textwidth}{!}{%
\begin{tabular}{l|cccccccccccccccccc}
\toprule
& \rotatebox{90}{$\text{Re}\,(h_0^{(0)})$} & \rotatebox{90}{$\text{Im}\,(h_0^{(0)})$} & \rotatebox{90}{$\text{Re}\,(h_+^{(0)})$} & \rotatebox{90}{$\text{Im}\,(h_+^{(0)})$} & \rotatebox{90}{$\text{Re}\,(h_-^{(0)})$} & \rotatebox{90}{$\text{Im}\,(h_-^{(0)})$} & \rotatebox{90}{$\text{Re}\,(h_0^{(1)})$} & \rotatebox{90}{$\text{Im}\,(h_0^{(1)})$} & \rotatebox{90}{$\text{Re}\,(h_+^{(1)})$} & \rotatebox{90}{$\text{Im}\,(h_+^{(1)})$} & \rotatebox{90}{$\text{Re}\,(h_-^{(1)})$} & \rotatebox{90}{$\text{Im}\,(h_-^{(1)})$} & \rotatebox{90}{$\text{Re}\,(h_+^{(2)})$} & \rotatebox{90}{$\text{Im}\,(h_+^{(2)})$} & \rotatebox{90}{$\text{Re}\,(h_-^{(2)})$} & \rotatebox{90}{$\text{Im}\,(h_-^{(2)})$} & \rotatebox{90}{$\text{Re}\,(h_K^{(1)})$} & \rotatebox{90}{$\text{Re}\,(h_K^{(2)})$} \\
\midrule
$\text{Re}\,(h_0^{(0)})$  & 1.00 & -0.12 & 0.04 & 0.01 & 0.54 & 0.15 & -0.09 & 0.06 & 0.25 & -0.06 & -0.82 & 0.06 & -0.01 & 0.06 & 0.74 & -0.06 & 0.80 & 0.69 \\
$\text{Im}\,(h_0^{(0)})$  & -0.12 & 1.00 & 0.11 & -0.22 & 0.14 & 0.62 & -0.01 & 0.01 & -0.07 & 0.56 & 0.14 & -0.79 & -0.01 & -0.35 & -0.28 & 0.68 & -0.17 & -0.17 \\
$\text{Re}\,(h_+^{(0)})$  & 0.04 & 0.11 & 1.00 & -0.10 & 0.07 & 0.20 & 0.04 & -0.02 & -0.41 & -0.03 & -0.05 & -0.10 & 0.32 & 0.07 & 0.03 & 0.01 & 0.04 & 0.05 \\
$\text{Im}\,(h_+^{(0)})$  & 0.01 & -0.22 & -0.10 & 1.00 & 0.00 & -0.19 & -0.02 & -0.02 & 0.02 & -0.34 & -0.04 & 0.19 & -0.02 & 0.16 & 0.06 & -0.14 & 0.04 & 0.06 \\
$\text{Re}\,(h_-^{(0)})$  & 0.54 & 0.14 & 0.07 & 0.00 & 1.00 & 0.48 & 0.09 & 0.02 & 0.09 & 0.10 & -0.69 & -0.12 & -0.04 & -0.02 & 0.31 & 0.04 & 0.56 & 0.51 \\
$\text{Im}\,(h_-^{(0)})$  & 0.15 & 0.62 & 0.20 & -0.19 & 0.48 & 1.00 & 0.12 & 0.08 & 0.12 & -0.02 & -0.16 & -0.65 & 0.01 & 0.22 & -0.10 & 0.28 & 0.09 & 0.07 \\
$\text{Re}\,(h_0^{(1)})$  & -0.09 & -0.01 & 0.04 & -0.02 & 0.09 & 0.12 & 1.00 & -0.13 & -0.14 & -0.08 & -0.23 & 0.02 & 0.55 & 0.10 & 0.19 & -0.12 & 0.22 & 0.21 \\
$\text{Im}\,(h_0^{(1)})$  & 0.06 & 0.01 & -0.02 & -0.02 & 0.02 & 0.08 & -0.13 & 1.00 & 0.12 & -0.16 & 0.01 & -0.48 & -0.17 & 0.31 & -0.02 & 0.62 & -0.02 & 0.00 \\
$\text{Re}\,(h_+^{(1)})$  & 0.25 & -0.07 & -0.41 & 0.02 & 0.09 & 0.12 & -0.14 & 0.12 & 1.00 & -0.18 & -0.13 & -0.03 & -0.71 & 0.14 & 0.03 & 0.01 & 0.13 & 0.09 \\
$\text{Im}\,(h_+^{(1)})$  & -0.06 & 0.56 & -0.03 & -0.34 & 0.10 & -0.02 & -0.08 & -0.16 & -0.18 & 1.00 & 0.06 & -0.22 & -0.04 & -0.88 & -0.13 & 0.46 & -0.07 & -0.07 \\
$\text{Re}\,(h_-^{(1)})$  & -0.82 & 0.14 & -0.05 & -0.04 & -0.69 & -0.16 & -0.23 & 0.01 & -0.13 & 0.06 & 1.00 & -0.12 & -0.12 & -0.02 & -0.86 & 0.10 & -0.96 & -0.85 \\
$\text{Im}\,(h_-^{(1)})$  & 0.06 & -0.79 & -0.10 & 0.19 & -0.12 & -0.65 & 0.02 & -0.48 & -0.03 & -0.22 & -0.12 & 1.00 & 0.04 & 0.04 & 0.24 & -0.83 & 0.14 & 0.15 \\
$\text{Re}\,(h_+^{(2)})$  & -0.01 & -0.01 & 0.32 & -0.02 & -0.04 & 0.01 & 0.55 & -0.17 & -0.71 & -0.04 & -0.12 & 0.04 & 1.00 & 0.04 & 0.24 & -0.13 & 0.13 & 0.14 \\
$\text{Im}\,(h_+^{(2)})$  & 0.06 & -0.35 & 0.07 & 0.16 & -0.02 & 0.22 & 0.10 & 0.31 & 0.14 & -0.88 & -0.02 & 0.04 & 0.04 & 1.00 & 0.05 & -0.35 & 0.01 & 0.03 \\
$\text{Re}\,(h_-^{(2)})$  & 0.74 & -0.28 & 0.03 & 0.06 & 0.31 & -0.10 & 0.19 & -0.02 & 0.03 & -0.13 & -0.86 & 0.24 & 0.24 & 0.05 & 1.00 & -0.18 & 0.88 & 0.76 \\
$\text{Im}\,(h_-^{(2)})$  & -0.06 & 0.68 & 0.01 & -0.14 & 0.04 & 0.28 & -0.12 & 0.62 & 0.01 & 0.46 & 0.10 & -0.83 & -0.13 & -0.35 & -0.18 & 1.00 & -0.12 & -0.11 \\
$\text{Re}\,(h_K^{(1)})$  & 0.80 & -0.17 & 0.04 & 0.04 & 0.56 & 0.09 & 0.22 & -0.02 & 0.13 & -0.07 & -0.96 & 0.14 & 0.13 & 0.01 & 0.88 & -0.12 & 1.00 & 0.76 \\
$\text{Re}\,(h_K^{(2)})$  & 0.69 & -0.17 & 0.05 & 0.06 & 0.51 & 0.07 & 0.21 & 0.00 & 0.09 & -0.07 & -0.85 & 0.15 & 0.14 & 0.03 & 0.76 & -0.11 & 0.76 & 1.00 \\
\bottomrule
\end{tabular}}
\caption{Correlation matrix of the hadronic parameters for the LQCD DM \emph{Data Driven} fit.}
\label{tab:corr_dm}
\end{table*}

\begin{table*}[!t]
\centering
\renewcommand{\arraystretch}{1.1}
\setlength{\tabcolsep}{2.5pt}
\resizebox{\textwidth}{!}{%
\begin{tabular}{l|cccccccccccccccccc}
\toprule
& \rotatebox{90}{$\text{Re}\,(h_0^{(0)})$} & \rotatebox{90}{$\text{Im}\,(h_0^{(0)})$} & \rotatebox{90}{$\text{Re}\,(h_+^{(0)})$} & \rotatebox{90}{$\text{Im}\,(h_+^{(0)})$} & \rotatebox{90}{$\text{Re}\,(h_-^{(0)})$} & \rotatebox{90}{$\text{Im}\,(h_-^{(0)})$} & \rotatebox{90}{$\text{Re}\,(h_0^{(1)})$} & \rotatebox{90}{$\text{Im}\,(h_0^{(1)})$} & \rotatebox{90}{$\text{Re}\,(h_+^{(1)})$} & \rotatebox{90}{$\text{Im}\,(h_+^{(1)})$} & \rotatebox{90}{$\text{Re}\,(h_-^{(1)})$} & \rotatebox{90}{$\text{Im}\,(h_-^{(1)})$} & \rotatebox{90}{$\text{Re}\,(h_+^{(2)})$} & \rotatebox{90}{$\text{Im}\,(h_+^{(2)})$} & \rotatebox{90}{$\text{Re}\,(h_-^{(2)})$} & \rotatebox{90}{$\text{Im}\,(h_-^{(2)})$} & \rotatebox{90}{$\text{Re}\,(h_K^{(1)})$} & \rotatebox{90}{$\text{Re}\,(h_K^{(2)})$} \\
\midrule
$\text{Re}\,(h_0^{(0)})$  & 1.00 & -0.25 & 0.01 & 0.06 & 0.24 & -0.26 & -0.31 & 0.07 & -0.02 & -0.11 & -0.76 & 0.15 & 0.07 & 0.07 & 0.74 & -0.09 & 0.69 & 0.45 \\
$\text{Im}\,(h_0^{(0)})$  & -0.25 & 1.00 & 0.11 & -0.25 & -0.44 & 0.70 & -0.05 & -0.17 & -0.06 & 0.56 & 0.34 & -0.73 & -0.01 & -0.46 & -0.27 & 0.62 & -0.30 & -0.19 \\
$\text{Re}\,(h_+^{(0)})$  & 0.01 & 0.11 & 1.00 & -0.13 & -0.09 & 0.20 & -0.01 & 0.01 & -0.50 & -0.01 & 0.01 & -0.12 & 0.33 & 0.04 & 0.01 & 0.05 & 0.00 & 0.00 \\
$\text{Im}\,(h_+^{(0)})$  & 0.06 & -0.25 & -0.13 & 1.00 & 0.20 & -0.32 & 0.04 & -0.02 & 0.03 & -0.38 & -0.12 & 0.24 & 0.00 & 0.21 & 0.08 & -0.18 & 0.09 & 0.09 \\
$\text{Re}\,(h_-^{(0)})$  & 0.24 & -0.44 & -0.09 & 0.20 & 1.00 & -0.57 & 0.17 & -0.09 & 0.01 & 0.00 & -0.46 & 0.49 & 0.03 & -0.07 & 0.27 & -0.31 & 0.36 & 0.24 \\
$\text{Im}\,(h_-^{(0)})$  & -0.26 & 0.70 & 0.20 & -0.32 & -0.57 & 1.00 & -0.04 & 0.09 & 0.00 & 0.00 & 0.43 & -0.74 & -0.02 & 0.10 & -0.33 & 0.47 & -0.36 & -0.26 \\
$\text{Re}\,(h_0^{(1)})$  & -0.31 & -0.05 & -0.01 & 0.04 & 0.17 & -0.04 & 1.00 & -0.17 & -0.23 & -0.02 & -0.16 & 0.12 & 0.43 & 0.00 & 0.13 & -0.16 & 0.14 & 0.11 \\
$\text{Im}\,(h_0^{(1)})$  & 0.07 & -0.17 & 0.01 & -0.02 & -0.09 & 0.09 & -0.17 & 1.00 & 0.08 & -0.29 & 0.03 & -0.44 & -0.12 & 0.44 & -0.02 & 0.59 & -0.02 & -0.02 \\
$\text{Re}\,(h_+^{(1)})$  & -0.02 & -0.06 & -0.50 & 0.03 & 0.01 & 0.00 & -0.23 & 0.08 & 1.00 & -0.08 & 0.27 & -0.02 & -0.88 & 0.05 & -0.30 & 0.04 & -0.25 & -0.17 \\
$\text{Im}\,(h_+^{(1)})$  & -0.11 & 0.56 & -0.01 & -0.38 & 0.00 & 0.00 & -0.02 & -0.29 & -0.08 & 1.00 & 0.08 & -0.19 & 0.01 & -0.90 & -0.09 & 0.30 & -0.09 & -0.04 \\
$\text{Re}\,(h_-^{(1)})$  & -0.76 & 0.34 & 0.01 & -0.12 & -0.46 & 0.43 & -0.16 & 0.03 & 0.27 & 0.08 & 1.00 & -0.33 & -0.31 & -0.03 & -0.93 & 0.23 & -0.89 & -0.58 \\
$\text{Im}\,(h_-^{(1)})$  & 0.15 & -0.73 & -0.12 & 0.24 & 0.49 & -0.74 & 0.12 & -0.44 & -0.02 & -0.19 & -0.33 & 1.00 & 0.06 & 0.10 & 0.25 & -0.89 & 0.28 & 0.19 \\
$\text{Re}\,(h_+^{(2)})$  & 0.07 & -0.01 & 0.33 & 0.00 & 0.03 & -0.02 & 0.43 & -0.12 & -0.88 & 0.01 & -0.31 & 0.06 & 1.00 & 0.00 & 0.38 & -0.12 & 0.29 & 0.20 \\
$\text{Im}\,(h_+^{(2)})$  & 0.07 & -0.46 & 0.04 & 0.21 & -0.07 & 0.10 & 0.00 & 0.44 & 0.05 & -0.90 & -0.03 & 0.10 & 0.00 & 1.00 & 0.05 & -0.23 & 0.04 & 0.02 \\
$\text{Re}\,(h_-^{(2)})$  & 0.74 & -0.27 & 0.01 & 0.08 & 0.27 & -0.33 & 0.13 & -0.02 & -0.30 & -0.09 & -0.93 & 0.25 & 0.38 & 0.05 & 1.00 & -0.19 & 0.84 & 0.54 \\
$\text{Im}\,(h_-^{(2)})$  & -0.09 & 0.62 & 0.05 & -0.18 & -0.31 & 0.47 & -0.16 & 0.59 & 0.04 & 0.30 & 0.23 & -0.89 & -0.12 & -0.23 & -0.19 & 1.00 & -0.20 & -0.13 \\
$\text{Re}\,(h_K^{(1)})$  & 0.69 & -0.30 & 0.00 & 0.09 & 0.36 & -0.36 & 0.14 & -0.02 & -0.25 & -0.09 & -0.89 & 0.28 & 0.29 & 0.04 & 0.84 & -0.20 & 1.00 & 0.30 \\
$\text{Re}\,(h_K^{(2)})$  & 0.45 & -0.19 & 0.00 & 0.09 & 0.24 & -0.26 & 0.11 & -0.02 & -0.17 & -0.04 & -0.58 & 0.19 & 0.20 & 0.02 & 0.54 & -0.13 & 0.30 & 1.00 \\
\bottomrule
\end{tabular}}
\caption{Correlation matrix of the hadronic parameters for the LQCD+LCSR DM \emph{Data Driven} fit.}
\label{tab:corr_dm_lcsr}
\end{table*}

\section{NP fit results for extended NP basis}
\label{app:AppD}

\begin{figure}[!t!]
    \centering
    \includegraphics[width=0.98\linewidth]{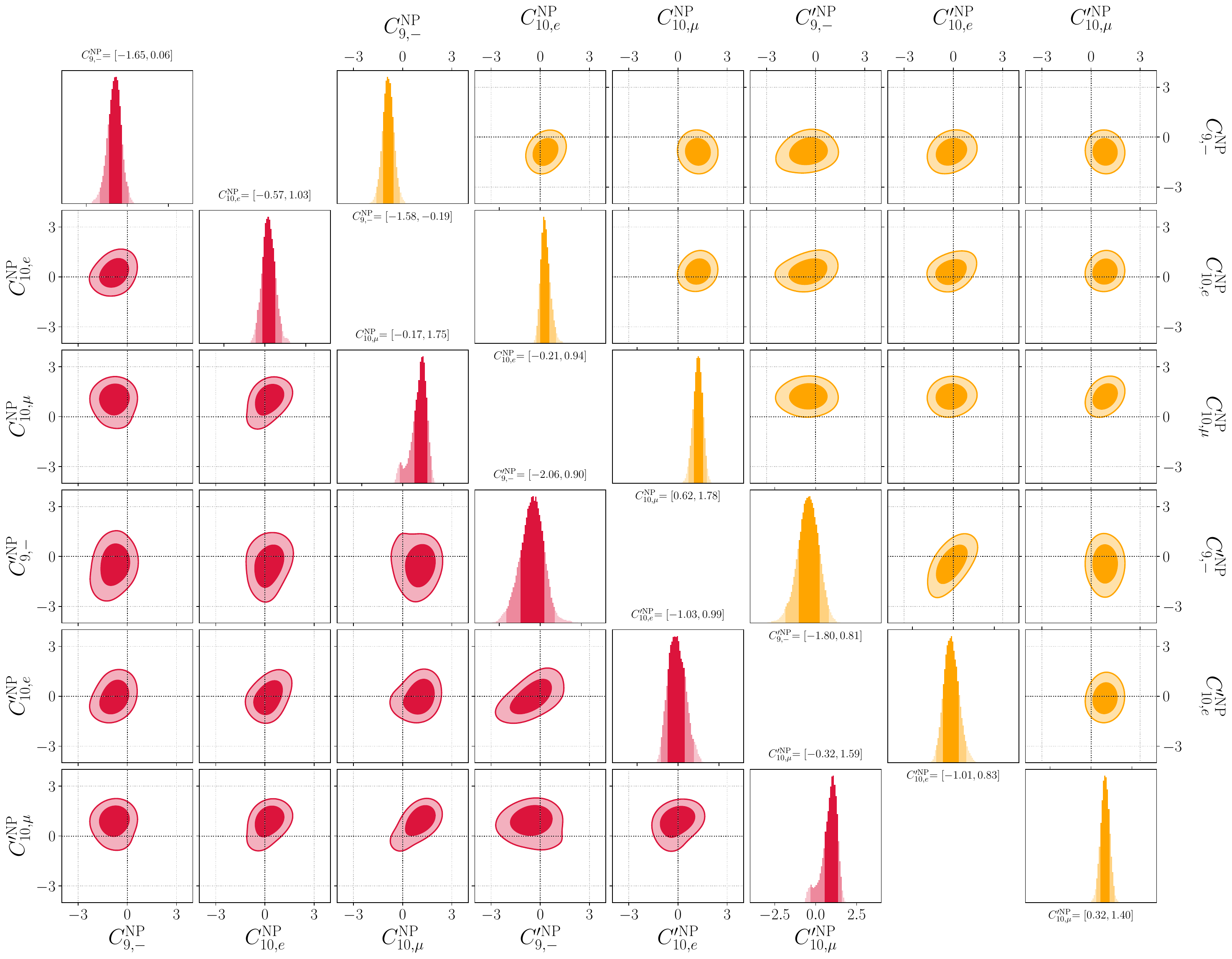}
    \caption{Two- and one-dimensional marginalized joint p.d.f. for the set of  6 WCs $C_{9,-}^{\rm NP}$, $C_{10,e}^{\rm NP}$, $C_{10,\mu}^{\rm NP}$, $C_{9,-}^{\prime\ \rm NP}$, $C_{10,e}^{\prime\ \rm NP}$ and $C_{10,\mu}^{\prime\ \rm NP}$ obtained from a fit employing the \emph{Data Driven} approach concerning non-local hadronic effects (see text for details). Results concerning the $C_{9,+}^{\rm NP}$ and $C_{9,+}^{\prime\ \rm NP}$ WCs are not reported since these coefficients are flatly distributed. The results obtained employing LQCD DM FFs are shown in red, while the ones obtained with LQCD+LCSR DM FFs are shown in orange. For both scenarios, we show the 68\% and 95\% probability regions and we quote 95\% probability regions numbers.}
    \label{fig:8D_FDD}
\end{figure}

\begin{figure}[!t!]
    \centering
    \includegraphics[width=0.98\linewidth]{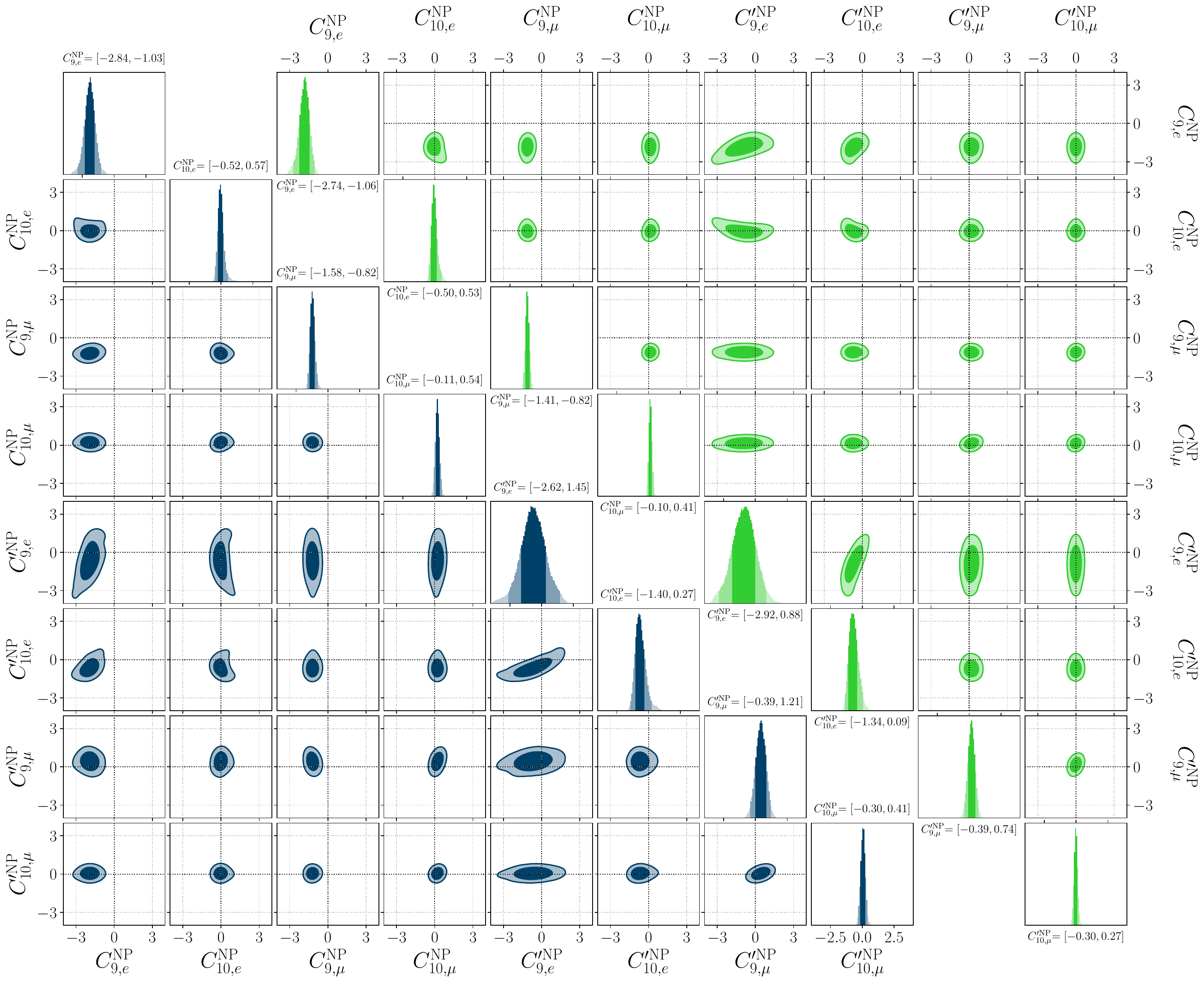}
    \caption{Two- and one-dimensional marginalized joint p.d.f. for the set of 8 WCs $C^{(\prime)\ \rm NP}_{i,j}$ with $i=9,10$ and $j=e,\mu$, obtained from a fit employing the \emph{Model Dependent} approach concerning non-local hadronic effects (see text for details). The results obtained employing LQCD DM FFs are shown in blue, while the ones obtained with LQCD+LCSR DM FFs are shown in green. For both scenarios, we show the 68\% and 95\% probability regions and we quote 95\% probability regions numbers.}
    \label{fig:8D_PMD}
\end{figure}

The results of the extended NP fits are presented in
Tables~\ref{tab:8WC_FDD} and~\ref{tab:8WC_PMD}, with the corresponding marginalized
distributions shown in Figs.~\ref{fig:8D_FDD} and~\ref{fig:8D_PMD}.
Compared to the case studied in Sec.~\ref{subsec:NP_b2sll}, this scenarios introduce additional freedom through the
inclusion of right-handed operators, allowing for a more general description of
potential NP effects.

\begin{table*}[!b!]
\centering
\setlength{\tabcolsep}{10pt}
\renewcommand{\arraystretch}{1.35}
\begin{tabular}{|c|c|cc|}
\hline
\textbf{Wilson coefficient} & \textbf{Form Factors} & \textbf{68\% HPDI} & \textbf{95\% HPDI} \\
\hline

\multirow{2}{*}{$C_{9,-}^{\rm NP}$}
& LQCD DM & $[-1.16, -0.34]$ & $[-1.65, 0.06]$ \\
& LQCD+LCSR DM & $[-1.22, -0.55]$ & $[-1.58, -0.19]$ \\
\hline

\multirow{2}{*}{$C_{10,e}^{\rm NP}$}
& LQCD DM & $[-0.16, 0.62]$ & $[-0.57, 1.03]$ \\
& LQCD+LCSR DM & $[0.00, 0.55]$ & $[-0.21, 0.94]$ \\
\hline

\multirow{2}{*}{$C_{10,\mu}^{\rm NP}$}
& LQCD DM & $[0.73, 1.56]$ & $[-0.17, 1.75]$ \\
& LQCD+LCSR DM & $[0.93, 1.51]$ & $[0.62, 1.78]$ \\
\hline

\multirow{2}{*}{$C_{9,-}^{\prime\ \rm NP}$}
& LQCD DM & $[-1.18, 0.51]$ & $[-2.39, 1.57]$ \\
& LQCD+LCSR DM & $[-1.17, 0.25]$ & $[-2.06, 0.90]$ \\
\hline

\multirow{2}{*}{$C_{10,e}^{\prime\ \rm NP}$}
& LQCD DM & $[-0.48, 0.67]$ & $[-1.06, 1.18]$ \\
& LQCD+LCSR DM & $[-0.62, 0.43]$ & $[-1.03, 0.99]$ \\
\hline

\multirow{2}{*}{$C_{10,\mu}^{\prime\ \rm NP}$}
& LQCD DM & $[0.62, 1.34]$ & $[0.09, 1.64]$ \\
& LQCD+LCSR DM & $[0.57, 1.38]$ & $[-0.32, 1.59]$ \\
\hline

\end{tabular}
\caption{68\% and 95\% HPDI of the posterior distribution of the WCs $C_{9,-}^{\rm NP}$, $C_{10,e}^{\rm NP}$, $C_{10,\mu}^{\rm NP}$, $C_{9,-}^{\prime\ \rm NP}$, $C_{10,e}^{\prime\ \rm NP}$ and $C_{10,\mu}^{\prime\ \rm NP}$ obtained from a fit employing the \emph{Data Driven} approach concerning non-local hadronic effects (see text for details). These results have been obtained employing either the FFs based on Lattice results only (LQCD DM), or the ones obtained when also LCSR input are included in the FF determination (LQCD+LCSR DM). Results concerning the $C_{9,+}^{\rm NP}$ and $C_{9,+}^{\prime\ \rm NP}$ WCs are not reported since these coefficients are flatly distributed.}
\label{tab:8WC_FDD}
\end{table*}

\begin{table*}[!b!]
\centering
\setlength{\tabcolsep}{10pt}
\renewcommand{\arraystretch}{1.35}
\begin{tabular}{|c|c|cc|}
\hline
\textbf{Wilson coefficient} & \textbf{Form Factors} & \textbf{68\% HPDI} & \textbf{95\% HPDI} \\
\hline

\multirow{2}{*}{$C_{9,e}^{\rm NP}$}
& LQCD DM & $[-2.30, -1.48]$ & $[-2.84, -1.03]$ \\
& LQCD+LCSR DM & $[-2.23, -1.44]$ & $[-2.74, -1.06]$ \\
\hline

\multirow{2}{*}{$C_{10,e}^{\rm NP}$}
& LQCD DM & $[-0.27, 0.18]$ & $[-0.52, 0.57]$ \\
& LQCD+LCSR DM & $[-0.29, 0.15]$ & $[-0.50, 0.53]$ \\
\hline

\multirow{2}{*}{$C_{9,\mu}^{\rm NP}$}
& LQCD DM & $[-1.41, -1.04]$ & $[-1.58, -0.82]$ \\
& LQCD+LCSR DM & $[-1.27, -0.98]$ & $[-1.41, -0.82]$ \\
\hline

\multirow{2}{*}{$C_{10,\mu}^{\rm NP}$}
& LQCD DM & $[0.05, 0.37]$ & $[-0.11, 0.54]$ \\
& LQCD+LCSR DM & $[0.01, 0.27]$ & $[-0.10, 0.41]$ \\
\hline

\multirow{2}{*}{$C_{9,e}^{\prime\ \rm NP}$}
& LQCD DM & $[-1.59, 0.33]$ & $[-2.62, 1.45]$ \\
& LQCD+LCSR DM & $[-1.83, 0.02]$ & $[-2.92, 0.88]$ \\
\hline

\multirow{2}{*}{$C_{10,e}^{\prime\ \rm NP}$}
& LQCD DM & $[-1.08, -0.31]$ & $[-1.40, 0.27]$\\
& LQCD+LCSR DM & $[-1.11, -0.38]$ & $[-1.34, 0.09]$ \\
\hline

\multirow{2}{*}{$C_{9,\mu}^{\prime\ \rm NP}$}
& LQCD DM & $[0.04, 0.85]$ & $[-0.39, 1.21]$ \\
& LQCD+LCSR DM & $[-0.10, 0.45]$ & $[-0.39, 0.74]$ \\
\hline

\multirow{2}{*}{$C_{10,\mu}^{\prime\ \rm NP}$}
& LQCD DM & $[-0.14, 0.23]$ & $[-0.30, 0.41]$ \\
& LQCD+LCSR DM & $[-0.16, 0.12]$ & $[-0.30, 0.27]$ \\
\hline

\end{tabular}
\caption{68\% and 95\% HPDI of the posterior distribution of the WCs $C_{9,e}^{\rm NP}$, $C_{10,e}^{\rm NP}$, $C_{9,\mu}^{\rm NP}$, $C_{10,\mu}^{\rm NP}$, $C_{9,e}^{\prime\ \rm NP}$, $C_{10,e}^{\prime\ \rm NP}$, $C_{9,\mu}^{\prime\ \rm NP}$ and $C_{10,\mu}^{\prime\ \rm NP}$ obtained from a fit employing the \emph{Model Dependent} approach concerning non-local hadronic effects (see text for details). These results have been obtained employing either the FFs based on Lattice results only (LQCD DM), or the ones obtained when also LCSR input are included in the FF determination (LQCD+LCSR DM). }
\label{tab:8WC_PMD}
\end{table*}

\bibliography{hepbiblio}

\end{document}